\documentclass[longauth]{aa} 

\usepackage{subcaption}
\usepackage{graphicx}
\usepackage{txfonts}
\usepackage[colorlinks=true,citecolor=blue]{hyperref}
\defcitealias{Lubin_HD191939}{L22}
\defcitealias{Mariona_2020}{BA20}

\DeclareUnicodeCharacter{2212}{-}

\begin{document}

   \title{HD 191939 revisited: New and refined planet mass determinations, and a new planet in the habitable zone.}

   \titlerunning{HD 191939 revisited}
   

   \author{
J.~Orell-Miquel\inst{\ref{ins:iac},\ref{ins:ull}} \and
G.~Nowak\inst{\ref{ins:iac},\ref{ins:ull}} \and
F.~Murgas\inst{\ref{ins:iac},\ref{ins:ull}} \and
E.~Palle\inst{\ref{ins:iac},\ref{ins:ull}} \and
G.~Morello\inst{\ref{ins:iac},\ref{ins:ull}} \and
R.~Luque\inst{\ref{ins:IAA},\ref{ins:Chicago}} \and
M.~Badenas-Agusti\inst{\ref{ins:MIT},\ref{ins:MIT2}} \and
I.~Ribas\inst{\ref{ins:ICE},\ref{ins:IEEC}} \and
M.~Lafarga\inst{\ref{ins:warwick}} \and
N.~Espinoza\inst{\ref{ins:Heidelberg}} \and
J.\,C.~Morales\inst{\ref{ins:ICE},\ref{ins:IEEC}} \and
M.~Zechmeister\inst{\ref{ins:Gotingen}} \and
A.~Alqasim\inst{\ref{ins:VV}} \and
W.\,D.~Cochran\inst{\ref{ins:Mcdonald},\ref{ins:austin}} \and
D.~Gandolfi\inst{\ref{ins:italia}}\and
E.~Goffo\inst{\ref{ins:italia},\ref{ins:tautenburg}} \and
P.~Kab\'{a}th\inst{\ref{ins:CR}} \and
J.~Korth\inst{\ref{Chalmers}} \and
K.\,W.\,F.~Lam\inst{\ref{ins:Berlin}} \and
J.~Livingston\inst{\ref{ins:japan1},\ref{ins:japan2},\ref{ins:japan3}} \and
A.~Muresan\inst{\ref{ins:gothernburg}} \and
C.\,M.~Persson\inst{\ref{ins:Sweden}} \and
V.~Van Eylen\inst{\ref{ins:VV}}
}

   \institute{
\label{ins:iac}Instituto de Astrof\'isica de Canarias (IAC), 38205 La Laguna, Tenerife, Spain\\
\email{jom@iac.es}
\and
\label{ins:ull}Departamento de Astrof\'isica, Universidad de La Laguna (ULL), 38206 La Laguna, Tenerife, Spain.
\and
\label{ins:IAA}Instituto de Astrof\'isica de Andaluc\'ia (IAA-CSIC), Glorieta de la Astronom\'ia s/n, 18008 Granada, Spain.
\and
\label{ins:Chicago}Department of Astronomy \& Astrophysics, University of Chicago, Chicago, IL 60637, USA.
\and
\label{ins:MIT}Department of Earth, Atmospheric and Planetary Sciences, Massachusetts Institute of Technology, Cambridge, MA 02139, USA.
\and
\label{ins:MIT2}Department of Physics, and Kavli Institute for Astrophysics and Space Research, Massachusetts Institute of Technology, Cambridge, MA 02139, USA.
\and
\label{ins:ICE}Institut de Ci\`encies de l’Espai (CSIC), Campus UAB, c/ de Can Magrans s/n, E-08193 Bellaterra, Barcelona, Spain.
\and
\label{ins:IEEC}Institut d’Estudis Espacials de Catalunya, E-08034 Barcelona, Spain.
\and
\label{ins:warwick}Department of Physics, University of Warwick, Gibbet Hill Road, Coventry CV4 7AL, United Kingdom.
\and
\label{ins:Heidelberg}Max-Planck-Institut f\"ur Astronomie, K\"onigstuhl 17, 69117 Heidelberg, Germany.
\and
\label{ins:Gotingen}Institut f\"ur Astrophysik, Georg-August-Universit\"at, Friedrich-Hund-Platz 1, 37077 G\"ottingen, Germany.
\and
\label{ins:Mcdonald}McDonald Observatory and Center for Planetary Systems Habitability.
\and
\label{ins:austin}The University of Texas, Austin, Texas, USA.
\and
\label{ins:italia}Dipartimento di Fisica, Universitá di Torino, via P. Giuria 1, I-10125 Torino, Italy.
\and
\label{ins:tautenburg}Thüringer Landessternwarte Tautenburg, Sternwarte 5, D-07778 Tautenburg, Germany.
\and
\label{ins:CR}Astronomical Institute of the Czech Academy of Sciences, Fri\v{c}ova
298, 25165, Ond\v{r}ejov, Czech Republic.
\and
\label{Chalmers}Department of Space, Earth and Environment, Astronomy and Plasma Physics, Chalmers University of Technology, Chalmersplatsen 4, 412 96 Gothenburg, Sweden.
\and
\label{ins:Berlin}Institute of Planetary Research, German Aerospace Center (DLR), Rutherfordstrasse 2, D-12489 Berlin, Germany.
\and
\label{ins:japan1}Astrobiology Center, 2-21-1 Osawa, Mitaka, Tokyo 181-8588, Japan.
\and
\label{ins:japan2}National Astronomical Observatory of Japan, 2-21-1 Osawa, Mitaka, Tokyo 181-8588, Japan.
\and
\label{ins:japan3}Department of Astronomy, The Graduate University for Advanced Studies (SOKENDAI), 2-21-1 Osawa, Mitaka, Tokyo, Japan.
\and
\label{ins:gothernburg}Department of Space, Earth and Environment, Astronomy and Plasma Physics, Chalmers University of Technology, 412 96 Gothenburg, Sweden.
\and
\label{ins:Sweden}Department of Space, Earth and Environment, Chalmers University of Technology, Onsala Space Observatory, SE-439 92 Onsala, Sweden.
\and
\label{ins:VV}Mullard Space Science Laboratory, University College London, Holmbury St Mary, Dorking, Surrey RH5 6NT, UK.
}

   \date{Received 27 May 2022 / Accepted 16 October 2022}

 
  \abstract
  {HD 191939 (TOI-1339) is a nearby (d\,$=$\,54\,pc), bright (V\,$=$\,9\,mag), and inactive Sun-like star (G9\,V) known to host a multi-planet transiting system. Ground-based spectroscopic observations confirmed the planetary nature of the three transiting sub-Neptunes (HD 191939\,b, c, and d) originally detected by \textit{TESS} and were used to measure the masses for planets b and c with $3\sigma$ precision. These previous observations also reported the discovery of an additional Saturn-mass planet (HD 191939\,e) and evidence for a further, very long-period companion (HD 191939\,f). Here, we report the discovery of a new non-transiting planet in the system and a refined mass determination of HD 191939\,d. The new planet, HD 191939\,g, has a minimum mass of 13.5$\pm$2.0\,$M_\oplus$ and a period of about 280\,d. This period places the planet within the conservative habitable zone of the host star, and near a 1:3 resonance with HD 191939\,e.
  The compilation of 362 radial velocity measurements with a baseline of 677 days from four different high-resolution spectrographs also allowed us to refine the properties of the previously known planets, including a $4.6\sigma$ mass determination for planet d, for which only a $2\sigma$ upper limit had been set until now. We confirm the previously suspected low density of HD 191939\,d, which makes it an attractive target for attempting atmospheric characterisation. 
  Overall, the planetary system consists of three sub-Neptunes interior to a Saturn-mass and a Uranus-mass planet plus a high-mass long-period companion. This particular configuration has no counterpart in the literature and makes HD 191939 an exceptional multi-planet transiting system with an unusual planet demographic worthy of future observation.
  } 


   \maketitle
   
%

\section{Introduction}
\label{sect: INTRODUCTION}

\begin{table}
\caption{\label{table - stellar parameters} Stellar parameters of HD 191939.}
\centering
\resizebox{\columnwidth}{!}{
\begin{tabular}{lcr}
\hline
\hline
\noalign{\smallskip} 
Parameter & Value & Reference \\
\hline
\noalign{\smallskip} 

Name & HD 191939 & HD \\
 & TIC 269701147 & \textit{TESS}\\
 & TOI-1339 & TOI \\
 & HIP 99175 & HIP \\

\hline

\noalign{\smallskip} 
\multicolumn{3}{c}{ \textit{Coordinates and spectral type}} \\
\noalign{\smallskip} 

$\alpha$\,(J2000) & 20$^\mathrm{h}$\,08$^\mathrm{m}$\,05$^\mathrm{s}$.755 & \textit{Gaia} \\
$\delta$\,(J2000) & $+$66º\,51'\,02".077 & \textit{Gaia} \\
Spectral type& G9\,V & L22 \\

\hline

\noalign{\smallskip} 
\multicolumn{3}{c}{ \textit{Parallax and kinematics}} \\
\noalign{\smallskip} 

$\mu_\alpha$ ~[mas\,yr$^{-1}$] & 150.256 $\pm$ 0.044 & \textit{Gaia} \\
$\mu_\delta$ ~[mas\,yr$^{-1}$] & $-$63.909 $\pm$ 0.047 & \textit{Gaia} \\
Parallax ~[mas] &  18.706 $\pm$ 0.071 & \textit{Gaia} \\
Distance ~[pc] &  53.48$\pm$ $−$0.20 & BA20 \\
$V_{\mathrm{r}}$ ~[$\mathrm{km\,s^{-1}}$] & $-$9.266 $\pm$ 0.002 & \textit{Gaia} \\

\hline

\noalign{\smallskip} 
\multicolumn{3}{c}{ \textit{Magnitudes}} \\
\noalign{\smallskip} 

$B$ ~[mag] & 9.720 $\pm$ 0.038 & TYC \\
$V$ ~[mag] &  8.97 $\pm$ 0.03  & HIP \\
$G$ ~[mag] & 8.7748 $\pm$ 0.0002 & \textit{Gaia} \\ 
$T$ ~[mag] &  8.292 $\pm$ 0.006 & \textit{TESS} \\
$J$ ~[mag] & 7.597 $\pm$ 0.029 & 2MASS \\
$H$ ~[mag] & 7.215 $\pm$ 0.023 & 2MASS \\
$K$ ~[mag] & 7.180 $\pm$ 0.021 & 2MASS \\

\hline
\noalign{\smallskip} 
\multicolumn{3}{c}{ \textit{Stellar parameters}} \\
\noalign{\smallskip} 

Radius ~$R_{\star}$ [$R_{\odot}$] & 0.94 $\pm$ 0.02 & BA20, L22\\

Mass ~$M_{\star}$ [$M_{\odot}$] & 0.92 $\pm$ 0.06 & BA20 \\
                                & 0.81 $\pm$ 0.04 & L22 \\

$\rho_{\star}$ ~[g\,cm$^{-3}$] &  1.56 $\pm$ 0.15 & BA20 \\
                              &  1.37 $\pm$ 0.11 & L22 \\

$L_{\star}$ ~[$L_{\odot}$] & 0.69 $\pm$ 0.01 & BA20 \\
                                      & 0.65 $\pm$  0.02 & L22 \\

$T_{\mathrm{eff}}$ ~[K] &  5427 $\pm$ 50 & BA20 \\
                       &  5348 $\pm$ 100 & L22 \\

$\log( g\,[$cm\,s$^{-2}] )$ & 4.3 $\pm$ 0.1 & L22 \\

Metallicity ~[Fe/H] & $-$0.15 $\pm$ 0.06 & L22 \\

Age ~[Gyr] &  7 $\pm$ 3 & BA20 \\

$\log( $R$^{'}_{HK} )$ & $-$5.11 $\pm$ 0.05 & L22 \\

$v\,\sin{i}$ ~[km\,s$^{-1}$] &  0.6 $\pm$ 0.5 & BA20 \\
                            &  $<$2.0 & L22 \\

\noalign{\smallskip} 
\hline

\end{tabular}}

\tablebib{
HD: \citet{HD_Catalog};
\textit{TESS}: \citet{TIC_Catalog_Stassun2018};
TOI: \citet{Guerrero_2021};
HIP: \citet{HIP_Catalog};
\textit{Gaia}: \citet{GaiaDR2}, \citet{GAIA_RV_2018};
L22: \citet{Lubin_HD191939};
BA20: \citet{Mariona_2020};
TYC: \citet{TYC_Catalog};
2MASS: \cite{2MASS_Catalog}.
Parallax is corrected for a systematic
offset of $+$0.082$\pm$0.033\,mas, as described in \citet{offset_DR3}.}

\end{table}

The \textit{Transiting Exoplanet Survey Satellite} (\textit{TESS}; \citealp{TESS_Ricker}) is a NASA-sponsored space telescope whose original mission was a two-year full-sky survey searching for transiting planets. One of \textit{TESS} main scientific goals is to look for small planets ($R_\mathrm{P} < 4\,R_\oplus$) around bright stars suitable for radial velocity (RV) follow up and atmospheric characterisation. Since the beginning of operations in 2018, \textit{TESS} has discovered several multi-planetary transiting systems around bright host stars (e.g. \citealp{Dragomir_2019, Gunther_2019, Quinn_2019, Rafa_2021}). This type of system is an excellent laboratory for planetary astrophysics. Multi-planet systems share the same initial conditions (e.g. protoplanetary disc), allowing for comparison between sibling planets, and also the testing of planet formation and evolution processes and theories.

Here, we focus on the HD 191939 system, one of those multi-planetary transiting systems with small planets discovered by \textit{TESS}. HD 191939 is a bright (V\,=\,9\,mag), nearby (d\,=\,54\,pc), inactive solar-like star (G9\,V). \citet{Mariona_2020}, hereafter \citetalias{Mariona_2020}; confirmed the presence of three transiting sub-Neptune-sized planets, HD 191939\,b, c, and d, with periods of 8.88, 28.58, and 38.35\,d, respectively. Their radii are very similar, ranging 3.16--3.42\,$R_\oplus$. The brightness of the host star made long-term RV monitoring campaigns with different high-resolution spectroscopy facilities feasible, allowing the determination of the planetary masses. The physical properties of the planetary system were studied by \citet{Lubin_HD191939}, hereafter \citetalias{Lubin_HD191939}; finding masses of 10.4$\pm$0.9\,$M_\oplus$ and 7.2$\pm$1.4\,$M_\oplus$ for planets b and c, respectively. \citetalias{Lubin_HD191939} only presented an upper limit of 5.8\,$M_\oplus$ ($2\sigma$ confidence) for planet d. Furthermore, \citetalias{Lubin_HD191939} found evidence for two extra planets via RV measurements: planets e and f. HD 191939\,e has a period of 101\,d and a minimum mass of 108$\pm$3\,$M_\oplus$. The long-term trend showed by the RVs was related to the presence of HD 191939\,f, a high-mass planet with an unconstrained period of between 1700 and 7200\,d. In Table\,\ref{table - stellar parameters}, we compile a comprehensive list of HD 191939 stellar properties from the literature.

In this work, we combined the previously published RV observations with new time series obtained with the CARMENES and HARPS-N spectrographs, and with the existing \textit{TESS} photometric data for this target. In particular, the combination of four RV datasets allowed us to refine the physical properties of the system and all the planetary masses. Furthermore, the significant increase in the number of observations and its baseline permitted a check for extra planetary signals beyond the period of planet e. This is also the case for the newly discovered HD 191939\,g: a Uranus-mass planet in the habitable zone with an orbital period of $\sim$300\,d.

\section{Observations}

\subsection{\textit{TESS} photometry}
\label{sect: TESS photometry}

Listed as TIC 269701147 in the \textit{TESS} Input Catalog (TIC; \citealp{TIC_Catalog_Stassun2018}), HD 191939 was observed by \textit{TESS} in two-minute short-cadence integrations in Sectors\,15--19 from 15 August 2019 to 24 December 2019, Sectors\,21--22 from 21 January 2020 to 18 March 2020, Sectors\,24--25 from 16 April 2020 to 8 June 2020, Sector\,41 from 23 July 2021 to 20 August 2021, and Sector\,48 from 28 January 2022 to 26 February 2022.

In this work, we made use of the Presearch Data Conditioning-corrected simple aperture photometry (PDC-SAP; \citealp{PDC_1, stumpe_2012, PDC_2, SAP}) reduced by the Science Processing Operations Center (SPOC; \citealp{SPOC}) at the NASA Ames Research Center and publicly available at the Mikulski Archive for Space Telescopes (MAST\footnote{\url{https://mast.stsci.edu/portal/Mashup/Clients/Mast/Portal.html}}).

To remove some additional variability present in the PDC-SAP and to save computational time fitting the 11 \textit{TESS} sectors, we detrended the PDC-SAP light curves, with the planetary transits masked, performing a Gaussian process (GP) regression model using a Matern kernel 3/2 from \texttt{celerite} \citep{celerite}. The model only considered a relative flux offset and a jitter term, and two GP hyperparameters, which are shared between the different \textit{TESS} sectors. The priors and posteriors from the detrended process are shown in Table\,\ref{table - TESS GP}. We obtained the detrending model by evaluating the GP component at each time point, which includes the transit times. Finally, we divided the PDC-SAP light curve by the detrending model.

\begin{figure*}
\includegraphics[width=1\linewidth]{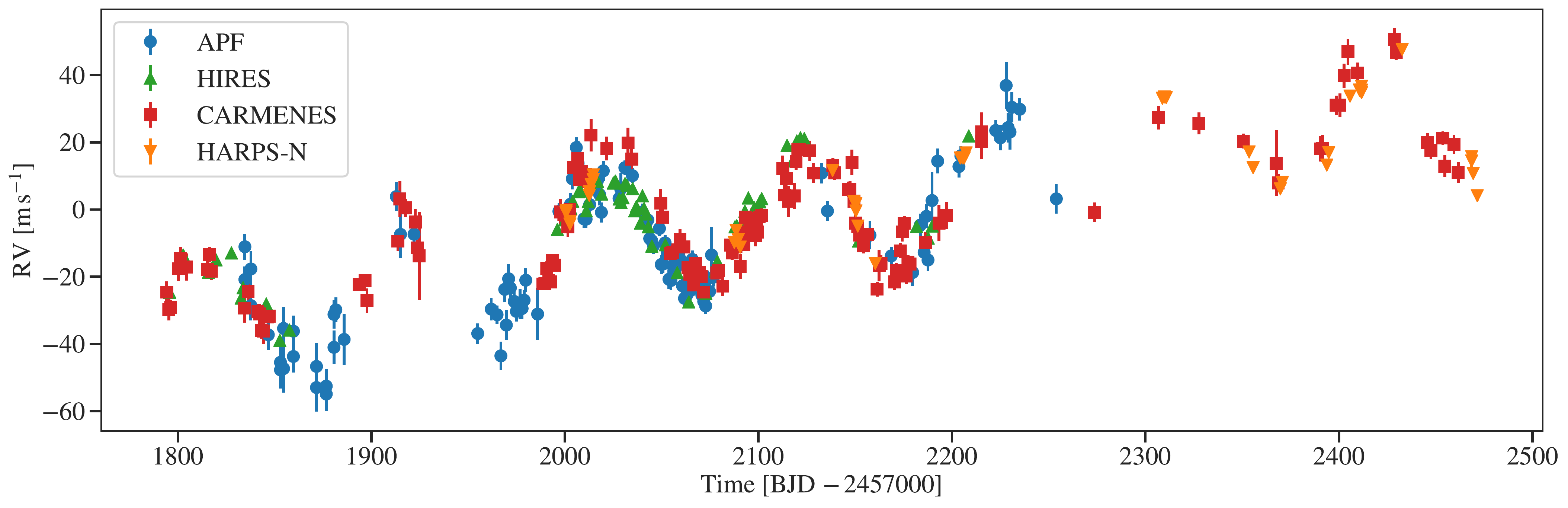}
\caption{\label{fig: RV DATA}
Time series of RV measurements taken by APF (blue circles), HIRES (green up triangles), CARMENES (red squares), and HARPS-N (orange down triangles).}
\end{figure*}

\subsection{High-resolution spectroscopy follow up}
\label{sect: RV observations}

\subsubsection{High-resolution spectroscopy with CARMENES}
\label{sect: CARMENES}

HD 191939 was observed with the Calar Alto high-Resolution search for M dwarfs with Exoearths with Near-infrared and optical \'Echelle Spectrographs (CARMENES; \citealp{Quirrenbach_2014, Quirrenbach_2020}) located at the Calar Alto Observatory, Almer\'ia, Spain. CARMENES has two spectral channels: the optical channel (VIS), which covers the wavelength range from 0.52 to 0.96\,$\mu$m with a resolving power of $\mathcal{R}$\,=\,94\,600, and the near-infrared
(NIR) channel, which goes from 0.96 to 1.71\,$\mu$m with a resolving power of $\mathcal{R}$\,=\,80\,400. The star was monitored from 6 November 2019 to 4 September 2021. During this time, we obtained 138 high-resolution spectra.

The observations were carried out as part of observing programs F19-3.5-014, S20-3.5-011 (PI: Nowak), and F20-3.5-013 (PI: Luque). The exposure times were set to 900\,s, leading to a signal-to-noise ratio (S/N) per pixel of 41--188 at 7370\,\AA. The observations were reduced using the CARMENES pipeline \texttt{caracal} \citep{Caballero2016} and we processed the VIS and NIR spectra with \texttt{serval}\footnote{\url{https://github.com/mzechmeister/serval}} \citep{SERVAL}, which is the standard CARMENES pipeline to derive relative RVs and several activity indicators using template matching: chromatic RV index (CRX), differential line width (dLW), and H$\alpha$, Na\,D1 and Na\,D2, and \ion{Ca}{II}\,IRT line indices. We also used the {\tt RACCOON} code\footnote{\url{https://github.com/mlafarga/raccoon}} \citep{2020A&A...636A..36L} to measure the CCF\_FWHM, CCF\_CTR, and CCF\_BIS spectral activity indicators via cross-correlation. In the analysis presented here, we used the activity indicators from VIS and NIR extracted with {\tt serval} and {\tt RACOON}, and the RVs measured from CARMENES VIS spectra with {\tt serval}. Because the precision in the RVs obtained from the VIS is higher than that obtained with the NIR, we only used the CARMENES VIS RVs, which have smaller error bars. CARMENES VIS RVs are corrected using measured nightly zero-point corrections as discussed in \citealt{Trifonov2020_nzp} (shown in Figure\,\ref{fig: RV DATA}).

\subsubsection{High-resolution spectroscopy with HARPS-N}
\label{sect: HARPS-N}

HD 191939 was observed with the High Accuracy Radial velocity Planet Searcher for the Northern hemisphere (HARPS-N; \citealp{HARPS-N}) mounted on the 3.6m \textit{Telescopio Nazionale Galileo} (TNG) in Roque de los Muchachos Observatory, La Palma. The star was monitored from 30 May 2020 to 13 September 2021. During this time, we obtained 42 high-resolution ($\mathcal{R}$\,=\,115\,000) spectra.

The observations were carried out as part of observing programs CAT19A\_162 program (PI: Nowak), ITP19\_1 (PI: Pall\'{e}) and CAT21A\_119 (PI: Nowak). The exposure times varied from 284 to 1800 seconds, depending on weather conditions and scheduling constraints, leading to a S/N per pixel of 27--134 at 5500\,\AA. The spectra were extracted using the off-line version 3.7 of the HARPS-N DRS pipeline \citep{2014SPIE.9147E..8CC}. Doppler measurements (absolute RVs) and spectral activity indicators (CCF\_FWHM, CCF\_CTR, CCF\_BIS, and Mount-Wilson S-index) were measured using an online version of the DRS, the YABI tool\footnote{Available at \url{http://ia2-harps.oats.inaf.it:8000}.}, by cross-correlating the extracted spectra with a G2 mask \citep{1996A&AS..119..373B}. We also used the {\tt serval} code to measure relative RVs by template matching, CRX, dLW, and H$\alpha$ and sodium Na\,D1 and Na\,D2 indexes. The uncertainties of the RVs measured with {\tt serval} are in the range of 0.5--3.1\,$\mathrm{m\,s^{-1}}$, with a mean value of 1.07\,$\mathrm{m\,s^{-1}}$. In the analysis presented here, we used the relative RVs measured from HARPS-N spectra with {\tt serval} (shown in Figure\,\ref{fig: RV DATA}).

\subsubsection{High-resolution spectroscopy with APF and HIRES}
\label{sect: APF + HIRES}

\citetalias{Lubin_HD191939} also performed a ground-based follow-up campaign with two different high-resolution spectrographs. They obtained 73 RV measurements with HIRES and 107 RV measurements with the Automated Planet Finder (APF, \citealp{APF_Vogt2014})  telescope. In total, these observations covered a baseline of 415 days, and their details are explained in Sect.\,2.2 of \citetalias{Lubin_HD191939}. The time series of APF and HIRES are displayed in Figure\,\ref{fig: RV DATA} along with the CARMENES and HARPS-N RVs.

\section{Analysis and results}
\label{Sect: Analysis}

\subsection{Stellar rotation and activity indicators}
\label{sect: Activity indicators}

\citetalias{Mariona_2020} and \citetalias{Lubin_HD191939} reported that HD 191939 is a slow rotator star with low or null stellar and chromospherical activity. \citetalias{Mariona_2020} derived a $P_{\mathrm{rot}}/\sin{i}$\,=\,79$\pm$66\,d, where the large uncertainties come from the large error on the $v \sin{i}$.

We searched for modulations in the different activity indices derived from CARMENES and HARPS-N spectra using the generalised Lomb-Scargle (GLS; \citealp{GLS_paper}) periodogram\footnote{\url{https://github.com/mzechmeister/GLS}}. We computed the theoretical false-alarm probability (FAP) as it is described in \citet{GLS_paper}. The GLS periodograms of the activity indices are shown in Figure\,\ref{fig: GLS Activity}.
None of the indices present a significant peak at either the periodicities of the known planets or at the $\sim$300\,d signal, attributed to HD 191939\,g. As is expected for a low-activity star, the GLS periodograms remain below the 10\% level of significance. Only some CARMENES indices show a broad peak near 400\,d that is not detected in the HARPS-N indices, and CARMENES \ion{Ca}{II}\,IRT$_1$ and HARPS-N H$\alpha$ display significant periodicities near 100\,d and 200\,d, respectively. However, none of those peaks have a counterpart in the RVs periodogram analyses.

\subsection{Radial velocity analysis}
\label{sect: RV ANALISIS}

\begin{figure}
    \centering
    \includegraphics[width=\hsize]{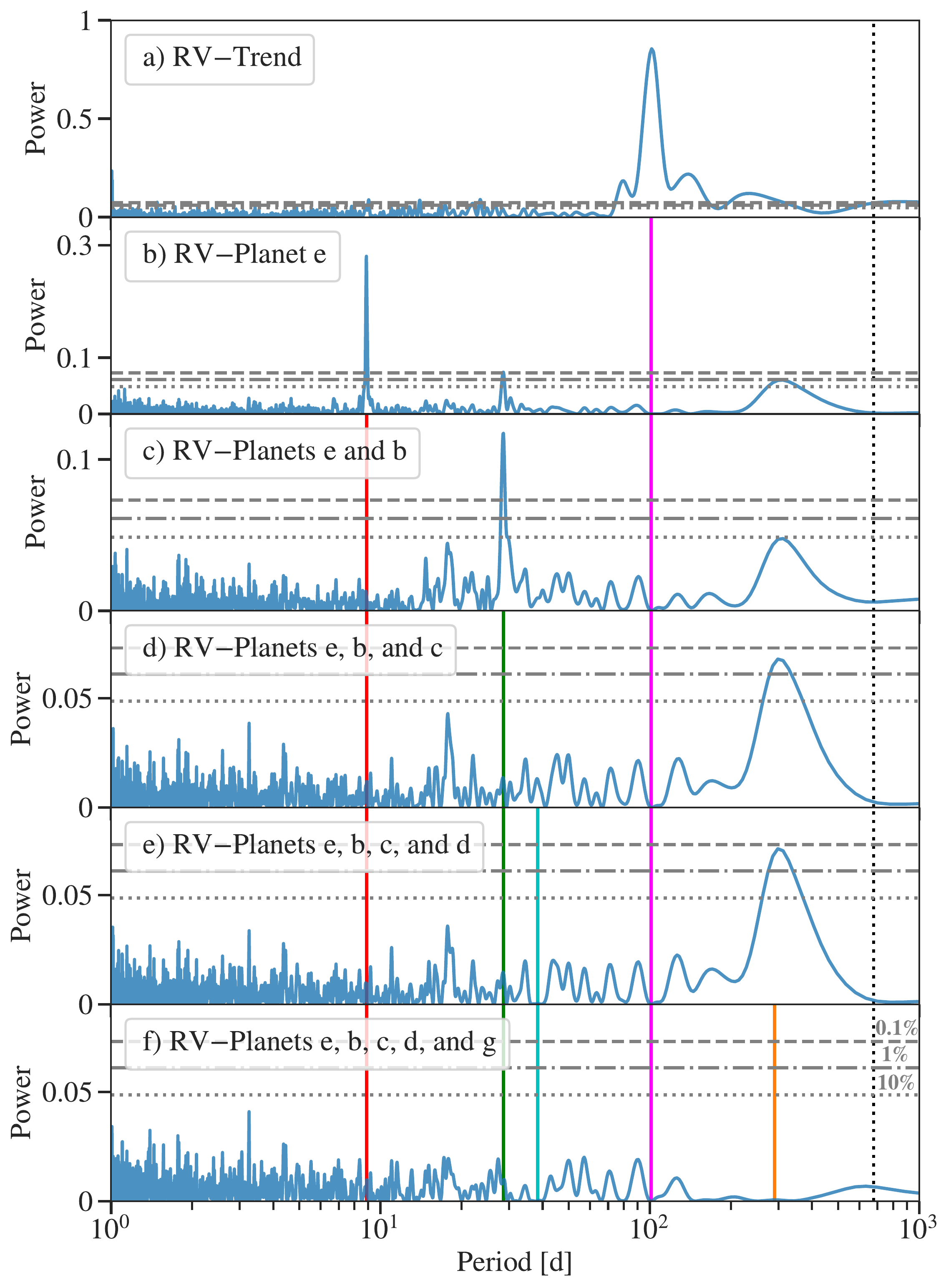}
    \caption{\label{fig: GLS Models}
    GLS periodograms of the time series of APF, HIRES, CARMENES, and HARPS-N RV measurements and the residuals after subtraction of different models. All the models include quadratic and linear terms to account for the long-term trend detected in the RV time series.
    $(a)$ GLS periodogram of RVs after removing the long-term trend.
    $(b)$ GLS periodogram of the RV residuals after fitting the 101\,d signal (vertical magenta line).
    $(c)$ GLS periodogram of the RV residuals after simultaneously fitting the 8.8\,d (vertical red line) and 101\,d signals.
    $(d)$ GLS periodogram of the RV residuals after simultaneously fitting the 8.8\,d, 28.6\,d (vertical green line), and 101\,d signals.
($e)$ GLS periodogram of the RV residuals after simultaneously fitting the 8.8\,d, 28.6\,d, 38\,d (vertical cyan line), and 101\,d signals.
    $(f)$ GLS periodogram of the RV residuals after simultaneously fitting the 8.8\,d, 28.6\,d, 38\,d, 101\,d, and 300\,d (vertical orange line) signals.
    In all panels, the 10\%, 1\%, and 0.1\% FAP levels are indicated by dotted, dash-dotted, and dashed 
grey horizontal lines, respectively. The vertical black dotted line indicates the dataset baseline. We highlight the different scale of the y axis in each panel.
    }
\end{figure}

The analyses presented in this work combine the APF and HIRES RVs from \citetalias{Lubin_HD191939} with those obtained with CARMENES and HARPS-N. With a total of 362 RVs covering a baseline of 677 days (Fig.\,\ref{fig: RV DATA}), we were able to improve the planetary mass determination but also look for the presence of planets with periods between those of planets e (101\,d) and f (>1700\,d).

We analysed the planetary signals in the RVs computing the GLS periodogram and modelling the detected signals with \texttt{juliet}\footnote{\url{https://juliet.readthedocs.io/en/latest/index.html}} (\citealp{juliet}). This \texttt{python} library is based on other public packages for transit light curve (\texttt{batman}, \citealp{batman}), RV (\texttt{radvel}, \citealp{radvel}), and GP (\texttt{george}, \citealp{george}; \texttt{celerite}, \citealp{celerite}) modelling. \texttt{Juliet} uses nested sampling algorithms (\texttt{dynesty}, \citealp{dynesty}; \texttt{MultiNest}, \citealp{MultiNest, PyMultiNest}) to explore all the parameter space and compute the Bayesian model log-evidence ($\ln Z$), which allows us to compare models with different numbers of free parameters. If the difference between two models, for example M1 and M2, is $\Delta \ln{Z} = \ln{Z_{\mathrm{M2}}} - \ln{Z_{\mathrm{M1}}} > 5 $, then the M2 model is strongly preferred statistically over the M1 model \citep{Trotta_2008}. If $\Delta \ln{Z} \leq 1 $, the two models are statistically indistinguishable and the preferred one is the simplest model with the least free parameters. We consider that M2 has moderate evidence over M1 for intermediate cases ($\Delta \ln{Z} \sim 2.5$).

Because the main purpose of this preliminary study is to look for additional signals, we considered circular orbits for simplicity. Eccentric models are explored during the joint fit (see Sect.\,\ref{sect: JOINT FIT ANALISIS}) after exploring the signals present in the data. For this RV-only analysis, we fixed the period ($P$) and the central time of transit ($t_0$) for the three transiting planets based on a photometric-only analysis. The precision derived for $P$ and $t_0$ from the light curves is significantly higher than from the RVs alone. Fixing these two parameters saves computational time without an impact on the RV modeling. For the signals with no counterpart in the photometry, we set normal priors for $P$ and uniform priors for $t_0$ and we set uninformative priors for the semi-amplitude ($K$) of the fitted signals.  For each spectrograph, we also included an instrumental jitter term ($\sigma$) and a systemic velocity ($\gamma$) term.
The procedure described below to fit the periodicities detected in the periodograms is illustrated in Figure\,\ref{fig: GLS Models}. The following points also refer to the panels of  Fig.\,\ref{fig: GLS Models}.

{\textit{(a)}} As in \citetalias{Lubin_HD191939} and to save computational time, we modelled and subtracted the long-term RV trend detected by the four instruments with a linear ({$\dot \gamma$}) term and a quadratic ({$\ddot \gamma$}) term. The linear and quadratic term model is statistically preferred ($\Delta \ln{Z} > 5 $) over an only linear or only quadratic term model. The conspicuous signal at 101\,d due to planet e dominates the RV periodogram.

{\textit{(b)}} After fitting planet e model, the planet b periodicity is clearly seen in the periodogram of the residuals.

{\textit{(c)}} When planet b is removed, the planet c signal is the most significant peak in the periodogram.

{\textit{(d)}} After fitting planet c, all periodicities in the periodogram of the residuals remain well below the 10\% level of significance, except for a peak at $\sim$300\,d (FAP\,$<$\,1\%). As in Fig.\,1 from \citetalias{Lubin_HD191939}, the signature of planet d at 38.35\,d is not detected or significant in the RVs.

{\textit{(e)}} When all the previously known planets are removed, the periodogram of the residuals is still dominated by the signal at $\sim$300\,d which slightly increased its significance until FAP\,$\sim$\,0.1\%. The $\sim$300\,d peak is already visible and significant after removing planet e.

{\textit{(f)}} The periodogram after fitting the $\sim$300\,d signal is flat without significant peaks. We refer to the $\sim$300\,d signal as HD 191939\,g. We analyse this signal in detail below.

\begin{table*}
\caption[width=\textwidth]{
\label{table - Models logZ}
Comparative between Bayesian log-evidences ($\Delta \ln{Z}$) and planet semi-amplitudes ($K_{\mathrm{p}}$) for the different explored models. We used the $\ln{Z}$ from the four-planet model as a reference.
The adopted model used in the joint fit is marked in boldface (see Sect.\,\ref{sect: RV ANALISIS} for details about the selection of the final model). The last row shows the $K_{\mathrm{p}}$ from the joint fit in Sect.\,\ref{sect: JOINT FIT ANALISIS} for illustrative purposes.
}
\centering

\begin{tabular}{lcccccc}

\hline \hline 
\noalign{\smallskip} 

Model & $\Delta\ln{Z}$ & $K_{\mathrm{b}}$ ($P=8.8$\,d) & $K_{\mathrm{c}}$ ($P=28.6$\,d) & $K_{\mathrm{d}}$ ($P=38$\,d) & $K_{\mathrm{e}}$ ($P=101$\,d) & $K_{\mathrm{g}}$ ($P \simeq 300$\,d) \vspace{0.05cm}\\
\hline
\noalign{\smallskip}

4 planets       & 0.0 & 3.66$\pm$0.25 & 1.75$\pm$0.25 & 0.57$\pm$0.24 & 17.7$\pm$0.3 & -- \vspace{0.05cm}\\

GP$_{\mathrm{exponential}}$   & 3.1 & 3.62$\pm$0.30 & 1.76$\pm$0.30 & 0.58$\pm$0.28 & 17.8$\pm$0.3 & -- \vspace{0.05cm}\\
GP$_{\mathrm{Matern}}$ & 4.2 & 3.58$\pm$0.25 & 1.93$\pm$0.32 & 0.67$\pm$0.33 & 17.8$\pm$0.5 & -- \vspace{0.05cm}\\

GP$_{\mathrm{qp}}$\,$^{(a)}$&9.2&3.55$\pm$0.26 & 1.96$\pm$0.34 & 0.70$\pm$0.35 & 17.8$\pm$0.5 & -- \vspace{0.05cm}\\
5 planets\,$^{(a)}$&7.7& 3.56$\pm$0.25 & 1.91$\pm$0.26 & 0.61$\pm$0.25 & 17.6$\pm$0.3 & 1.3$\pm$0.3 \vspace{0.05cm}\\

GP$_{\mathrm{qp}}$\,$^{(b)}$&10.0&3.54$\pm$0.26& 1.96$\pm$0.33 & 0.72$\pm$0.35 & 17.8$\pm$0.5 & -- \vspace{0.05cm}\\
\textbf{5 planets}\,$^{(b)}$&\textbf{9.4}& 3.55$\pm$0.25 & 1.91$\pm$0.26 & 0.61$\pm$0.25 & 17.6$\pm$0.3 & 1.3$\pm$0.3 \vspace{0.05cm}\\

\noalign{\smallskip}
\noalign{\smallskip}
\multicolumn{2}{c}{Joint fit}  & 3.56$\pm$0.24 & 1.93$\pm$0.24 & 0.61$\pm$0.13 & 17.75$\pm$0.15 & 1.53$\pm$0.23 \vspace{0.05cm}\\

\noalign{\smallskip}
\noalign{\smallskip}
\hline
\end{tabular}

\tablefoot{
$^{(a)}$ Using an uninformative prior for the periodic parameter.
$^{(b)}$ Using a normal prior for the periodic parameter centred at 300\,d.
}

\end{table*}

To crosscheck our results, we also analysed the RV dataset using \texttt{Exo-Striker}\footnote{\url{https://github.com/3fon3fonov/exostriker}} \citep{2019ascl.soft06004T}, obtaining similar results. After fitting the transiting planets b, c, and d as well as planets e and f, the GLS periodogram only showed a peak around $\sim$300\,days, which is the signal we named HD 191939\,g. Furthermore, the signal at 17.7\,d found by \citetalias{Lubin_HD191939} (Sect.\,8 therein) is observed in the residuals in Fig.\,\ref{fig: GLS Models} but at a very low significance level (FAP$\gg$10\%). Our RV dataset does not support the scenario of a non-transiting planet with a period between transiting planets c and d.

We repeated the analysis presented above only with the APF and HIRES datasets, obtaining similar results to \citetalias{Lubin_HD191939}; that is, there is no statistically significant evidence (FAP>10\%) for an additional signal at $\sim$300\,d (see Fig.\,\ref{fig: GLS LUBIN}). The non-detection of that signal in \citetalias{Lubin_HD191939} may be due to the lower number of RVs used and the shorter baseline of those observations. The baseline of the 180 APF and HIRES RVs is 415 days, which is less than 1.5 periods of the $\sim$300d signal. Here, we used 362 RVs with a baseline of 677 days, which cover two complete periods.

To unveil the nature of the $\sim$300\,d signal, we computed a set of models that simultaneously fit the known planets and also account for the additional signal with two different strategies: fitting the signal by adding a GP term or with a Keplerian orbit. We considered three different GP kernels: exponential (GP$_{\mathrm{exponential}}$), Matern 3/2 (GP$_{\mathrm{Matern}}$), and quasi-periodic (GP$_{\mathrm{qp}}$). When fitting the signal with a Keplerian orbit or with a GP$_{\mathrm{qp}}$ kernel, we tested two different priors for the period parameter or rotational hyperparameter: an uninformative prior and a normal prior centred at 300\,d. The priors and posteriors of the hyperparameters used in the GP models are shown in Table\,\ref{table - GP hyperparameters}. Table\,\ref{table - Models logZ} presents the Bayesian log-evidence for the explored models and the measured $K$ for each of the fitted signals. The derived $K$s for the different models are consistent within errors, ensuring that the planet mass is not model-dependent or affected by the $\sim$300\,d signal.

The five-planet model and the GP$_{\mathrm{qp}}$ model are statistically preferred over the four-planet model, the GP$_{\mathrm{exponential}}$ model, and the GP$_{\mathrm{Matern}}$ model. The $\Delta \ln{Z}$ between the five-planet model and its analogous GP$_{\mathrm{qp}}$ model is less than 2. Therefore, the GP$_{\mathrm{qp}}$ model is moderately preferred over the five-planet model. However, the GP$_{\mathrm{qp}}$ model has one free parameter more than the five-planet model. Moreover, closer inspection of the posterior distributions of the GP$_{\mathrm{qp}}$ hyperparameters shows that the P$_{\mathrm{rot}}$ is not constrained (see Fig.\,\ref{fig: GP CORNER PLOTS}). In the uninformative GP$_{\mathrm{qp}}$ model, the P$_{\mathrm{rot}}$ posterior distribution is mainly flat. When we used a normal prior for P$_{\mathrm{rot}}$, the posterior distribution is equal to the input prior, which is in contrast to the period of the additional signal, which is well determined in both five-planet models.

Therefore, we chose the five-planet model over the GP$_{\mathrm{qp}}$ model due to its more physical plausibility. The complete RV model includes quadratic and linear terms, planets HD 191939\,b, c, d, and e, and the new planet HD 191939\,g.

\subsection{Joint fit}
\label{sect: JOINT FIT ANALISIS}

\begin{table*}[ht!]
\caption[width=\textwidth]{
\label{table - Joint Fit RESULTS}
Parameters and $1\sigma$ uncertainties for the \texttt{juliet} joint fit model for HD 191939 planetary system. Priors and description for each parameter are presented in Table\,\ref{table - Joint Fit PRIORS}. The adopted stellar properties used to derive the planetary parameters are the ones from \citetalias{Lubin_HD191939} in Table\,\ref{table - stellar parameters}. The model to estimate planet f limits is explained in Sect.\,\ref{sect: Planet f}.
}
\centering

\begin{tabular}{lccccccc}
\hline \hline 
\noalign{\smallskip} 
\multicolumn{1}{c}{Parameter} & \textbf{Planet b}  & \textbf{Planet c}  & \textbf{Planet d}  & \textbf{Planet e}  & \textbf{Planet g} & \textbf{Planet f} \\ 
\noalign{\smallskip} 
\hline 
\noalign{\smallskip} 
\noalign{\smallskip} 
$P$ [d]  & 8.8803256\,(47)  & 28.579743\,(45) & 38.353037\,(60) & 101.12$\pm$0.13  & 284$^{+10}_{-8}$ & >2200 \vspace{0.05cm} \\  
$t_{0}$\,$^{(a)}$  & 2443.54236$^{+0.00025}_{-0.00023}$  & 2440.5491$^{+0.0011}_{-0.0007}$  & 2433.906$^{+0.0005}_{-0.0009}$  & 2348.12$\pm$0.50  & 2385$^{+13}_{-10}$ & --\vspace{0.05cm} \\  
$K$ [$\mathrm{m\,s^{-1}}$]  & 3.56$^{+0.21}_{-0.24}$  & 1.93$^{+0.23}_{-0.24}$  & 0.61$\pm$0.13  & 17.73$^{+0.14}_{-0.16}$  & 1.53$^{+0.21}_{-0.23}$ &  >36 \vspace{0.05cm} \\  
\textit{ecc}  & 0.031$^{+0.010}_{-0.011}$  & 0.034$^{+0.034}_{-0.013}$  & 0.031$^{+0.018}_{-0.012}$  & 0.031$^{+0.008}_{-0.016}$  & 0.030$^{+0.025}_{-0.011}$ & -- \vspace{0.05cm} \\  
$\omega$ (deg)  & 5$^{+40}_{-35}$  & $-$90$^{+220}_{-40}$  & 15$^{+90}_{-160}$  & $-$130$^{+180}_{-16}$  & 18$\pm$65 & -- \vspace{0.05cm} \\  
$r_{1}$  & 0.739$^{+0.016}_{-0.020}$  & 0.754$^{+0.031}_{-0.020}$  & 0.626$^{+0.035}_{-0.033}$  & --  & -- & -- \vspace{0.05cm} \\  
$r_{2}$  & 0.03319$^{+0.00032}_{-0.00017}$  & 0.03118$^{+0.00027}_{-0.00034}$  & 0.02916$^{+0.00034}_{-0.00021}$  & --  & -- & -- \vspace{0.05cm} \\  
 
\noalign{\smallskip} 
\hline 
\noalign{\smallskip} 
\multicolumn{7}{c}{\textit{ Derived parameters }} \\ 
\noalign{\smallskip} 
$p = {R}_{\mathrm{p}}/{R}_{\star}$  & 0.03319$^{+0.00032}_{-0.00017}$  & 0.03118$^{+0.00027}_{-0.00034}$  & 0.02916$^{+0.00034}_{-0.00021}$  & --  & -- & -- \vspace{0.05cm} \\  
$b =(a/{R}_{\star}) \cos{ i_{\mathrm{p}} }$  & 0.610$^{+0.025}_{-0.030}$  & 0.630$^{+0.050}_{-0.030}$  & 0.440$\pm$0.050  & --  & -- & -- \vspace{0.05cm} \\  
$a/{R}_{\star}$  & 18.36$^{+0.50}_{-0.33}$  & 40.0$^{+1.1}_{-0.7}$  & 48.7$^{+1.3}_{-0.9}$  & 92.9$^{+2.5}_{-1.7}$  & 186.0$^{+5.0}_{-5.5}$ & >730 \vspace{0.05cm} \\  
$i_{\mathrm{p}}$ (deg)  & 88.10$^{+0.14}_{-0.10}$  & 89.10$^{+0.06}_{-0.08}$  & 89.49$^{+0.05}_{-0.08}$  & --  & -- & -- \vspace{0.05cm} \\  
$t_{\mathrm{T}}$ [h]  & 2.939$^{+0.036}_{-0.051}$  & 4.162$^{+0.24}_{-0.096}$  & 5.36$^{+0.19}_{-0.11}$  & --  & -- & -- \vspace{0.05cm} \\  
$R_{\mathrm{p}}$ [${R}_\oplus$]  & 3.410$\pm$0.075  & 3.195$\pm$0.075  & 2.995$\pm$0.070  & --  & -- & -- \vspace{0.05cm} \\  
$M_{\mathrm{p}}$ [${M}_\oplus$]\,$^{(b)}$  & 10.00$\pm$0.70 & 8.0$\pm$1.0  & 2.80$\pm$0.60  & >112.2$\pm$4.0  & >13.5$\pm$2.0 & >660 \vspace{0.05cm} \\  
$\rho_{\mathrm{p}}$ [$\mathrm{g\,cm^{-3}}$]  & 1.40$^{+0.15}_{-0.13}$  & 1.35$\pm$0.20  & 0.57$\pm$0.13  & --  & -- & -- \vspace{0.05cm} \\  
$g_{\mathrm{p}}$ [$\mathrm{m\,s^{-2}}$]  & 8.4$^{+0.8}_{-0.7}$  & 7.7$^{+1.1}_{-1.0}$  & 3.1$\pm$0.7  & --  & -- & -- \vspace{0.05cm} \\  
$a_{\mathrm{p}}$ [AU]  & 0.0804$^{+0.0025}_{-0.0023}$  & 0.1752$^{+0.0055}_{-0.0050}$  & 0.2132$^{+0.0065}_{-0.0061}$  & 0.407$\pm$0.012  & 0.812$\pm$0.028 & >3.2 \vspace{0.05cm} \\  
$T_{\mathrm{eq}}$ [K]\,$^{(c)}$  & 880$\pm$20  & 600$\pm$13  & 540$\pm$11  & 390$\pm$8  & 278$\pm$6 & <125 \vspace{0.05cm} \\  
${S}$ [$\mathrm{S}_\oplus$]  & 100$\pm$7 & 21.0$\pm$1.4  & 14.3$\pm$1.0  & 3.9$\pm$0.3  & 0.99$\pm$0.08 & -- \vspace{0.05cm} \\  
\noalign{\smallskip} 
\hline \hline 
\noalign{\smallskip} 
\multicolumn{7}{c}{\textbf{ Model Parameters }} \\ 
\noalign{\smallskip} 
 & \multicolumn{4}{c}{\textit{ Stellar density }} \\ 
$\rho_{\star}$ [kg\,m$^{-3}$] & \multicolumn{4}{c}{ 1485$^{+120}_{-80}$ } \vspace{0.20cm} \\ 
 
\noalign{\smallskip}
 & \multicolumn{4}{c}{\textit{ Photometry parameters }} \\ \noalign{\smallskip} 
$\mu_{\textit{TESS}}$ (ppm) & \multicolumn{4}{c}{ $-$26.2$\pm$1.0 } \vspace{0.05cm} \\ 
$\sigma_{\textit{TESS}}$ (ppm) & \multicolumn{4}{c}{ 152.2$^{+1.9}_{-1.3}$ } \vspace{0.05cm} \\ 
$q_{1,\textit{TESS}}$ & \multicolumn{4}{c}{ 0.304$^{+0.051}_{-0.052}$ } \vspace{0.05cm} \\ 
$q_{2,\textit{TESS}}$ & \multicolumn{4}{c}{ 0.4$^{+0.18}_{-0.12}$ } \vspace{0.20cm} \\ 
 
\noalign{\smallskip} 
 & \multicolumn{4}{c}{\textit{ RV parameters }} \\ \noalign{\smallskip} 
Slope {$\dot \gamma$} [$\mathrm{m\,s^{-1}\,d^{-1}}$] & \multicolumn{4}{c}{ $0.1902^{+0.0022}_{-0.0020}$ } \vspace{0.05cm} \\ 
Curve {$\ddot \gamma$} [$\mathrm{m\,s^{-1}\,d^{-2}}$] & \multicolumn{4}{c}{ $-2.27^{+0.07}_{-0.08} \times 10^{-5}$ } \vspace{0.05cm} \\ 
$\gamma_{\mathrm{APF}}$ [$\mathrm{m\,s^{-1}}$] & \multicolumn{4}{c}{ $-6.1^{+1.6}_{-1.4}$ } \vspace{0.05cm} \\ 
$\sigma_{\mathrm{APF}}$ [$\mathrm{m\,s^{-1}}$] & \multicolumn{4}{c}{ $3.48^{+0.35}_{-0.38}$ } \vspace{0.05cm} \\ 
$\gamma_{\mathrm{HIRES}}$ [$\mathrm{m\,s^{-1}}$] & \multicolumn{4}{c}{ $-1.4^{+1.7}_{-1.6}$ } \vspace{0.05cm} \\ 
$\sigma_{\mathrm{HIRES}}$ [$\mathrm{m\,s^{-1}}$] & \multicolumn{4}{c}{ $1.95^{+0.16}_{-0.15}$ } \vspace{0.05cm} \\ 
$\gamma_{\mathrm{CARMENES}}$ [$\mathrm{m\,s^{-1}}$] & \multicolumn{4}{c}{ $4.5\pm$1.4 } \vspace{0.05cm} \\ 
$\sigma_{\mathrm{CARMENES}}$ [$\mathrm{m\,s^{-1}}$] & \multicolumn{4}{c}{ $4.24^{+0.33}_{-0.26}$ } \vspace{0.05cm} \\ 
$\gamma_{\mathrm{HARPS-N}}$ [$\mathrm{m\,s^{-1}}$] & \multicolumn{4}{c}{ $-5.0^{+1.4}_{-1.5}$ } \vspace{0.05cm} \\ 
$\sigma_{\mathrm{HARPS-N}}$ [$\mathrm{m\,s^{-1}}$] & \multicolumn{4}{c}{ $1.76^{+0.15}_{-0.22}$ } \vspace{0.05cm} \\ 
 
\noalign{\smallskip} 
\hline \hline

\end{tabular}

\tablefoot{$^{(a)}$ Central time of transit ($t_0$) units are BJD\,$-$\,2\,457\,000. $^{(b)}$ The masses for planets e, g, and f are a lower limit ($M_{\mathrm{p}} \sin{i_{\mathrm{p}}}$) because they are only detected in the RV data. $^{(c)}$ Equilibrium temperatures were calculated assuming zero Bond albedo.}
\end{table*}

\begin{figure*}[ht!]
    \includegraphics[width=1\linewidth]{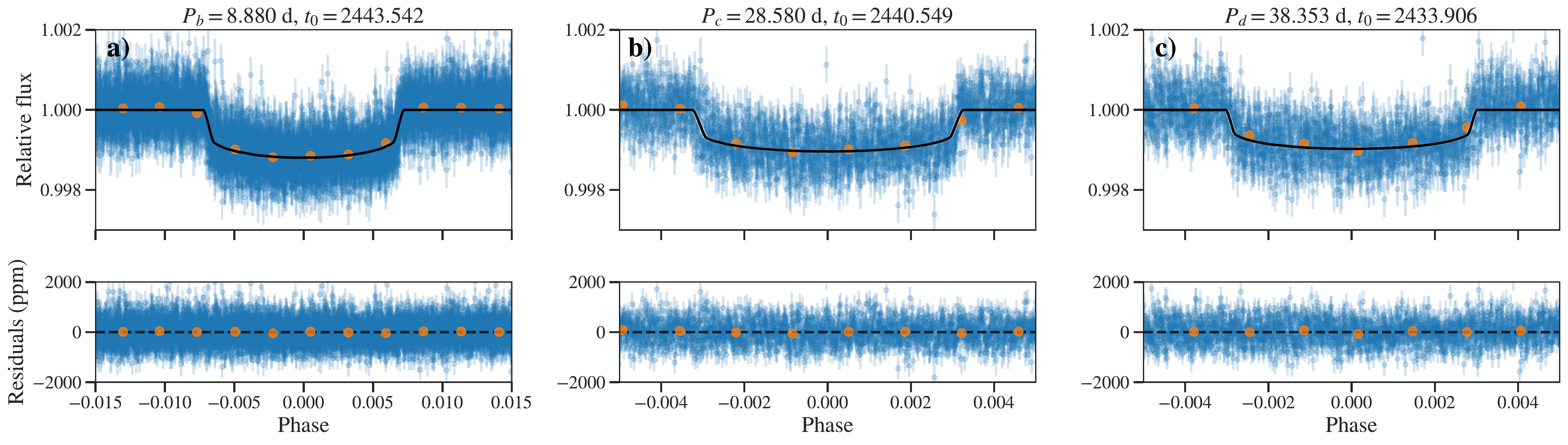}
    \caption{\label{Fig: PHASE FOLDED LC JF} Photometry data phase-folded to the period $P$ and central time of transit $t_0$ (shown above each panel, $t_0$ units are BJD\,$-$\,2\,457\,000) derived from the joint fit model. Two-minute cadence \textit{TESS} phase-folded photometry for HD 191939\,b (panel \textit{a}), c (panel \textit{b}), and d (panel \textit{c}) along with the best-fit model. Orange points show binned photometry for visualisation. The error bars include the photometric jitter term added in quadrature.
    }
\end{figure*}

\begin{figure*}[ht!]
     \centering
     \begin{subfigure}{\textwidth}
         \centering
         \includegraphics[width=1\linewidth]{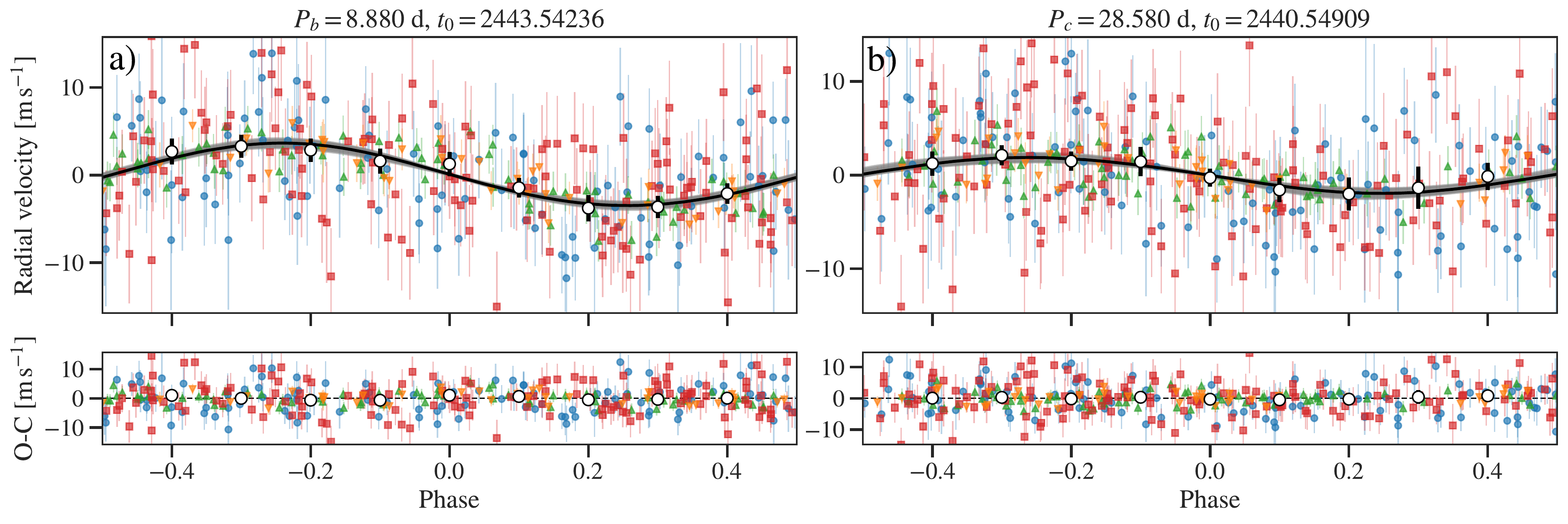}
     \end{subfigure}
     \begin{subfigure}{\textwidth}
         \centering
         \includegraphics[width=1\linewidth]{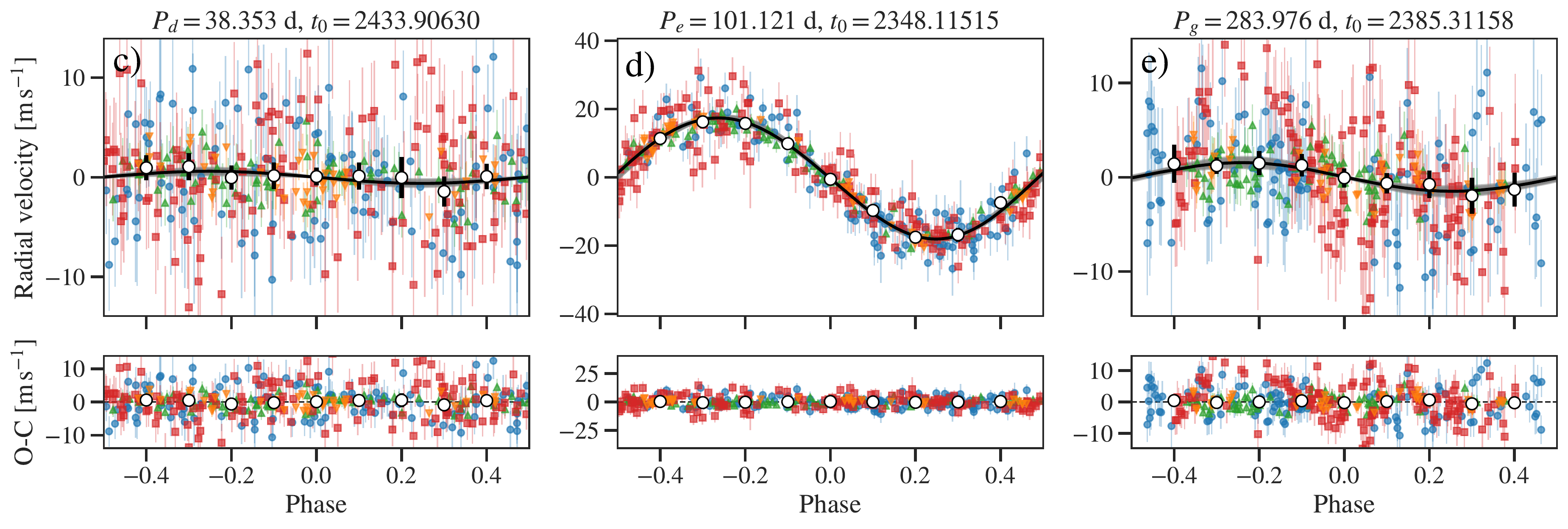}
     \end{subfigure}
        \caption{\label{Fig: PHASE FOLDED RV JF} RV data phase-folded to the period $P$ and central time of transit $t_0$ (shown above each panel, $t_0$ units are BJD\,$-$\,2\,457\,000) derived from the joint-fit model. Here, we show APF (blue circles), HIRES (green up triangles), CARMENES (red squares), and HARPS-N (orange down triangles) RVs phase-folded for HD 191939\,b (panel \textit{a}), c (panel \textit{b}), d (panel \textit{c}), e (panel \textit{d}), and g (panel \textit{e}) along with the best-fit model (black line) and the $3\sigma$ confidence interval (shaded grey area). The error bars include the instrumental jitter term added in quadrature.
        }
\end{figure*}

We simultaneously modeled the detrended two-minute cadence \textit{TESS} photometry and the APF, HIRES, CARMENES, and HARPS-N RVs using \texttt{juliet} to refine the parameters for the HD 191939 system. For the joint fit, we considered transits for the planets b, c, and d, and we adopted a five-planet model with a quadratic and linear trend for the RVs.

We adopted a quadratic limb-darkening law for the \textit{TESS} light curve. The limb-darkening coefficients were parameterised with the uniform sampling prior ($q_1$,$q_2$) introduced by \citet{Kipping2013}. Additionally, rather than directly fitting the impact parameter of the orbit ($b$) and the planet-to-star radius ratio ($p$\,$=$\,${R}_{\mathrm{p}}/{R}_{\star}$), we considered the uninformative sample ($r_1$,$r_2$) parameterization introduced in \citet{Espinoza2018}. The parameters $r_1$ and $r_2$ ensure a full exploration of the physically plausible values of $p$ and $b$, with uniform priors sampling. We used the value and uncertainties of the stellar density ($\rho_{\star}$) from \citetalias{Lubin_HD191939} in Table\,\ref{table - stellar parameters} to set a normal prior for $\rho_{\star}$. We fixed the dilution factor to 1 based on \citetalias{Mariona_2020} and we added a relative flux offset ($\mu$) and a jitter term ($\sigma$) to \textit{TESS} data.

Systems with multiple transiting planets, such as HD 191939, normally present low eccentricities, but not necessarily zero (\citealp{VanEylen_ecc}; \citealp{Xie_ecc}; \citealp{Hadden_ecc}). We therefore considered Keplerian orbits for the five planets with a beta prior distribution with $\alpha$\,$=$\,$1.52$ and $\beta$\,$=$\,$29$ for the orbital eccentricity \textit{ecc} (\citealp{VanEylen_ecc, VanEylen_ecc_2019}). We also computed a joint fit with circular orbits. However, the eccentric joint-fit model is statistically preferred over the non-eccentric one ($\Delta \ln{Z}$\,=\,$\ln{Z_{\rm ecc}} - \ln{Z_{\rm no ~ ecc}} > 8 $) and the results from both models are coincident within their uncertainties.

The priors used in the joint fit are listed in Table\,\ref{table - Joint Fit PRIORS}. The posterior distributions and the derived parameters for the planetary system are reported in Table\,\ref{table - Joint Fit RESULTS}. As is shown in Table\,\ref{table - Models logZ}, the semi-amplitudes obtained from the joint fit are consistent with the RV-only models explored in Sect.\,\ref{sect: RV ANALISIS}. The stellar density is consistent with those derived by \citetalias{Lubin_HD191939} and by \citetalias{Mariona_2020} within $1\sigma$. The adopted stellar properties used to derive the planetary parameters are the ones from \citetalias{Lubin_HD191939} (see Table\,\ref{table - stellar parameters}) for a better comparison with the results presented there. The best-fit models for phase-folded light curves and phase-folded RVs are shown in Figures\,\ref{Fig: PHASE FOLDED LC JF} and \ref{Fig: PHASE FOLDED RV JF}, respectively. Photometric and RV time series along with the best-fit models are shown in Figures\,\ref{fig: TESS SECTORS} and \ref{fig: TESS SECTORS 41 48}, and Figure\,\ref{fig: RV MODEL JF}, respectively.

We checked for transit-like events in the \textit{TESS} PDC-SAP and SAP for the non-transiting planets with no results. We estimated the flux decrease ($\Delta F \simeq ({R}_{\mathrm{p}}/{R}_{\star})^2$) and transit duration ${t}_{\mathrm{T}}$ of planets e and g with their predicted radii (see Sect.\,\ref{sect: HD 191939 g}). HD 191939\,e would produce a flux decrease of $\sim$1.6\% over $\sim$8\,h. According to the derived ephemeris, a transit of planet e was expected during Sector\,16 (see Fig.\,\ref{fig: TESS SECTORS}) and is clearly not detected. Because the predicted radius for HD 191939\,g is similar to that of the inner planets, the flux decrease for planet g  would be similar ($\sim$1300\,ppm). However, the transit would span $\sim$12\,h due to its large period. Planet g was expected to transit at some point in Sectors\,18 and 19 (see Fig.\,\ref{fig: TESS SECTORS}). Although there is no clear evidence of a complete transit, ingress, or egress during these sectors, the $1\sigma$ uncertainty for the transit midpoint ($\sim$16\,d) comprises observing gaps where the transit could have happened. These observing gaps represent 18\% of the $\pm\,1\sigma$ expected transit time region. Upcoming \textit{TESS} sectors will allow us to confirm whether we were unlucky or HD 191939\,g does not transit, as in HD 191939\,e.

\subsection{Planet-candidate f constraints}
\label{sect: Planet f}

Based on high-angular-resolution imaging, \citetalias{Mariona_2020} (Sect.\,2.5 therein) did not find companions 5\,mag fainter than HD 191939 at 200\,mas or 8.4\,mag fainter at 1". However, the RV time series present a long-term trend that \citetalias{Lubin_HD191939} cautiously referred to as a planetary object, HD 191939\,f, because its mass is most likely below 13\,${M}_{\mathrm{J}}$. \citetalias{Lubin_HD191939} performed a joint RV and astrometric analysis to impose some limits on the properties of planet-candidate f. From that analysis, the period should be between 1700 and 7200\,d, which is still much longer than the baseline of our new combined RV dataset.

We also tried to analyse the properties of planet-candidate f by fitting the RV long-term trend with a Keplerian signal instead of a quadratic term plus a linear term. Although the RV time series from this work have a longer base line, they do not show a peak to peak of HD 191939\,f signal that could help to constrain its period and mass. Therefore, the values reported for planet-candidate f (see last column in Table\,\ref{table - Joint Fit RESULTS}) should be considered as updated upper and lower limits. The best-fit model for the RVs (Fig.\,\ref{fig: PLANET F fit model}) and the phase-folded RVs for  planet-candidate f(Fig.\,\ref{fig: PLANET F phase-folded}) also confirm that HD 191939\,f properties are not constrained. The period of $\gtrsim$2200\,d is in agreement with the 1700--7200\,d period range, and the semi-amplitude of $\gtrsim$36\,m\,s$^{-1}$ is consistent with the previous lower limit of >23\,m\,s$^{-1}$ from APF and HIRES RVs.
The properties of the planetary candidate HD 191939\,f are still not well determined, and therefore further observations of HD 191939 are needed to sample its long period and better constrain its properties.

\section{Discussion}
\label{sect: Discussion}

\begin{figure}
    \centering
    \begin{subfigure}{\hsize}
     \centering
     \includegraphics[width=\hsize]{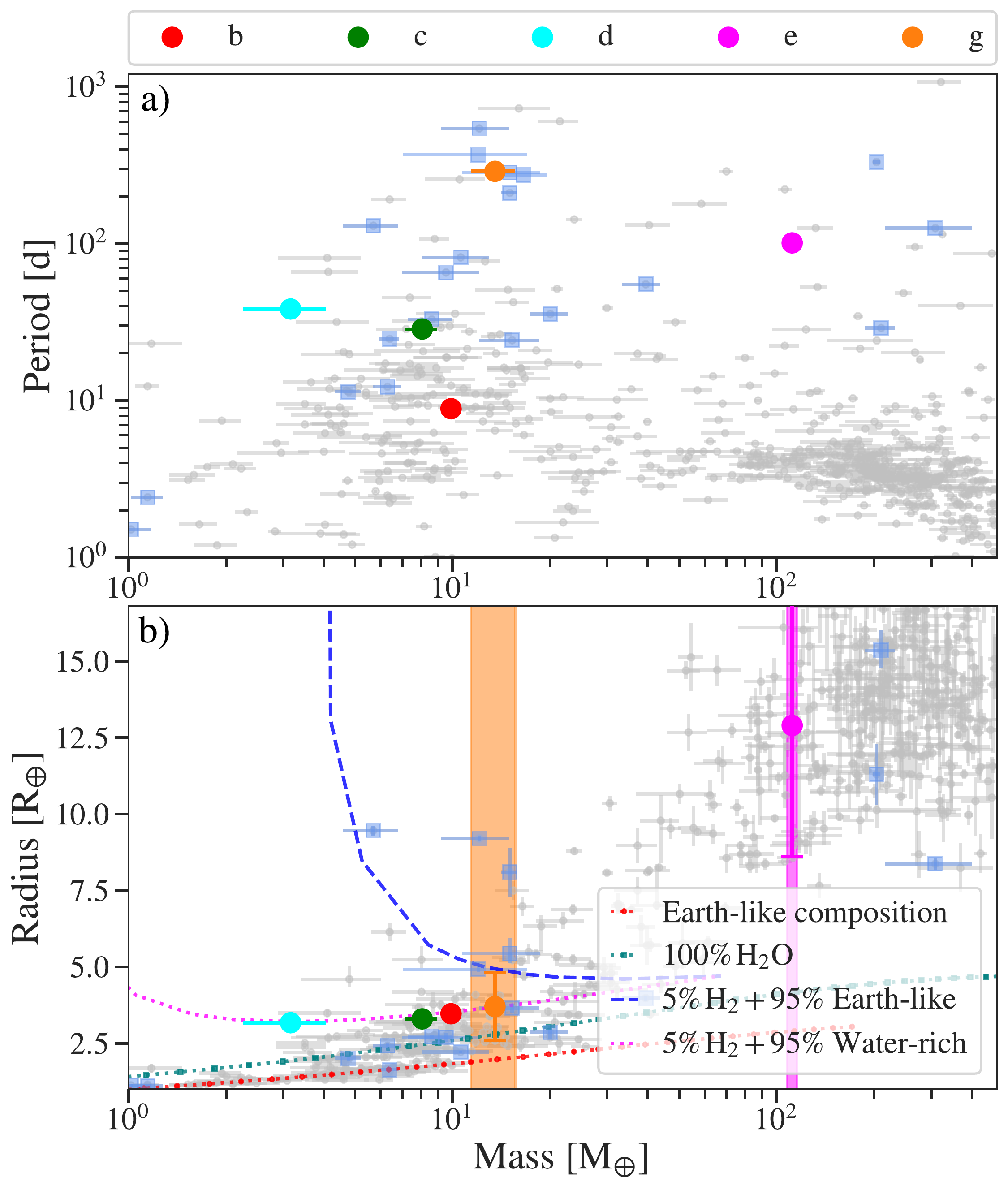}
    \end{subfigure}
    \caption{\label{fig: SYSTEM DIAGRAM PLOT}
    Mass--period (\textit{top panel}) and mass--radius (\textit{bottom panel}) diagrams for well-characterised planets with masses and radii measured with a precision better than 30\%, from 1\,${M}_\oplus$ up to 1.5\,${M}_{\mathrm{J}}$ and 1\,${R}_\oplus$ up to 1.5\,${R}_{\mathrm{J}}$, from the TEPCat database (February 2022; \citealp{Southworth_database}) and \url{http://exoplanet.eu}. HD 191939 planets are colour coded and marked with a filled circle with error bars (the radii  of HD 191939\,e and g are forecasted). Vertical colour bands indicate the $\pm 1\sigma$ mass regions of HD 191939\,g and e. Temperate planets with $T_{\mathrm{eq}}$\,=\,250--395\,K are marked by blue squares. The mass--radius panel also shows theoretical composition models at 300\,K from \citet{Zeng_models}.}
\end{figure}

By compiling several RV observations, we detect the signal of a new, likely non-transiting planet, namely HD 191939\,g, with a period of 284$^{+10}_{-8}$\,d. We derived a minimum mass of 13.5$\pm$2.0\,${M}_\oplus$ with a precision of 15\%. Moreover, we refined the planet properties of the previously known planets, HD 191939\,b, c, d, and e. In particular, we were able to determine the semi-amplitude and planetary mass of HD 191939\,d 
with a $4.6\sigma$ level of significance and confirm the low density of this sub-Neptune planet.
Figure\,\ref{fig: SYSTEM DIAGRAM PLOT} puts in context the different planets of the HD 191939 system as compared to the known exoplanets with masses and radii measured with a precision of better than 30\%, from 1\,${M}_\oplus$ up to 1.5\,${M}_{\mathrm{J}}$, and 1\,${R}_\oplus$ up to 1.5\,${R}_{\mathrm{J}}$, along with theoretical composition models of \citet{Zeng_models}\footnote{\url{ https://lweb.cfa.harvard.edu/~lzeng/planetmodels.html}}.

\subsection{HD 191939\,g: a new planet}
\label{sect: HD 191939 g}

\begin{figure}
    \centering
    \includegraphics[width=\hsize]{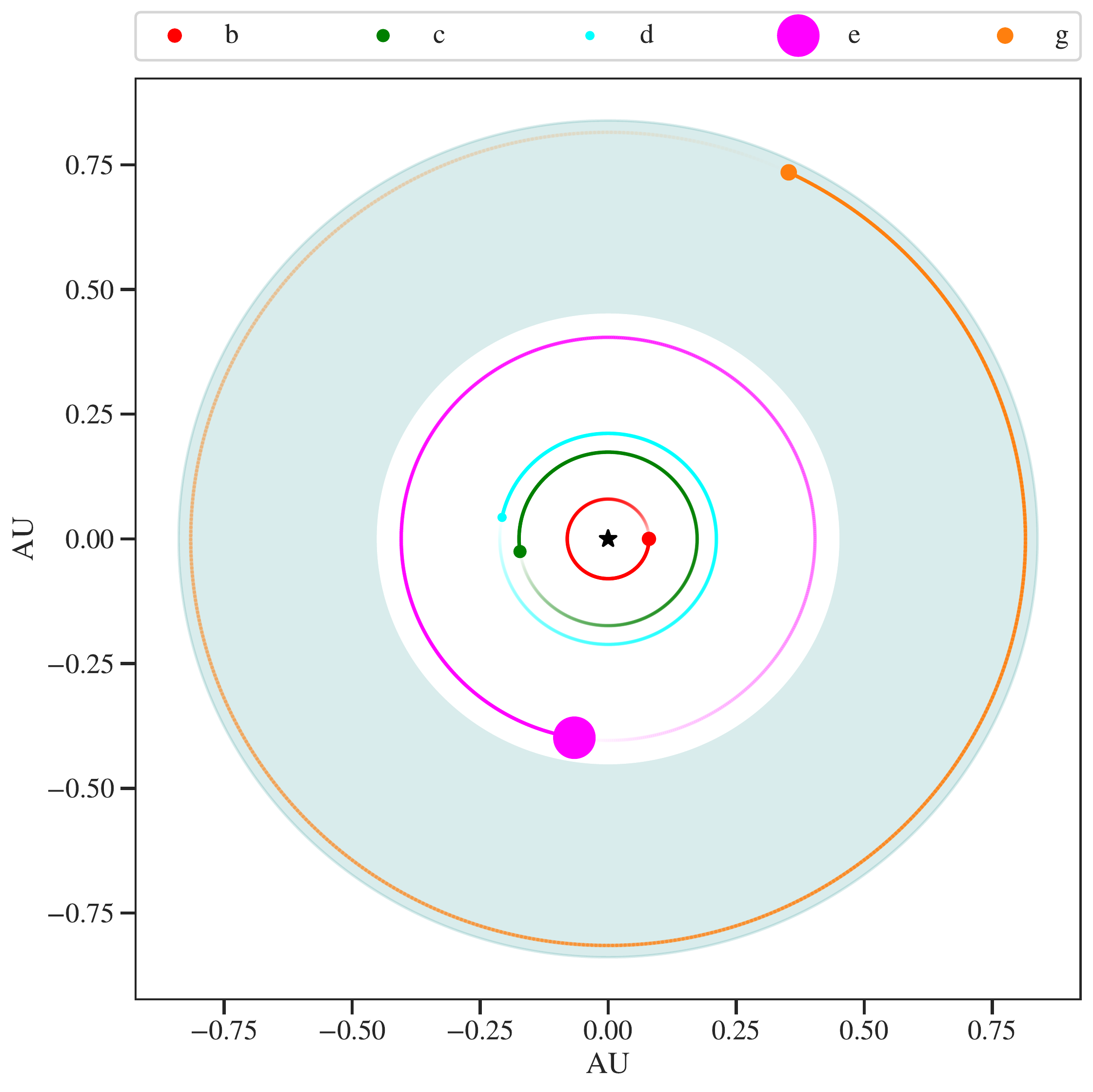}
    \caption{\label{fig: HZ diagram}
    Diagram of the planetary system HD 191939. Planetary orbits are colour coded and planets (filled circles) are scaled according to their mass. The shaded area marks the region of $T_{\mathrm{eq}}$\,=\,273--373\,K. HD 191939 is marked at the centre with a black star. A diagram including planet f is shown in Fig.\,\ref{fig: HZ planet F}.
    }
\end{figure}

With an orbital period of $\sim$280\,d and a minimum mass of 13.5\,${M}_\oplus$, HD 191939\,g joins the selected group of exoplanets that could
only be detected thanks to a large number of RV measurements spanning a relatively wide time interval. Because the RV method is more sensitive to shorter period planets and also to the more massive ones, there are only a few long-period planets with intermediate ($\sim$5--20\,${M}_\oplus$) and well-determined masses, namely:
HD 31527\,d (\textit{P}$\simeq$274\,d, 16$\pm$3\,${M}_\oplus$; \citealp{HD31527d_2011}),
HD 10180\,g (\textit{P}$\simeq$602\,d, 21$\pm$3\,${M}_\oplus$; \citealp{HD10180g_2011}),
GJ 3138\,d (\textit{P}$\simeq$257\,d, 10$\pm$2\,${M}_\oplus$; \citealp{GJ3138d_2017}),
Barnard\,b (\textit{P}$\simeq$232\,d, 3.2$\pm$0.4\,${M}_\oplus$; \citealp{Ribas_2018Natur}),
GJ 273\,d (\textit{P}$\sim$414\,d, 11$\pm$4\,${M}_\oplus$) and e (\textit{P}$\sim$542\,d, 9$\pm$4\,${M}_\oplus$; \citealp{GJ273_de_Tuomi}), and
GJ 687\,c (\textit{P}$\simeq$727\,d, 16$\pm$4\,${M}_\oplus$; \citealp{GJ687_2020}).
The intermediate- and long-period planet group is also supplemented with some transiting planets discovered by the \textit{Kepler} space mission \citep{Kepler_telescope}, as spectroscopic observations confirmed the planetary nature of HIP 41378\,d ($P$=278\,d, $<$4.6\,${M}_\oplus$),
e (\textit{P}$\simeq$369\,d, 12$\pm$5\,${M}_\oplus$), and
f ($P$=542\,d, 12$\pm$3\,${M}_\oplus$; \citealp{HIP41378f, Density_HIP_f}). However, the faintness of some of the \textit{Kepler} targets complicates RV follow-up campaigns. Finally, transit-time-variation analyses helped to compute the planetary masses for Kepler-87\,c ($P$\,=\,191\,d, 6.4$\pm$0.8\,${M}_\oplus$; \citealp{Kepler87}), KOI-1783\,c ($P$\,=\,284\,d, 15.0$^{+4.3}_{-3.6}$\,${M}_\oplus$; \citealp{KOI1783c_Vissa2020}), and Kepler-90\,g ($P$\,=\,210\,d, 15$\pm$1\,${M}_\oplus$; \citealp{Cabrera_2014_Kepler-90, Density_K90_g}).

Because HD 191939\,g and e are only detected in the RV measurements and \textit{TESS} photometry does not show evidence of their transits, we were not able to determine radii. We therefore marked their positions with vertical bands in the mass--radius diagram (Fig.\,\ref{fig: SYSTEM DIAGRAM PLOT} \textit{bottom}). However, their radii can be forecasted using empirical mass--radius relations for planets. We used the probabilistic planet mass--radius relation given in \citet{Forecaster} via its \texttt{python} implementation\footnote{\url{https://github.com/chenjj2/forecaster}}. The code predicted planetary radii for planets g and e of 3.7$^{+1.5}_{-1.1}$\,${R}_\oplus$ and 12.9$^{+4.7}_{-3.9}$\,${R}_\oplus$, respectively. From their minimum masses and forecasted radius, the expected mean bulk densities for planets g and e are 1.5$^{+2.7}_{-0.8}$\,$\mathrm{g\,cm^{-3}}$ and 0.30$^{+0.70}_{-0.16}$\,$\mathrm{g\,cm^{-3}}$, respectively. To crosscheck these forecasted results for planet g, we also estimated its radius using the mass--radius relation for sub-Neptune-sized planets of \citet{Wolfgang_2016}. This method predicts a planetary radius of $\sim$3.4\,${R}_\oplus$ for planet g, which is consistent with the above estimation. If we extrapolate this mass--radius relation to planet e, the predicted radius is about $\sim$17\,${R}_\oplus$, also falling within the uncertainties.
However, we stress that these estimated values are merely illustrative, and should not be considered as the actual planetary radii and densities.

HD 191939\,g is the only planet in the system in the conservative habitable zone (HZ) of the star; that is, by definition its $T_{\mathrm{eq}}$ is compatible with the presence of liquid water ($T$\,$\in$\,$[273,373]$\,K). With a semi-major axis of $\sim$0.82\,AU, planet g is in the outer edge of the HZ, which we set at $\sim$0.44--0.84\,AU. Figure\,\ref{fig: HZ diagram} displays a face-on view of the HD 191939 system, where the HZ of the star is marked. However, we stress that despite being in the HZ, HD 191939\,g, being a gaseous planet, cannot be considered as a habitable planet.

Long-period, intermediate-mass planets are located in a lonely region of the mass--period diagram, with HD 191939\,g at the centre of this group (see Fig.\,\ref{fig: SYSTEM DIAGRAM PLOT}). These objects have the commonality that they are outer planets in their respective planetary systems and their orbits are in or near the HZ of the host star. In the mass--radius diagram, planets with masses of $\sim$13.5\,${M}_\oplus$ are above the Earth-like composition line, supporting the idea that HD 191939\,g is likely a gaseous planet. Moreover, its $1\sigma$ mass uncertainty overlaps with Kepler-90\,g and HIP 41378\,f mass determinations (see Fig.\,\ref{fig: SYSTEM DIAGRAM PLOT}). These planets are two of the lowest density (puffy) planets  known, with $\rho_\mathrm{p}$\,=\,0.15$\pm$0.05\,$\mathrm{g\,cm^{-3}}$ \citep{Density_K90_g} and $\rho_\mathrm{p}$\,=\,0.09$\pm$0.02\,$\mathrm{g\,cm^{-3}}$ \citep{Density_HIP_f}, respectively.

\subsection{HD 191939\,d: a puffy planet}
\label{sect: HD 191939 d}

\citetalias{Lubin_HD191939} already noted that HD 191939\,d was probably a low-density planet. Here, we derived a bulk planetary density of $\rho_\mathrm{d}$\,=\,0.57$\pm$0.13\,$\mathrm{g\,cm^{-3}}$, confirming this previous result with better uncertainty. With such low density, planet d is at the edge of the planetary mass--radius distribution (Fig.\,\ref{fig: SYSTEM DIAGRAM PLOT} \textit{bottom}). Thermal expansion of the atmosphere is a possible mechanism to explain planet inflation leading to puffy atmospheres, such as in the case of ultra-hot Jupiters. However, given the relatively cold equilibrium temperature of HD 191939\,d ($T_{\mathrm{eq}}$\,$\simeq$\,540\,K), this explanation is unlikely.

The nearest planet to HD 191939\,d in the mass--radius diagram is Kepler-79\,e (KOI-152\,e; \citealp{Kepler79_System}), which has very similar properties (3.49$\pm$0.14\,${R}_\oplus$, $\rho_\mathrm{p}$\,=\,0.53$\pm$0.15\,$\mathrm{g\,cm^{-3}}$, $T_{\mathrm{eq}}$\,$\simeq$\,480\,K). The most relevant characteristic of the Kepler-79 system is the low density of its planets, whose masses were calculated from transit-time variations. Their densities range between $\rho_\mathrm{p}$\,=\,0.09 and 1.43\,$\mathrm{g\,cm^{-3}}$ and the densest planet is the innermost one. This similarity with the HD 191939 transiting planets reinforces the hypothesis that the non-transiting planets of the system are also
of a gaseous-like composition.

The brightness (J\,=\,7.6\,mag) and low level of stellar activity of the host star offer an excellent opportunity to inspect and study the atmosphere of a puffy planet. To quantify the viability of these observations, we computed the transmission spectroscopy metric (TSM) proposed by \citet{Kempton_2018_TSM}. The estimated TSM value for HD 191939\,d is 227, which is well above the threshold of 90 indicated by \citet{Kempton_2018_TSM} and planets b and c (TSM$_{\mathrm{b}}$\,=\,153; TSM$_{\mathrm{c}}$\,=\,107).
Moreover, HD 191939\,d has a much better TSM value than other known puffy planets such as HIP 41378\,d (TSM\,=\,71), HIP 41378\,e (TSM\,=\,57), Kepler-79 planets (TSM\,=\,7--60), or Kepler-90\,g (TSM\,=\,27). We note that the TSM is simply proportional to the expected transmission spectroscopy S/N, assuming standardised planetary atmosphere models (e.g. clear atmosphere with solar composition). Observational surveys do not support a strong correlation between expected transmission spectroscopy S/N and actual atmospheric detectability \citep{tsiaras2018}.

We searched for planets with a radius of $\sim$3\,${R}_\oplus$ and/or a $T_{\mathrm{eq}}$\,$\sim$\,500--600\,K in the database of exoplanet atmospheric observations of ExoAtmospheres\footnote{\url{http://research.iac.es/proyecto/exoatmospheres/index.php}}. Only the warm sub-Neptune GJ\,1214\,b (2.74$\pm$0.05\,${R}_\oplus$, $T_{\mathrm{eq}}$\,$\simeq$\,596\,K; \citealp{Cloutier_2021}) fitted our conditions, although it is a denser planet ($\rho_\mathrm{p}$\,=\,2.20$\pm$0.17\,$\mathrm{g\,cm^{-3}}$). For GJ\,1214\,b (TSM\,=\,440), only a tentative detection of \ion{He}{I} could be set recently \citep{Orell_2022} after many non-detection results \citep{Bean_2010, Kreidberg_2014, Petit_2020, Kasper_2020}.
When we looked for puffy planet observations, we found that the atmosphere of HIP 41378\,f was analysed via transmission spectroscopy at low resolution with the \textit{Hubble Space Telescope} (\textit{HST}). However, HIP 41378\,f, with a higher TSM (=\,342) than HD 191939\,d, showed a featureless NIR spectrum with a median precision of 84 ppm \citep{HIP_f_featureless}.

\begin{figure*}
    \sidecaption
    \includegraphics[width=12cm]{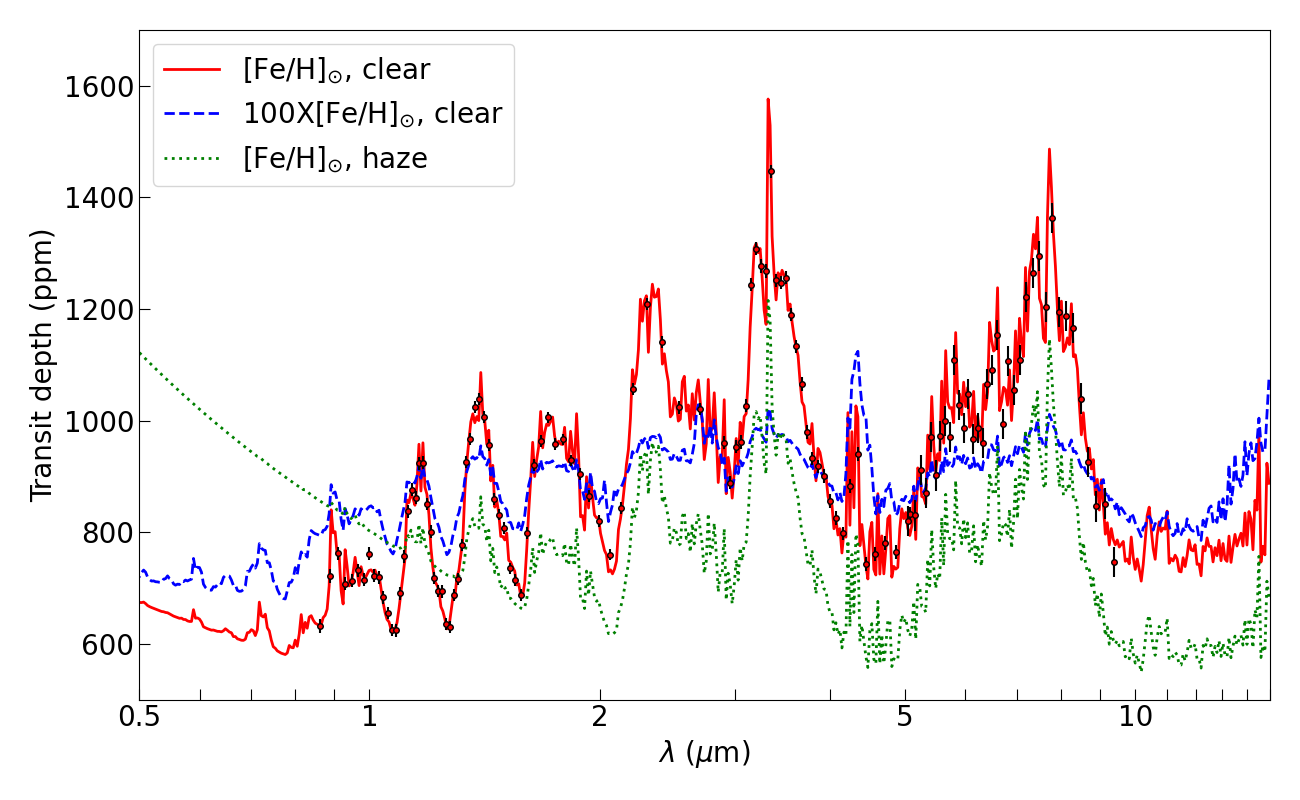}
    \caption{Synthetic transmission spectra for HD 191939\,d. Models assuming a clear atmosphere with solar abundances (solid red line), a clear atmosphere with metallicity enhanced by a factor of 100 (blue dashed line), and a hazy atmosphere with solar abundances (green dotted line). Simulated measurements with error bars are shown for the observation of one transit with \textit{JWST} NIRISS-SOSS, NIRSpec-G395M, and MIRI-LRS configurations. \label{fig:jwst_atmo}
    }
\end{figure*}

We explored the potential of HD 191939\,d for transmission spectroscopy with the \textit{James Webb Space Telescope} (\textit{JWST}) through spectral simulations for a range of atmospheric scenarios. We adopted \texttt{TauREx 3} \citep{al-refaie2021} to compute our set of model atmospheres using the atmospheric chemical equilibrium (ACE) module \citep{agundez2012}, including collisionally induced absorption by H$_2$–H$_2$ and H$_2$–He \citep{abel2011,abel2012,fletcher2018}, and Rayleigh scattering. The benchmark model assumes a clear atmosphere with solar composition, which displays the largest spectral features. The other models include the dampening effects on the transmission spectrum due to enhanced metallicity or haze in the HD 191939\,d atmosphere. The haze was modelled with \texttt{TauREx 3} using a Mie scattering contribution with the formalism of \cite{lee2013}. A super-solar metallicity is indeed predicted for low-mass, low-density planets such as HD 191939\,d based on the core accretion theory of planet formation \citep{fortney2013,thorngren2016}. The equilibrium temperature of 540\,K also favours the formation of high-altitude photochemical haze in the HD 191939\,d atmosphere \citep{gao2020,ohno2021,yu2021}.

We used \texttt{ExoTETHyS} \citep{morello2021} to simulate the corresponding \textit{JWST} spectra, as observed with the NIRISS-SOSS (0.6--2.8\,$\mu$m), NIRSpec-G395M (2.88--5.20\,$\mu$m), and MIRI-LRS (5--12\,$\mu$m) instrumental modes. The procedure to select the spectral bins and estimate the error bars was identical to that of previous papers (e.g. \citealp{espinoza2022,luque2022}). In particular, we obtained error bars of 10--12\,ppm per spectral point for the NIRISS-SOSS and NIRSpec-G395M modes at median resolving power of $\mathcal{R}$\,$\sim$\,50, and of 27\,ppm for the MIRI-LRS bins with sizes of 0.1--0.2\,$\mu$m. We note that the predicted error bars in the NIR spectrum of HD 191939\,d are seven times smaller than those reported for HIP 41378\,f by \cite{HIP_f_featureless}.

Figure\,\ref{fig:jwst_atmo} shows the synthetic transmission spectra for three selected atmospheric configurations, one of which with simulated \textit{JWST} observations overplotted. These spectra exhibit strong absorption features due to H$_2$O and CH$_4$, which are an order of magnitude larger than the predicted error bars. Table\,\ref{table:jwst_atmo} reports the amplitudes of spectral modulations at low resolution ($\mathcal{R}$\,$\sim$\,170) for the full set of synthetic spectra within the nominal wavelength ranges of \textit{HST} and \textit{JWST} instrumental modes. Higher metallicities lead to smaller absorption features over the entire spectral range, as they increase the mean molecular weight, thereby reducing the atmospheric scale height. The haze mostly affects the visible and NIR portion of the spectrum, flattening the absorption features and introducing a possible slope. The mid-IR spectrum is less severely affected by haze, but also depends on its physical properties. Based on the predicted spectroscopic amplitudes, even a clear atmosphere with 1\,000$\times$ solar metallicity or a hazy one with 100$\times$ solar metallicity surrounding HD 191939\,d would be detectable with a single \textit{JWST} visit. Multiple instruments can break the degeneracy between haze or clouds and metallicity effects. Similar considerations could also apply to other puffy planets with a flat near-infrared spectrum observed with \textit{HST} WFC3-G141 \citep{HIP_f_featureless,chachan2020,libby-roberts2020}, albeit with quantitative differences.

\begin{table*}
\caption[width=\textwidth]{Transmission spectroscopy amplitudes at low resolution ($\mathcal{R}$\,$\sim$\,170) for various models of the HD 191939\,d atmosphere and nominal wavelength ranges of \textit{HST} and \textit{JWST} instrumental modes. These include clear or hazy atmospheres with scaled solar metallicities. We adopted the formalisms of \cite{lee2013} for the hazy models, where $\alpha$ denotes the particle size in $\mu$m, the mixing ratio is $\chi_c$\,=\,10$^{-12}$, and the extinction coefficient is $Q_0$\,=\,40. Lines in bold correspond to the synthetic spectra shown in Figure\,\ref{fig:jwst_atmo}.
\label{table:jwst_atmo}
}
\centering

\begin{tabular}{cccccc}

\hline \hline 
\noalign{\smallskip} 

Haze & Met. & \multicolumn{4}{c}{Spectroscopic amplitudes (ppm)}\vspace{0.05cm} \\
 & ($[M/H]_{\oplus}$) & 1.1--1.7\,$^{(a)}$ & 0.6--2.8\,$^{(b)}$ & 2.88--5.20\,$^{(c)}$ & 5--12\,$^{(d)}$\,$\mu$m \\
\noalign{\smallskip}
\hline
\noalign{\smallskip}
\textbf{Clear} & \textbf{1} & \textbf{460} & \textbf{664} & \textbf{860} & \textbf{774} \vspace{0.05cm} \\
Clear & 10 & 490 & 665 & 670 & 643 \vspace{0.05cm} \\
\textbf{Clear} & \textbf{100} & \textbf{195} & \textbf{348} & \textbf{294} & \textbf{219} \vspace{0.05cm} \\
Clear & 1000 & 38 & 77 & 59 & 37 \vspace{0.05cm} \\
\textbf{$\alpha=$\,0.05} & \textbf{1} & \textbf{201} & \textbf{400} & \textbf{658} & \textbf{597} \vspace{0.05cm} \\
$\alpha=$\,0.05 & 100 & 144 & 213 & 276 & 206 \vspace{0.05cm} \\
$\alpha=$\,0.10 & 1 & 174 & 693 & 380 & 400 \vspace{0.05cm} \\
$\alpha=$\,0.10 & 100 & 57 & 181 & 250 & 190 \\
\noalign{\smallskip}
\hline
\end{tabular}
\tablefoot{Nominal wavelength ranges of $^{(a)}$ HST WFC3-G141 scanning mode, $^{(b)}$ JWST NIRISS-SOSS mode, $^{(c)}$ JWST NIRSpec-G395M mode, and $^{(d)}$ JWST MIRI-LRS mode.}
\end{table*}

\subsection{Architecture of the planetary system}
\label{sect: Architecture}

\citetalias{Mariona_2020} and \citetalias{Lubin_HD191939} already noted that the transiting planets are close to a near mean motion resonance of 1:3:4 (\textit{P}$_\mathrm{b}$=8.88\,d, \textit{P}$_\mathrm{c}$=28.58\,d, \textit{P}$_\mathrm{d}$=38.35\,d). HD 191939\,e, with a much longer period, seems disconnected from that resonance chain. However, the discovery of HD 191939\,g reveals that the non-transiting planets of the system appear to be in a period ratio of 1:3 ($P_\mathrm{e}$\,$=$\,101\,d, $P_\mathrm{g}$\,$\simeq$\,280\,d). There are other cases of multi-planetary systems where inner planets are gathered in a different resonance chain from the outer ones: Kepler-90 planets \citep{Cabrera_2014_Kepler-90} are in 2:3:4 and 4:5 periods, and HIP 41378 planets (\citealp{HIP41378f, Density_HIP_f}) are in 1:2:4 and 3:4:6 periods. Furthermore, \citetalias{Lubin_HD191939} noted that planetary systems hosting puffy planets tend to have their planets in resonance (e.g. Kepler-79, \citealp{Kepler79_System}; Kepler-51, \citealp{Kepler51_TTV}; Kepler-87, \citealp{Kepler87}), which seems to also be the case for HD 191939.

\citetalias{Lubin_HD191939} searched in the literature for planetary systems similar to that described in their work; their best match was Kepler-68. \citet{Kepler_25_65_68} described this system as two sub-Neptunes interior to a 634d period Jovian planet, and with strong evidence for an object with $>$0.6\,$M_{\mathrm{J}}$ in a very long-period orbit ($\gg$3000\,d). Another similar system analysed in \citet{Kepler_25_65_68} is Kepler-65: three sub-Neptunes near orbital resonance of 1:3:4 interior to a non-transiting planet with a mass of 212\,${M}_\oplus$ and a period of 259\,d. However, the detection of a Uranus-mass planet between the warm Saturn and the massive long-period planet makes the HD 191939  system more exceptional.

We performed a new search in the NASA Exoplanet Archive\footnote{\url{https://exoplanetarchive.ipac.caltech.edu/}} database looking for systems with 2--4 intermediate planets (2--25\,${M}_\oplus$ or 2--8\,${R}_\oplus$) interior to a gas-giant planet (>50\,${M}_\oplus$ or >8\,${R}_\oplus$) plus a planet with comparable properties to those of the inner ones. Although a system with three intermediate planets interior to a gas giant plus another intermediate planet was not found, our search returned one system that suited our initial conditions: KOI-94 \citep{KOI-94}.

\begin{figure}
    \centering
    \includegraphics[width=\hsize]{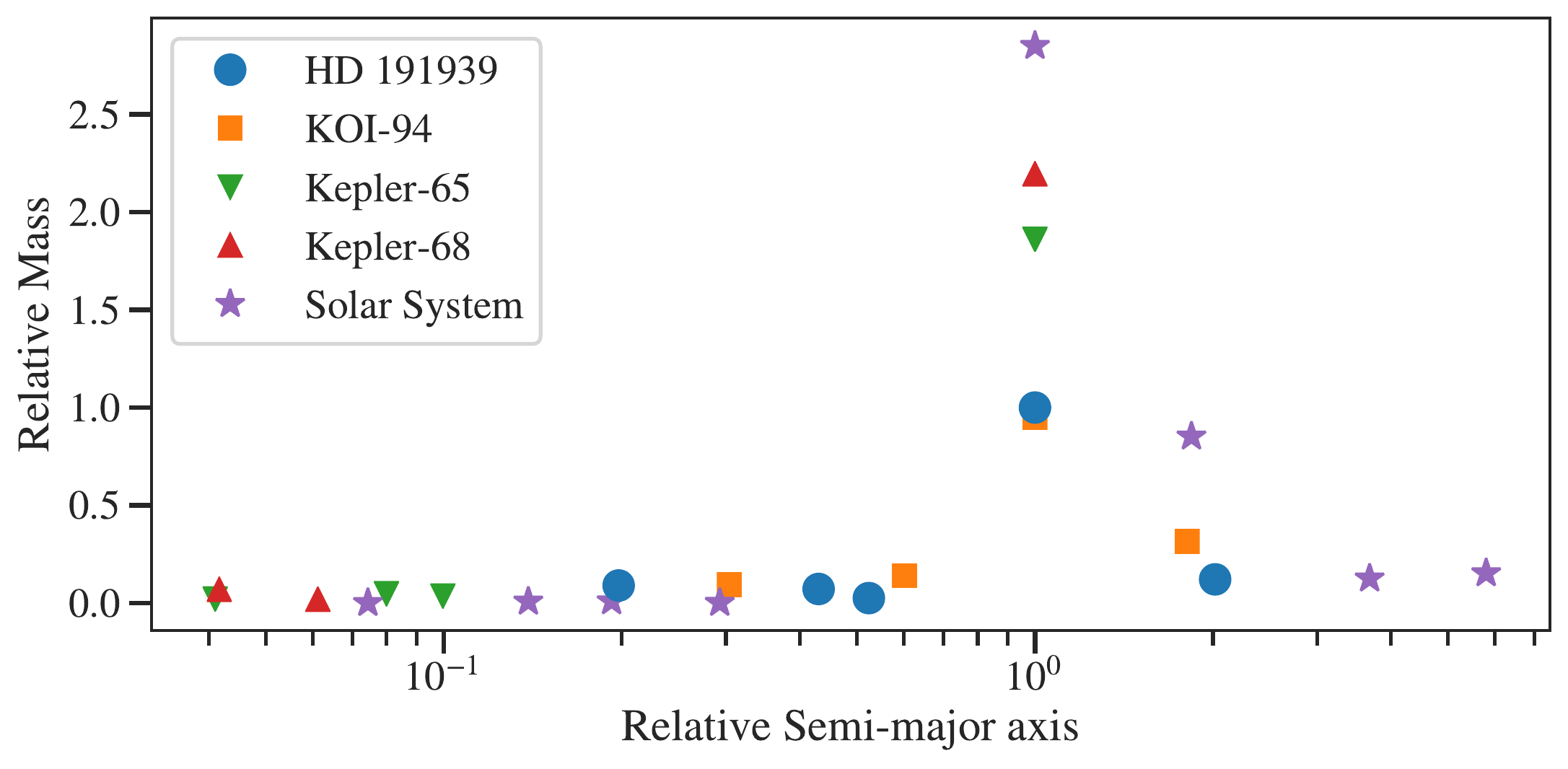}
    \caption{\label{fig: KOI94 comparison}
    Planet mass distribution across planets for HD 191939 (blue circles), KOI-94 (orange squares), Kepler-65 (green down triangles), Kepler-68 (red up triangles), and the Solar System (purple stars) systems. Masses are scaled to HD 191939\,e minimum mass, and semi-major axes are scaled to that of the massive planet for each system.
    }
\end{figure}

KOI-94 has two inner planets with masses of 10.5$\pm$4.6\,${M}_\oplus$ and <21.3\,${M}_\oplus$, and then a warm Saturn-like planet (106$\pm$11\,${M}_\oplus$, 11.2$\pm$1.1\,${R}_\oplus$) and an outer planet slightly more massive than the inner ones (35$^{+18}_{-28}$\,${M}_\oplus$, 6.6$\pm$0.6\,${R}_\oplus$), but without evidence for a long-period companion like HD 191939\,f. KOI-94 planets are gaseous with low densities ranging between $\sim$0.35 and 1\,$\mathrm{g\,cm^{-3}}$ (except for KOI-94\,b; 10.1$\pm$5.5\,$\mathrm{g\,cm^{-3}}$), and the mass of KOI-94\,d is consistent with the minimum mass derived for HD 191939\,e. KOI-94\,b, c, and d are close to a mean motion resonance of 1:3:6 (\textit{P}$_\mathrm{b}$\,$=$\,3.7\,d, \textit{P}$_\mathrm{c}$\,$=$\,10.4\,d, \textit{P}$_\mathrm{d}$\,$=$\,22.3\,d), with the giant planet also close to 5:2 with the outermost planet KOI-94\,e (\textit{P}$_\mathrm{e}$\,$=$\,54\,d).
In both systems, a more massive planet divides the planets with similar masses (and likely similar characteristics). This mass distribution across the planets is illustrated in Figure\,\ref{fig: KOI94 comparison}. Although the KOI-94 system is more compact than HD 191939, the semi-major axis ($a$) scaled to the semi-major axis of the most massive known planet (KOI-94\,d and HD 191939\,e, respectively) shows that the $a$ of the outer planet (KOI-94\,e and HD 191939\,g, respectively) is approximately twice  that of the massive one and the $a$ of the inner planets are $\sim$0.5 that of the massive one. The planets in these systems present resonances between periods with the massive planet linked to the outer one.

Moreover, in addition to the spectral type of the host star, the HD 191939 system presents some similarities (excluding planet f) to our own Solar System. The smaller planets are interior to the massive planet and the intermediate-mass planets are further out relative to the massive one. Still, the planetary
system HD 191939  is more compact, with all the constrained planets in the system confined within $\sim$0.82\,au (HD 191939\,g semi-major axis), which is comparable to the Venus orbital distance of 0.72\,au.

\section{Conclusions}
\label{sec: CONCLUSIONS}

The multi-planetary system around HD 191939 was previously known to host three transiting sub-Neptunes with very similar radii (HD 191939\,b, c, and d) and a non-transiting Saturn-mass planet (HD 191939\,e), and also showed evidence for an external long-period planet (HD 191939\,f). In this paper, we revisited the system using new RV data from CARMENES and HARPS-N spectrographs in addition to archival data from APF and HIRES. 
The combined dataset, containing 362 RV measurements spanning over $\sim$2 years, allowed the detection of a new non-transiting planetary signal (HD 191939\,g) with a period of $\sim$280\,d and a minimum mass of $M_\mathrm{p}\sin i$\,$=$\,13.5$\pm$2.0\,${M}_\oplus$. The planet-to-star distance of HD 191939\,g places this new planet in the conservative HZ around the host star. However, our measurements suggest HD 191939\,g is likely a gaseous planet.

We also present refined mass and bulk properties for planets HD 191939\,b, c, and e. Additionally, we improve the mass determination of HD 191939\,d at the $4.6\sigma$ level of significance, for which only an upper limit was known.

We determine a mass for HD 191939\,d of 2.80$\pm$0.60\,${M}_\oplus$, leading to a mean bulk density of $\rho_\mathrm{d}$\,$=$\,0.57\,$\mathrm{g\,cm^{-3}}$. Due to its low density and host-star brightness, HD 191939\,d is one of the best puffy targets for atmospheric exploration via transmission spectroscopy. Although the detection of spectral features in puffy atmospheres seems to be challenging, \textit{JWST} may be capable of detecting the atmosphere of HD 191939\,d  based on the predicted spectroscopic amplitudes. In particular, our simulations suggest that \textit{JWST} instruments may break the degeneracy between hazes, clouds, and metallicity effects with a single visit.

With a period of 101\,d, HD 191939\,e was disconnected from the near resonance chain of the three inner transiting planets (1:3:4). However, the detection of HD 191939\,g in a 280-day orbit indicates that these two outer non-transiting planets (e and g) are in a separate relation, close to a 1:3 period resonance. HD 191939 does not seem to be unique in this respect, as there are other multi-planetary systems in the literature where the inner and outer planets are in different resonance chains. Moreover, puffy planets tend to be in resonant orbits, which reinforces the hypothesis of a low mean density for planets e and g.

The singular system architecture of three sub-Neptunes interior to a Saturn-mass planet and a Uranus-mass planet, together with the existence of a very long-period massive companion, makes the HD 191929 system unique. The diversity of planets around this star makes this system a prime target for more follow-up observations.


\begin{acknowledgements}

This work was supported by the KESPRINT collaboration, an international consortium devoted to the characterization and research of exoplanets discovered with space-based missions (\url{www.kesprint.science}).

CARMENES is an instrument at the Centro Astron\'omico Hispano-Alem\'an (CAHA) at Calar Alto (Almer\'{\i}a, Spain), operated jointly by the Junta de Andaluc\'ia and the Instituto de Astrof\'isica de Andaluc\'ia (CSIC).

CARMENES was funded by the Max-Planck-Gesellschaft (MPG), the Consejo Superior de Investigaciones Cient\'{\i}ficas (CSIC), the Ministerio de Econom\'ia y Competitividad (MINECO) and the European Regional Development Fund (ERDF) through projects FICTS-2011-02, ICTS-2017-07-CAHA-4, and CAHA16-CE-3978,  and the members of the CARMENES Consortium
(Max-Planck-Institut f\"ur Astronomie,
Instituto de Astrof\'{\i}sica de Andaluc\'{\i}a,
Landessternwarte K\"onigstuhl,
Institut de Ci\`encies de l'Espai,
Institut f\"ur Astrophysik G\"ottingen,
Universidad Complutense de Madrid,
Th\"uringer Landessternwarte Tautenburg,
Instituto de Astrof\'{\i}sica de Canarias,
Hamburger Sternwarte,
Centro de Astrobiolog\'{\i}a and
Centro Astron\'omico Hispano-Alem\'an), 
with additional contributions by the MINECO, 
the Deutsche Forschungsgemeinschaft (DFG) through the Major Research Instrumentation Programme and Research Unit FOR2544 ``Blue Planets around Red Stars'', 
the Klaus Tschira Stiftung, 
the states of Baden-W\"urttemberg and Niedersachsen, 
and by the Junta de Andaluc\'{\i}a.

This research has made use of data obtained from or tools provided by the portal \url{exoplanet.eu} of The Extrasolar Planets Encyclopaedia.
JK gratefully acknowledge the support of the Swedish National Space Agency (SNSA; DNR 2020-00104). PK is acknowledges the support from the grant LTT-20015. K.W.F.L. was supported by Deutsche Forschungsgemeinschaft grants RA714/14-1 within the DFG Schwerpunkt SPP 1992, Exploring the Diversity of Extrasolar Planets. C.M.P. gratefully acknowledge the support of the  Swedish National Space Agency (DNR 65/19).
The first author acknowledges the special support received by P.\,Conxa, P.\,Merc\`e, Jeroni, and Merc\`e. This work has made use of resources from AstroPiso collaboration. J.O.M. gratefully acknowledge the inspiring discussions with Yess, Alejandro, his colleagues, and friends.

\end{acknowledgements}

%
%

\bibliographystyle{aa}
\bibliography{references}

\begin{thebibliography}{94}
\expandafter\ifx\csname natexlab\endcsname\relax\def\natexlab#1{#1}\fi

\bibitem[{{Abel} {et~al.}(2011){Abel}, {Frommhold}, {Li}, \& {Hunt}}]{abel2011}
{Abel}, M., {Frommhold}, L., {Li}, X., \& {Hunt}, K. L.~C. 2011, Journal of
  Physical Chemistry A, 115, 6805

\bibitem[{{Abel} {et~al.}(2012){Abel}, {Frommhold}, {Li}, \& {Hunt}}]{abel2012}
{Abel}, M., {Frommhold}, L., {Li}, X., \& {Hunt}, K. L.~C. 2012, \jcp, 136,
  044319

\bibitem[{{Ag{\'u}ndez} {et~al.}(2012){Ag{\'u}ndez}, {Venot}, {Iro}, {Selsis},
  {Hersant}, {H{\'e}brard}, \& {Dobrijevic}}]{agundez2012}
{Ag{\'u}ndez}, M., {Venot}, O., {Iro}, N., {et~al.} 2012, \aap, 548, A73

\bibitem[{{Al-Refaie} {et~al.}(2021){Al-Refaie}, {Changeat}, {Waldmann}, \&
  {Tinetti}}]{al-refaie2021}
{Al-Refaie}, A.~F., {Changeat}, Q., {Waldmann}, I.~P., \& {Tinetti}, G. 2021,
  \apj, 917, 37

\bibitem[{{Alam} {et~al.}(2022){Alam}, {Kirk}, {Dressing}, {L{\'o}pez-Morales},
  {Ohno}, {Gao}, {Akinsanmi}, {Santerne}, {Grouffal}, {Adibekyan}, {Barros},
  {Buchhave}, {Crossfield}, {Dai}, {Deleuil}, {Giacalone}, {Lillo-Box},
  {Marley}, {Mayo}, {Mortier}, {Santos}, {Sousa}, {Turtelboom}, {Wheatley}, \&
  {Vanderburg}}]{HIP_f_featureless}
{Alam}, M.~K., {Kirk}, J., {Dressing}, C.~D., {et~al.} 2022, \apjl, 927, L5

\bibitem[{{Ambikasaran} {et~al.}(2014){Ambikasaran}, {Foreman-Mackey},
  {Greengard}, {Hogg}, \& {O'Neil}}]{george}
{Ambikasaran}, S., {Foreman-Mackey}, D., {Greengard}, L., {Hogg}, D.~W., \&
  {O'Neil}, M. 2014

\bibitem[{{Astudillo-Defru} {et~al.}(2017){Astudillo-Defru}, {Forveille},
  {Bonfils}, {S{\'e}gransan}, {Bouchy}, {Delfosse}, {Lovis}, {Mayor}, {Murgas},
  {Pepe}, {Santos}, {Udry}, \& {W{\"u}nsche}}]{GJ3138d_2017}
{Astudillo-Defru}, N., {Forveille}, T., {Bonfils}, X., {et~al.} 2017, \aap,
  602, A88

\bibitem[{{Badenas-Agusti} {et~al.}(2020){Badenas-Agusti}, {G{\"u}nther},
  {Daylan}, {Mikal-Evans}, {Vanderburg}, {Huang}, {Matthews}, {Rackham},
  {Bieryla}, {Stassun}, {Kane}, {Shporer}, {Fulton}, {Hill}, {Nowak}, {Ribas},
  {Pall{\'e}}, {Jenkins}, {Latham}, {Seager}, {Ricker}, {Vanderspek}, {Winn},
  {Abril-Pla}, {Collins}, {Serra}, {Niraula}, {Rustamkulov}, {Barclay},
  {Crossfield}, {Howell}, {Ciardi}, {Gonzales}, {Schlieder}, {Caldwell},
  {Fausnaugh}, {McDermott}, {Paegert}, {Pepper}, {Rose}, \&
  {Twicken}}]{Mariona_2020}
{Badenas-Agusti}, M., {G{\"u}nther}, M.~N., {Daylan}, T., {et~al.} 2020, \aj,
  160, 113

\bibitem[{{Baranne} {et~al.}(1996){Baranne}, {Queloz}, {Mayor}, {Adrianzyk},
  {Knispel}, {Kohler}, {Lacroix}, {Meunier}, {Rimbaud}, \&
  {Vin}}]{1996A&AS..119..373B}
{Baranne}, A., {Queloz}, D., {Mayor}, M., {et~al.} 1996, \aaps, 119, 373

\bibitem[{{Bean} {et~al.}(2010){Bean}, {Miller-Ricci Kempton}, \&
  {Homeier}}]{Bean_2010}
{Bean}, J.~L., {Miller-Ricci Kempton}, E., \& {Homeier}, D. 2010, \nat, 468,
  669

\bibitem[{{Borucki} {et~al.}(2010){Borucki}, {Koch}, {Basri}, {Batalha},
  {Brown}, {Caldwell}, {Caldwell}, {Christensen-Dalsgaard}, {Cochran},
  {DeVore}, {Dunham}, {Dupree}, {Gautier}, {Geary}, {Gilliland}, {Gould},
  {Howell}, {Jenkins}, {Kondo}, {Latham}, {Marcy}, {Meibom}, {Kjeldsen},
  {Lissauer}, {Monet}, {Morrison}, {Sasselov}, {Tarter}, {Boss}, {Brownlee},
  {Owen}, {Buzasi}, {Charbonneau}, {Doyle}, {Fortney}, {Ford}, {Holman},
  {Seager}, {Steffen}, {Welsh}, {Rowe}, {Anderson}, {Buchhave}, {Ciardi},
  {Walkowicz}, {Sherry}, {Horch}, {Isaacson}, {Everett}, {Fischer}, {Torres},
  {Johnson}, {Endl}, {MacQueen}, {Bryson}, {Dotson}, {Haas}, {Kolodziejczak},
  {Van Cleve}, {Chandrasekaran}, {Twicken}, {Quintana}, {Clarke}, {Allen},
  {Li}, {Wu}, {Tenenbaum}, {Verner}, {Bruhweiler}, {Barnes}, \&
  {Prsa}}]{Kepler_telescope}
{Borucki}, W.~J., {Koch}, D., {Basri}, G., {et~al.} 2010, Science, 327, 977

\bibitem[{{Buchner} {et~al.}(2014){Buchner}, {Georgakakis}, {Nandra}, {Hsu},
  {Rangel}, {Brightman}, {Merloni}, {Salvato}, {Donley}, \&
  {Kocevski}}]{PyMultiNest}
{Buchner}, J., {Georgakakis}, A., {Nandra}, K., {et~al.} 2014, \aap, 564, A125

\bibitem[{{Caballero} {et~al.}(2016){Caballero}, {Gu{\`a}rdia}, {L{\'o}pez del
  Fresno}, {Zechmeister}, {de Juan}, {Alonso-Floriano}, {Amado}, {Colom{\'e}},
  {Cort{\'e}s-Contreras}, {Garc{\'\i}a-Piquer}, {Gesa}, {de Guindos}, {Hagen},
  {Helmling}, {Hern{\'a}ndez Casta{\~n}o}, {K{\"u}rster}, {L{\'o}pez-Santiago},
  {Montes}, {Morales Mu{\~n}oz}, {Pavlov}, {Quirrenbach}, {Reiners}, {Ribas},
  {Seifert}, \& {Solano}}]{Caballero2016}
{Caballero}, J.~A., {Gu{\`a}rdia}, J., {L{\'o}pez del Fresno}, M., {et~al.}
  2016, in Society of Photo-Optical Instrumentation Engineers (SPIE) Conference
  Series, Vol. 9910, Observatory Operations: Strategies, Processes, and Systems
  VI, ed. A.~B. {Peck}, R.~L. {Seaman}, \& C.~R. {Benn}, 99100E

\bibitem[{{Cabrera} {et~al.}(2014){Cabrera}, {Csizmadia}, {Lehmann}, {Dvorak},
  {Gandolfi}, {Rauer}, {Erikson}, {Dreyer}, {Eigm{\"u}ller}, \&
  {Hatzes}}]{Cabrera_2014_Kepler-90}
{Cabrera}, J., {Csizmadia}, S., {Lehmann}, H., {et~al.} 2014, \apj, 781, 18

\bibitem[{{Cannon} \& {Pickering}(1993)}]{HD_Catalog}
{Cannon}, A.~J. \& {Pickering}, E.~C. 1993, VizieR Online Data Catalog,
  III/135A

\bibitem[{{Chachan} {et~al.}(2020){Chachan}, {Jontof-Hutter}, {Knutson},
  {Adams}, {Gao}, {Benneke}, {Berta-Thompson}, {Dai}, {Deming}, {Ford}, {Lee},
  {Libby-Roberts}, {Madhusudhan}, {Wakeford}, \& {Wong}}]{chachan2020}
{Chachan}, Y., {Jontof-Hutter}, D., {Knutson}, H.~A., {et~al.} 2020, \aj, 160,
  201

\bibitem[{{Chen} \& {Kipping}(2017)}]{Forecaster}
{Chen}, J. \& {Kipping}, D. 2017, \apj, 834, 17

\bibitem[{{Cloutier} {et~al.}(2021){Cloutier}, {Charbonneau}, {Deming},
  {Bonfils}, \& {Astudillo-Defru}}]{Cloutier_2021}
{Cloutier}, R., {Charbonneau}, D., {Deming}, D., {Bonfils}, X., \&
  {Astudillo-Defru}, N. 2021, \aj, 162, 174

\bibitem[{{Cosentino} {et~al.}(2012){Cosentino}, {Lovis}, {Pepe}, {Collier
  Cameron}, {Latham}, {Molinari}, {Udry}, {Bezawada}, {Black}, {Born},
  {Buchschacher}, {Charbonneau}, {Figueira}, {Fleury}, {Galli}, {Gallie},
  {Gao}, {Ghedina}, {Gonzalez}, {Gonzalez}, {Guerra}, {Henry}, {Horne},
  {Hughes}, {Kelly}, {Lodi}, {Lunney}, {Maire}, {Mayor}, {Micela}, {Ordway},
  {Peacock}, {Phillips}, {Piotto}, {Pollacco}, {Queloz}, {Rice}, {Riverol},
  {Riverol}, {San Juan}, {Sasselov}, {Segransan}, {Sozzetti}, {Sosnowska},
  {Stobie}, {Szentgyorgyi}, {Vick}, \& {Weber}}]{HARPS-N}
{Cosentino}, R., {Lovis}, C., {Pepe}, F., {et~al.} 2012, in Society of
  Photo-Optical Instrumentation Engineers (SPIE) Conference Series, Vol. 8446,
  Ground-based and Airborne Instrumentation for Astronomy IV, ed. I.~S.
  {McLean}, S.~K. {Ramsay}, \& H.~{Takami}, 84461V

\bibitem[{{Cosentino} {et~al.}(2014){Cosentino}, {Lovis}, {Pepe}, {Collier
  Cameron}, {Latham}, {Molinari}, {Udry}, {Bezawada}, {Buchschacher},
  {Figueira}, {Fleury}, {Ghedina}, {Glenday}, {Gonzalez}, {Guerra}, {Henry},
  {Hughes}, {Maire}, {Motalebi}, \& {Phillips}}]{2014SPIE.9147E..8CC}
{Cosentino}, R., {Lovis}, C., {Pepe}, F., {et~al.} 2014, in Society of
  Photo-Optical Instrumentation Engineers (SPIE) Conference Series, Vol. 9147,
  Ground-based and Airborne Instrumentation for Astronomy V, 91478C

\bibitem[{{Cutri} {et~al.}(2003){Cutri}, {Skrutskie}, {van Dyk}, {Beichman},
  {Carpenter}, {Chester}, {Cambresy}, {Evans}, {Fowler}, {Gizis}, {Howard},
  {Huchra}, {Jarrett}, {Kopan}, {Kirkpatrick}, {Light}, {Marsh}, {McCallon},
  {Schneider}, {Stiening}, {Sykes}, {Weinberg}, {Wheaton}, {Wheelock}, \&
  {Zacarias}}]{2MASS_Catalog}
{Cutri}, R.~M., {Skrutskie}, M.~F., {van Dyk}, S., {et~al.} 2003, VizieR Online
  Data Catalog, II/246

\bibitem[{{Dragomir} {et~al.}(2019){Dragomir}, {Teske}, {G{\"u}nther},
  {S{\'e}gransan}, {Burt}, {Huang}, {Vanderburg}, {Matthews}, {Dumusque},
  {Stassun}, {Pepper}, {Ricker}, {Vanderspek}, {Latham}, {Seager}, {Winn},
  {Jenkins}, {Beatty}, {Bouchy}, {Brown}, {Butler}, {Ciardi}, {Crane},
  {Eastman}, {Fossati}, {Francis}, {Fulton}, {Gaudi}, {Goeke}, {James},
  {Klaus}, {Kuhn}, {Lovis}, {Lund}, {McDermott}, {Paegert}, {Pepe},
  {Rodriguez}, {Sha}, {Shectman}, {Shporer}, {Siverd}, {Garcia Soto},
  {Stevens}, {Twicken}, {Udry}, {Villanueva}, {Wang}, {Wohler}, {Yao}, \&
  {Zhan}}]{Dragomir_2019}
{Dragomir}, D., {Teske}, J., {G{\"u}nther}, M.~N., {et~al.} 2019, \apjl, 875,
  L7

\bibitem[{{Espinoza}(2018)}]{Espinoza2018}
{Espinoza}, N. 2018, Research Notes of the American Astronomical Society, 2,
  209

\bibitem[{{Espinoza} {et~al.}(2019){Espinoza}, {Kossakowski}, \&
  {Brahm}}]{juliet}
{Espinoza}, N., {Kossakowski}, D., \& {Brahm}, R. 2019, \mnras, 490, 2262

\bibitem[{{Espinoza} {et~al.}(2022){Espinoza}, {Pall{\'e}}, {Kemmer}, {Luque},
  {Caballero}, {Cifuentes}, {Herrero}, {S{\'a}nchez B{\'e}jar}, {Stock},
  {Molaverdikhani}, {Morello}, {Kossakowski}, {Schlecker}, {Amado}, {Bluhm},
  {Cort{\'e}s-Contreras}, {Henning}, {Kreidberg}, {K{\"u}rster}, {Lafarga},
  {Lodieu}, {Morales}, {Oshagh}, {Passegger}, {Pavlov}, {Quirrenbach},
  {Reffert}, {Reiners}, {Ribas}, {Rodr{\'\i}guez}, {L{\'o}pez}, {Schweitzer},
  {Trifonov}, {Chaturvedi}, {Dreizler}, {Jeffers}, {Kaminski},
  {L{\'o}pez-Gonz{\'a}lez}, {Lillo-Box}, {Montes}, {Nowak}, {Pedraz},
  {Vanaverbeke}, {Zapatero Osorio}, {Zechmeister}, {Collins}, {Girardin},
  {Guerra}, {Naves}, {Crossfield}, {Matthews}, {Howell}, {Ciardi}, {Gonzales},
  {Matson}, {Beichman}, {Schlieder}, {Barclay}, {Vezie}, {Villase{\~n}or},
  {Daylan}, {Mireies}, {Dragomir}, {Twicken}, {Jenkins}, {Winn}, {Latham},
  {Ricker}, \& {Seager}}]{espinoza2022}
{Espinoza}, N., {Pall{\'e}}, E., {Kemmer}, J., {et~al.} 2022, \aj, 163, 133

\bibitem[{{Feng} {et~al.}(2020){Feng}, {Shectman}, {Clement}, {Vogt}, {Tuomi},
  {Teske}, {Burt}, {Crane}, {Holden}, {Wang}, {Thompson}, {D{\'\i}az}, \&
  {Butler}}]{GJ687_2020}
{Feng}, F., {Shectman}, S.~A., {Clement}, M.~S., {et~al.} 2020, \apjs, 250, 29

\bibitem[{{Feroz} {et~al.}(2009){Feroz}, {Hobson}, \& {Bridges}}]{MultiNest}
{Feroz}, F., {Hobson}, M.~P., \& {Bridges}, M. 2009, \mnras, 398, 1601

\bibitem[{{Fletcher} {et~al.}(2018){Fletcher}, {Gustafsson}, \&
  {Orton}}]{fletcher2018}
{Fletcher}, L.~N., {Gustafsson}, M., \& {Orton}, G.~S. 2018, \apjs, 235, 24

\bibitem[{{Foreman-Mackey} {et~al.}(2017){Foreman-Mackey}, {Agol}, {Angus}, \&
  {Ambikasaran}}]{celerite}
{Foreman-Mackey}, D., {Agol}, E., {Angus}, R., \& {Ambikasaran}, S. 2017, AJ,
  154, 220

\bibitem[{{Fortney} {et~al.}(2013){Fortney}, {Mordasini}, {Nettelmann},
  {Kempton}, {Greene}, \& {Zahnle}}]{fortney2013}
{Fortney}, J.~J., {Mordasini}, C., {Nettelmann}, N., {et~al.} 2013, \apj, 775,
  80

\bibitem[{{Fulton} {et~al.}(2018){Fulton}, {Petigura}, {Blunt}, \&
  {Sinukoff}}]{radvel}
{Fulton}, B.~J., {Petigura}, E.~A., {Blunt}, S., \& {Sinukoff}, E. 2018, \pasp,
  130, 044504

\bibitem[{{Gaia Collaboration} {et~al.}(2018){Gaia Collaboration}, {Brown},
  {Vallenari}, {Prusti}, {de Bruijne}, {Babusiaux}, {Bailer-Jones}, {Biermann},
  {Evans}, {Eyer}, {Jansen}, {Jordi}, {Klioner}, {Lammers}, {Lindegren},
  {Luri}, {Mignard}, {Panem}, {Pourbaix}, {Randich}, {Sartoretti}, {Siddiqui},
  {Soubiran}, {van Leeuwen}, {Walton}, {Arenou}, {Bastian}, {Cropper},
  {Drimmel}, {Katz}, {Lattanzi}, {Bakker}, {Cacciari}, {Casta{\~n}eda},
  {Chaoul}, {Cheek}, {De Angeli}, {Fabricius}, {Guerra}, {Holl}, {Masana},
  {Messineo}, {Mowlavi}, {Nienartowicz}, {Panuzzo}, {Portell}, {Riello},
  {Seabroke}, {Tanga}, {Th{\'e}venin}, {Gracia-Abril}, {Comoretto},
  {Garcia-Reinaldos}, {Teyssier}, {Altmann}, {Andrae}, {Audard},
  {Bellas-Velidis}, {Benson}, {Berthier}, {Blomme}, {Burgess}, {Busso},
  {Carry}, {Cellino}, {Clementini}, {Clotet}, {Creevey}, {Davidson}, {De
  Ridder}, {Delchambre}, {Dell'Oro}, {Ducourant},
  {Fern{\'a}ndez-Hern{\'a}ndez}, {Fouesneau}, {Fr{\'e}mat}, {Galluccio},
  {Garc{\'\i}a-Torres}, {Gonz{\'a}lez-N{\'u}{\~n}ez}, {Gonz{\'a}lez-Vidal},
  {Gosset}, {Guy}, {Halbwachs}, {Hambly}, {Harrison}, {Hern{\'a}ndez},
  {Hestroffer}, {Hodgkin}, {Hutton}, {Jasniewicz}, {Jean-Antoine-Piccolo},
  {Jordan}, {Korn}, {Krone-Martins}, {Lanzafame}, {Lebzelter}, {L{\"o}ffler},
  {Manteiga}, {Marrese}, {Mart{\'\i}n-Fleitas}, {Moitinho}, {Mora}, {Muinonen},
  {Osinde}, {Pancino}, {Pauwels}, {Petit}, {Recio-Blanco}, {Richards},
  {Rimoldini}, {Robin}, {Sarro}, {Siopis}, {Smith}, {Sozzetti}, {S{\"u}veges},
  {Torra}, {van Reeven}, {Abbas}, {Abreu Aramburu}, {Accart}, {Aerts},
  {Altavilla}, {{\'A}lvarez}, {Alvarez}, {Alves}, {Anderson}, {Andrei},
  {Anglada Varela}, {Antiche}, {Antoja}, {Arcay}, {Astraatmadja}, {Bach},
  {Baker}, {Balaguer-N{\'u}{\~n}ez}, {Balm}, {Barache}, {Barata}, {Barbato},
  {Barblan}, {Barklem}, {Barrado}, {Barros}, {Barstow}, {Bartholom{\'e}
  Mu{\~n}oz}, {Bassilana}, {Becciani}, {Bellazzini}, {Berihuete}, {Bertone},
  {Bianchi}, {Bienaym{\'e}}, {Blanco-Cuaresma}, {Boch}, {Boeche}, {Bombrun},
  {Borrachero}, {Bossini}, {Bouquillon}, {Bourda}, {Bragaglia}, {Bramante},
  {Breddels}, {Bressan}, {Brouillet}, {Br{\"u}semeister}, {Brugaletta},
  {Bucciarelli}, {Burlacu}, {Busonero}, {Butkevich}, {Buzzi}, {Caffau},
  {Cancelliere}, {Cannizzaro}, {Cantat-Gaudin}, {Carballo}, {Carlucci},
  {Carrasco}, {Casamiquela}, {Castellani}, {Castro-Ginard}, {Charlot},
  {Chemin}, {Chiavassa}, {Cocozza}, {Costigan}, {Cowell}, {Crifo}, {Crosta},
  {Crowley}, {Cuypers}, {Dafonte}, {Damerdji}, {Dapergolas}, {David}, {David},
  {de Laverny}, {De Luise}, {De March}, {de Martino}, {de Souza}, {de Torres},
  {Debosscher}, {del Pozo}, {Delbo}, {Delgado}, {Delgado}, {Di Matteo},
  {Diakite}, {Diener}, {Distefano}, {Dolding}, {Drazinos}, {Dur{\'a}n},
  {Edvardsson}, {Enke}, {Eriksson}, {Esquej}, {Eynard Bontemps}, {Fabre},
  {Fabrizio}, {Faigler}, {Falc{\~a}o}, {Farr{\`a}s Casas}, {Federici},
  {Fedorets}, {Fernique}, {Figueras}, {Filippi}, {Findeisen}, {Fonti},
  {Fraile}, {Fraser}, {Fr{\'e}zouls}, {Gai}, {Galleti}, {Garabato},
  {Garc{\'\i}a-Sedano}, {Garofalo}, {Garralda}, {Gavel}, {Gavras}, {Gerssen},
  {Geyer}, {Giacobbe}, {Gilmore}, {Girona}, {Giuffrida}, {Glass}, {Gomes},
  {Granvik}, {Gueguen}, {Guerrier}, {Guiraud}, {Guti{\'e}rrez-S{\'a}nchez},
  {Haigron}, {Hatzidimitriou}, {Hauser}, {Haywood}, {Heiter}, {Helmi}, {Heu},
  {Hilger}, {Hobbs}, {Hofmann}, {Holland}, {Huckle}, {Hypki}, {Icardi},
  {Jan{\ss}en}, {Jevardat de Fombelle}, {Jonker}, {Juh{\'a}sz}, {Julbe},
  {Karampelas}, {Kewley}, {Klar}, {Kochoska}, {Kohley}, {Kolenberg},
  {Kontizas}, {Kontizas}, {Koposov}, {Kordopatis}, {Kostrzewa-Rutkowska},
  {Koubsky}, {Lambert}, {Lanza}, {Lasne}, {Lavigne}, {Le Fustec}, {Le
  Poncin-Lafitte}, {Lebreton}, {Leccia}, {Leclerc}, {Lecoeur-Taibi},
  {Lenhardt}, {Leroux}, {Liao}, {Licata}, {Lindstr{\o}m}, {Lister}, {Livanou},
  {Lobel}, {L{\'o}pez}, {Managau}, {Mann}, {Mantelet}, {Marchal}, {Marchant},
  {Marconi}, {Marinoni}, {Marschalk{\'o}}, {Marshall}, {Martino}, {Marton},
  {Mary}, {Massari}, {Matijevi{\v{c}}}, {Mazeh}, {McMillan}, {Messina},
  {Michalik}, {Millar}, {Molina}, {Molinaro}, {Moln{\'a}r}, {Montegriffo},
  {Mor}, {Morbidelli}, {Morel}, {Morris}, {Mulone}, {Muraveva}, {Musella},
  {Nelemans}, {Nicastro}, {Noval}, {O'Mullane}, {Ord{\'e}novic},
  {Ord{\'o}{\~n}ez-Blanco}, {Osborne}, {Pagani}, {Pagano}, {Pailler},
  {Palacin}, {Palaversa}, {Panahi}, {Pawlak}, {Piersimoni}, {Pineau}, {Plachy},
  {Plum}, {Poggio}, {Poujoulet}, {Pr{\v{s}}a}, {Pulone}, {Racero}, {Ragaini},
  {Rambaux}, {Ramos-Lerate}, {Regibo}, {Reyl{\'e}}, {Riclet}, {Ripepi}, {Riva},
  {Rivard}, {Rixon}, {Roegiers}, {Roelens}, {Romero-G{\'o}mez}, {Rowell},
  {Royer}, {Ruiz-Dern}, {Sadowski}, {Sagrist{\`a} Sell{\'e}s}, {Sahlmann},
  {Salgado}, {Salguero}, {Sanna}, {Santana-Ros}, {Sarasso}, {Savietto},
  {Schultheis}, {Sciacca}, {Segol}, {Segovia}, {S{\'e}gransan}, {Shih},
  {Siltala}, {Silva}, {Smart}, {Smith}, {Solano}, {Solitro}, {Sordo}, {Soria
  Nieto}, {Souchay}, {Spagna}, {Spoto}, {Stampa}, {Steele},
  {Steidelm{\"u}ller}, {Stephenson}, {Stoev}, {Suess}, {Surdej}, {Szabados},
  {Szegedi-Elek}, {Tapiador}, {Taris}, {Tauran}, {Taylor}, {Teixeira},
  {Terrett}, {Teyssand ier}, {Thuillot}, {Titarenko}, {Torra Clotet}, {Turon},
  {Ulla}, {Utrilla}, {Uzzi}, {Vaillant}, {Valentini}, {Valette}, {van Elteren},
  {Van Hemelryck}, {van Leeuwen}, {Vaschetto}, {Vecchiato}, {Veljanoski},
  {Viala}, {Vicente}, {Vogt}, {von Essen}, {Voss}, {Votruba}, {Voutsinas},
  {Walmsley}, {Weiler}, {Wertz}, {Wevers}, {Wyrzykowski}, {Yoldas},
  {{\v{Z}}erjal}, {Ziaeepour}, {Zorec}, {Zschocke}, {Zucker}, {Zurbach}, \&
  {Zwitter}}]{GaiaDR2}
{Gaia Collaboration}, {Brown}, A.~G.~A., {Vallenari}, A., {et~al.} 2018, \aap,
  616, A1

\bibitem[{{Gao} \& {Zhang}(2020)}]{gao2020}
{Gao}, P. \& {Zhang}, X. 2020, \apj, 890, 93

\bibitem[{{Guerrero} {et~al.}(2021){Guerrero}, {Seager}, {Huang}, {Vanderburg},
  {Garcia Soto}, {Mireles}, {Hesse}, {Fong}, {Glidden}, {Shporer}, {Latham},
  {Collins}, {Quinn}, {Burt}, {Dragomir}, {Crossfield}, {Vanderspek},
  {Fausnaugh}, {Burke}, {Ricker}, {Daylan}, {Essack}, {G{\"u}nther}, {Osborn},
  {Pepper}, {Rowden}, {Sha}, {Villanueva}, {Yahalomi}, {Yu}, {Ballard},
  {Batalha}, {Berardo}, {Chontos}, {Dittmann}, {Esquerdo}, {Mikal-Evans},
  {Jayaraman}, {Krishnamurthy}, {Louie}, {Mehrle}, {Niraula}, {Rackham},
  {Rodriguez}, {Rowden}, {Sousa-Silva}, {Watanabe}, {Wong}, {Zhan},
  {Zivanovic}, {Christiansen}, {Ciardi}, {Swain}, {Lund}, {Mullally},
  {Fleming}, {Rodriguez}, {Boyd}, {Quintana}, {Barclay}, {Col{\'o}n},
  {Rinehart}, {Schlieder}, {Clampin}, {Jenkins}, {Twicken}, {Caldwell},
  {Coughlin}, {Henze}, {Lissauer}, {Morris}, {Rose}, {Smith}, {Tenenbaum},
  {Ting}, {Wohler}, {Bakos}, {Bean}, {Berta-Thompson}, {Bieryla}, {Bouma},
  {Buchhave}, {Butler}, {Charbonneau}, {Doty}, {Ge}, {Holman}, {Howard},
  {Kaltenegger}, {Kane}, {Kjeldsen}, {Kreidberg}, {Lin}, {Minsky}, {Narita},
  {Paegert}, {P{\'a}l}, {Palle}, {Sasselov}, {Spencer}, {Sozzetti}, {Stassun},
  {Torres}, {Udry}, \& {Winn}}]{Guerrero_2021}
{Guerrero}, N.~M., {Seager}, S., {Huang}, C.~X., {et~al.} 2021, \apjs, 254, 39

\bibitem[{{G{\"u}nther} {et~al.}(2019){G{\"u}nther}, {Pozuelos}, {Dittmann},
  {Dragomir}, {Kane}, {Daylan}, {Feinstein}, {Huang}, {Morton}, {Bonfanti},
  {Bouma}, {Burt}, {Collins}, {Lissauer}, {Matthews}, {Montet}, {Vanderburg},
  {Wang}, {Winters}, {Ricker}, {Vanderspek}, {Latham}, {Seager}, {Winn},
  {Jenkins}, {Armstrong}, {Barkaoui}, {Batalha}, {Bean}, {Caldwell}, {Ciardi},
  {Collins}, {Crossfield}, {Fausnaugh}, {Furesz}, {Gan}, {Gillon}, {Guerrero},
  {Horne}, {Howell}, {Ireland}, {Isopi}, {Jehin}, {Kielkopf}, {Lepine},
  {Mallia}, {Matson}, {Myers}, {Palle}, {Quinn}, {Relles}, {Rojas-Ayala},
  {Schlieder}, {Sefako}, {Shporer}, {Su{\'a}rez}, {Tan}, {Ting}, {Twicken}, \&
  {Waite}}]{Gunther_2019}
{G{\"u}nther}, M.~N., {Pozuelos}, F.~J., {Dittmann}, J.~A., {et~al.} 2019,
  Nature Astronomy, 3, 1099

\bibitem[{{Hadden} \& {Lithwick}(2017)}]{Hadden_ecc}
{Hadden}, S. \& {Lithwick}, Y. 2017, \aj, 154, 5

\bibitem[{{H{\o}g} {et~al.}(2000){H{\o}g}, {Fabricius}, {Makarov}, {Urban},
  {Corbin}, {Wycoff}, {Bastian}, {Schwekendiek}, \& {Wicenec}}]{TYC_Catalog}
{H{\o}g}, E., {Fabricius}, C., {Makarov}, V.~V., {et~al.} 2000, \aap, 355, L27

\bibitem[{{Jenkins} {et~al.}(2016){Jenkins}, {Twicken}, {McCauliff},
  {Campbell}, {Sanderfer}, {Lung}, {Mansouri-Samani}, {Girouard}, {Tenenbaum},
  {Klaus}, {Smith}, {Caldwell}, {Chacon}, {Henze}, {Heiges}, {Latham},
  {Morgan}, {Swade}, {Rinehart}, \& {Vanderspek}}]{SPOC}
{Jenkins}, J.~M., {Twicken}, J.~D., {McCauliff}, S., {et~al.} 2016, in Society
  of Photo-Optical Instrumentation Engineers (SPIE) Conference Series, Vol.
  9913, Software and Cyberinfrastructure for Astronomy IV, ed. G.~{Chiozzi} \&
  J.~C. {Guzman}, 99133E

\bibitem[{{Jontof-Hutter} {et~al.}(2014){Jontof-Hutter}, {Lissauer}, {Rowe}, \&
  {Fabrycky}}]{Kepler79_System}
{Jontof-Hutter}, D., {Lissauer}, J.~J., {Rowe}, J.~F., \& {Fabrycky}, D.~C.
  2014, \apj, 785, 15

\bibitem[{{Kasper} {et~al.}(2020){Kasper}, {Bean}, {Oklop{\v{c}}i{\'c}},
  {Malsky}, {Kempton}, {D{\'e}sert}, {Rogers}, \& {Mansfield}}]{Kasper_2020}
{Kasper}, D., {Bean}, J.~L., {Oklop{\v{c}}i{\'c}}, A., {et~al.} 2020, \aj, 160,
  258

\bibitem[{{Kempton} {et~al.}(2018){Kempton}, {Bean}, {Louie}, {Deming}, {Koll},
  {Mansfield}, {Christiansen}, {L{\'o}pez-Morales}, {Swain}, {Zellem},
  {Ballard}, {Barclay}, {Barstow}, {Batalha}, {Beatty}, {Berta-Thompson},
  {Birkby}, {Buchhave}, {Charbonneau}, {Cowan}, {Crossfield}, {de Val-Borro},
  {Doyon}, {Dragomir}, {Gaidos}, {Heng}, {Hu}, {Kane}, {Kreidberg}, {Mallonn},
  {Morley}, {Narita}, {Nascimbeni}, {Pall{\'e}}, {Quintana}, {Rauscher},
  {Seager}, {Shkolnik}, {Sing}, {Sozzetti}, {Stassun}, {Valenti}, \& {von
  Essen}}]{Kempton_2018_TSM}
{Kempton}, E. M.~R., {Bean}, J.~L., {Louie}, D.~R., {et~al.} 2018, \pasp, 130,
  114401

\bibitem[{{Kipping}(2013)}]{Kipping2013}
{Kipping}, D.~M. 2013, \mnras, 435, 2152

\bibitem[{{Kreidberg}(2015)}]{batman}
{Kreidberg}, L. 2015, Publications of the Astronomical Society of the Pacific,
  127, 1161

\bibitem[{{Kreidberg} {et~al.}(2014){Kreidberg}, {Bean}, {D{\'e}sert},
  {Benneke}, {Deming}, {Stevenson}, {Seager}, {Berta-Thompson}, {Seifahrt}, \&
  {Homeier}}]{Kreidberg_2014}
{Kreidberg}, L., {Bean}, J.~L., {D{\'e}sert}, J.-M., {et~al.} 2014, \nat, 505,
  69

\bibitem[{{Lafarga} {et~al.}(2020){Lafarga}, {Ribas}, {Lovis}, {Perger},
  {Zechmeister}, {Bauer}, {K{\"u}rster}, {Cort{\'e}s-Contreras}, {Morales},
  {Herrero}, {Rosich}, {Baroch}, {Reiners}, {Caballero}, {Quirrenbach},
  {Amado}, {Alacid}, {B{\'e}jar}, {Dreizler}, {Hatzes}, {Henning}, {Jeffers},
  {Kaminski}, {Montes}, {Pedraz}, {Rodr{\'\i}guez-L{\'o}pez}, \&
  {Schmitt}}]{2020A&A...636A..36L}
{Lafarga}, M., {Ribas}, I., {Lovis}, C., {et~al.} 2020, \aap, 636, A36

\bibitem[{{Lee} {et~al.}(2013){Lee}, {Heng}, \& {Irwin}}]{lee2013}
{Lee}, J.-M., {Heng}, K., \& {Irwin}, P. G.~J. 2013, \apj, 778, 97

\bibitem[{{Liang} {et~al.}(2021){Liang}, {Robnik}, \& {Seljak}}]{Density_K90_g}
{Liang}, Y., {Robnik}, J., \& {Seljak}, U. 2021, \aj, 161, 202

\bibitem[{{Libby-Roberts} {et~al.}(2020){Libby-Roberts}, {Berta-Thompson},
  {D{\'e}sert}, {Masuda}, {Morley}, {Lopez}, {Deck}, {Fabrycky}, {Fortney},
  {Line}, {Sanchis-Ojeda}, \& {Winn}}]{libby-roberts2020}
{Libby-Roberts}, J.~E., {Berta-Thompson}, Z.~K., {D{\'e}sert}, J.-M., {et~al.}
  2020, \aj, 159, 57

\bibitem[{{Lovis} {et~al.}(2011){Lovis}, {S{\'e}gransan}, {Mayor}, {Udry},
  {Benz}, {Bertaux}, {Bouchy}, {Correia}, {Laskar}, {Lo Curto}, {Mordasini},
  {Pepe}, {Queloz}, \& {Santos}}]{HD10180g_2011}
{Lovis}, C., {S{\'e}gransan}, D., {Mayor}, M., {et~al.} 2011, \aap, 528, A112

\bibitem[{{Lubin} {et~al.}(2022){Lubin}, {Van Zandt}, {Holcomb}, {Weiss},
  {Petigura}, {Robertson}, {Akana Murphy}, {Scarsdale}, {Batygin}, {Polanski},
  {Batalha}, {Crossfield}, {Dressing}, {Fulton}, {Howard}, {Huber}, {Isaacson},
  {Kane}, {Roy}, {Beard}, {Blunt}, {Chontos}, {Dai}, {Dalba}, {Gary},
  {Giacalone}, {Hill}, {Mayo}, {Mo{\v{c}}nik}, {Kosiarek}, {Rice}, {Rubenzahl},
  {Latham}, {Seager}, {Winn}, \& {Gary}}]{Lubin_HD191939}
{Lubin}, J., {Van Zandt}, J., {Holcomb}, R., {et~al.} 2022, \aj, 163, 101

\bibitem[{{Luque} {et~al.}(2022){Luque}, {Fulton}, {Kunimoto}, {Amado},
  {Gorrini}, {Dreizler}, {Hellier}, {Henry}, {Molaverdikhani}, {Morello},
  {Pe{\~n}a-Mo{\~n}ino}, {P{\'e}rez-Torres}, {Pozuelos}, {Shan},
  {Anglada-Escud{\'e}}, {B{\'e}jar}, {Bergond}, {Boyle}, {Caballero},
  {Charbonneau}, {Ciardi}, {Dufoer}, {Espinoza}, {Everett}, {Fischer},
  {Hatzes}, {Henning}, {Hesse}, {Howard}, {Howell}, {Isaacson}, {Jeffers},
  {Jenkins}, {Kane}, {Kemmer}, {Khalafinejad}, {Kidwell}, {Kossakowski},
  {Latham}, {Lillo-Box}, {Lissauer}, {Montes}, {Orell-Miquel}, {Pall{\'e}},
  {Pollacco}, {Quirrenbach}, {Reffert}, {Reiners}, {Ribas}, {Ricker}, {Rogers},
  {Sanz-Forcada}, {Schlecker}, {Schweitzer}, {Seager}, {Shporer}, {Stassun},
  {Stock}, {Tal-Or}, {Ting}, {Trifonov}, {Vanaverbeke}, {Vanderspek},
  {Villase{\~n}or}, {Winn}, {Winters}, \& {Zapatero Osorio}}]{luque2022}
{Luque}, R., {Fulton}, B.~J., {Kunimoto}, M., {et~al.} 2022, \aap, 664, A199

\bibitem[{{Luque} {et~al.}(2021){Luque}, {Serrano}, {Molaverdikhani}, {Nixon},
  {Livingston}, {Guenther}, {Pall{\'e}}, {Madhusudhan}, {Nowak}, {Korth},
  {Cochran}, {Hirano}, {Chaturvedi}, {Goffo}, {Albrecht}, {Barrag{\'a}n},
  {Brice{\~n}o}, {Cabrera}, {Charbonneau}, {Cloutier}, {Collins}, {Collins},
  {Col{\'o}n}, {Crossfield}, {Csizmadia}, {Dai}, {Deeg}, {Esposito},
  {Fridlund}, {Gandolfi}, {Georgieva}, {Glidden}, {Goeke}, {Grziwa}, {Hatzes},
  {Henze}, {Howell}, {Irwin}, {Jenkins}, {Jensen}, {K{\'a}bath}, {Kidwell},
  {Kielkopf}, {Knudstrup}, {Lam}, {Latham}, {Lissauer}, {Mann}, {Matthews},
  {Mireles}, {Narita}, {Paegert}, {Persson}, {Redfield}, {Ricker}, {Rodler},
  {Schlieder}, {Scott}, {Seager}, {{\v{S}}ubjak}, {Tan}, {Ting}, {Vanderspek},
  {Van Eylen}, {Winn}, \& {Ziegler}}]{Rafa_2021}
{Luque}, R., {Serrano}, L.~M., {Molaverdikhani}, K., {et~al.} 2021, \aap, 645,
  A41

\bibitem[{{Masuda}(2014)}]{Kepler51_TTV}
{Masuda}, K. 2014, \apj, 783, 53

\bibitem[{{Mayor} {et~al.}(2011){Mayor}, {Marmier}, {Lovis}, {Udry},
  {S{\'e}gransan}, {Pepe}, {Benz}, {Bertaux}, {Bouchy}, {Dumusque}, {Lo Curto},
  {Mordasini}, {Queloz}, \& {Santos}}]{HD31527d_2011}
{Mayor}, M., {Marmier}, M., {Lovis}, C., {et~al.} 2011, arXiv e-prints,
  arXiv:1109.2497

\bibitem[{{Mills} {et~al.}(2019){Mills}, {Howard}, {Weiss}, {Steffen},
  {Isaacson}, {Fulton}, {Petigura}, {Kosiarek}, {Hirsch}, \&
  {Boisvert}}]{Kepler_25_65_68}
{Mills}, S.~M., {Howard}, A.~W., {Weiss}, L.~M., {et~al.} 2019, \aj, 157, 145

\bibitem[{{Morello} {et~al.}(2021){Morello}, {Zingales}, {Martin-Lagarde},
  {Gastaud}, \& {Lagage}}]{morello2021}
{Morello}, G., {Zingales}, T., {Martin-Lagarde}, M., {Gastaud}, R., \&
  {Lagage}, P.-O. 2021, \aj, 161, 174

\bibitem[{{Morris} {et~al.}(2017){Morris}, {Twicken}, {Smith}, {Clarke},
  {Jenkins}, {Bryson}, {Girouard}, \& {Klaus}}]{SAP}
{Morris}, R.~L., {Twicken}, J.~D., {Smith}, J.~C., {et~al.} 2017, {Kepler Data
  Processing Handbook: Photometric Analysis}, Kepler Science Document
  KSCI-19081-002

\bibitem[{{Ofir} {et~al.}(2014){Ofir}, {Dreizler}, {Zechmeister}, \&
  {Husser}}]{Kepler87}
{Ofir}, A., {Dreizler}, S., {Zechmeister}, M., \& {Husser}, T.-O. 2014, \aap,
  561, A103

\bibitem[{{Ohno} \& {Tanaka}(2021)}]{ohno2021}
{Ohno}, K. \& {Tanaka}, Y.~A. 2021, \apj, 920, 124

\bibitem[{{Orell-Miquel} {et~al.}(2022){Orell-Miquel}, {Murgas}, {Pall{\'e}},
  {Lamp{\'o}n}, {L{\'o}pez-Puertas}, {Sanz-Forcada}, {Nagel}, {Kaminski},
  {Casasayas-Barris}, {Nortmann}, {Luque}, {Molaverdikhani}, {Sedaghati},
  {Caballero}, {Amado}, {Bergond}, {Czesla}, {Hatzes}, {Henning},
  {Khalafinejad}, {Montes}, {Morello}, {Quirrenbach}, {Reiners}, {Ribas},
  {S{\'a}nchez-L{\'o}pez}, {Schweitzer}, {Stangret}, {Yan}, \& {Zapatero
  Osorio}}]{Orell_2022}
{Orell-Miquel}, J., {Murgas}, F., {Pall{\'e}}, E., {et~al.} 2022, \aap, 659,
  A55

\bibitem[{{Petit dit de la Roche} {et~al.}(2020){Petit dit de la Roche}, {van
  den Ancker}, \& {Miles-Paez}}]{Petit_2020}
{Petit dit de la Roche}, D.~J.~M., {van den Ancker}, M.~E., \& {Miles-Paez},
  P.~A. 2020, Research Notes of the American Astronomical Society, 4, 231

\bibitem[{{Quinn} {et~al.}(2019){Quinn}, {Becker}, {Rodriguez}, {Hadden},
  {Huang}, {Morton}, {Adams}, {Armstrong}, {Eastman}, {Horner}, {Kane},
  {Lissauer}, {Twicken}, {Vanderburg}, {Wittenmyer}, {Ricker}, {Vanderspek},
  {Latham}, {Seager}, {Winn}, {Jenkins}, {Agol}, {Barkaoui}, {Beichman},
  {Bouchy}, {Bouma}, {Burdanov}, {Campbell}, {Carlino}, {Cartwright},
  {Charbonneau}, {Christiansen}, {Ciardi}, {Collins}, {Collins}, {Conti},
  {Crossfield}, {Daylan}, {Dittmann}, {Doty}, {Dragomir}, {Ducrot}, {Gillon},
  {Glidden}, {Goeke}, {Gonzales}, {He{\l}miniak}, {Horch}, {Howell}, {Jehin},
  {Jensen}, {Kielkopf}, {Kristiansen}, {Law}, {Mann}, {Marmier}, {Matson},
  {Matthews}, {Mazeh}, {Mori}, {Murgas}, {Murray}, {Narita}, {Nielsen},
  {Ottoni}, {Palle}, {Paw{\l}aszek}, {Pepe}, {Pitogo de Leon}, {Pozuelos},
  {Relles}, {Schlieder}, {Sebastian}, {S{\'e}gransan}, {Shporer}, {Stassun},
  {Tamura}, {Udry}, {Waite}, {Winters}, \& {Ziegler}}]{Quinn_2019}
{Quinn}, S.~N., {Becker}, J.~C., {Rodriguez}, J.~E., {et~al.} 2019, \aj, 158,
  177

\bibitem[{{Quirrenbach} {et~al.}(2014){Quirrenbach}, {Amado}, {Caballero},
  {Mundt}, {Reiners}, {Ribas}, {Seifert}, {Abril}, {Aceituno},
  {Alonso-Floriano}, {Ammler-von Eiff}, {Antona Jim{\'e}nez},
  {Anwand-Heerwart}, {Azzaro}, {Bauer}, {Barrado}, {Becerril}, {B{\'e}jar},
  {Ben{\'\i}tez}, {Berdi{\~n}as}, {C{\'a}rdenas}, {Casal}, {Claret},
  {Colom{\'e}}, {Cort{\'e}s-Contreras}, {Czesla}, {Doellinger}, {Dreizler},
  {Feiz}, {Fern{\'a}ndez}, {Galad{\'\i}}, {G{\'a}lvez-Ortiz},
  {Garc{\'\i}a-Piquer}, {Garc{\'\i}a-Vargas}, {Garrido}, {Gesa}, {G{\'o}mez
  Galera}, {Gonz{\'a}lez {\'A}lvarez}, {Gonz{\'a}lez Hern{\'a}ndez},
  {Gr{\"o}zinger}, {Gu{\`a}rdia}, {Guenther}, {de Guindos},
  {Guti{\'e}rrez-Soto}, {Hagen}, {Hatzes}, {Hauschildt}, {Helmling}, {Henning},
  {Hermann}, {Hern{\'a}ndez Casta{\~n}o}, {Herrero}, {Hidalgo}, {Holgado},
  {Huber}, {Huber}, {Jeffers}, {Joergens}, {de Juan}, {Kehr}, {Klein},
  {K{\"u}rster}, {Lamert}, {Lalitha}, {Laun}, {Lemke}, {Lenzen}, {L{\'o}pez del
  Fresno}, {L{\'o}pez Mart{\'\i}}, {L{\'o}pez-Santiago}, {Mall}, {Mandel},
  {Mart{\'\i}n}, {Mart{\'\i}n-Ruiz}, {Mart{\'\i}nez-Rodr{\'\i}guez}, {Marvin},
  {Mathar}, {Mirabet}, {Montes}, {Morales Mu{\~n}oz}, {Moya}, {Naranjo},
  {Ofir}, {Oreiro}, {Pall{\'e}}, {Panduro}, {Passegger}, {P{\'e}rez-Calpena},
  {P{\'e}rez Medialdea}, {Perger}, {Pluto}, {Ram{\'o}n}, {Rebolo}, {Redondo},
  {Reffert}, {Reinhardt}, {Rhode}, {Rix}, {Rodler}, {Rodr{\'\i}guez},
  {Rodr{\'\i}guez-L{\'o}pez}, {Rodr{\'\i}guez-P{\'e}rez}, {Rohloff}, {Rosich},
  {S{\'a}nchez-Blanco}, {S{\'a}nchez Carrasco}, {Sanz-Forcada}, {Sarmiento},
  {Sch{\"a}fer}, {Schiller}, {Schmidt}, {Schmitt}, {Solano}, {Stahl}, {Storz},
  {St{\"u}rmer}, {Su{\'a}rez}, {Ulbrich}, {Veredas}, {Wagner}, {Winkler},
  {Zapatero Osorio}, {Zechmeister}, {Abell{\'a}n de Paco},
  {Anglada-Escud{\'e}}, {del Burgo}, {Klutsch}, {Lizon}, {L{\'o}pez-Morales},
  {Morales}, {Perryman}, {Tulloch}, \& {Xu}}]{Quirrenbach_2014}
{Quirrenbach}, A., {Amado}, P.~J., {Caballero}, J.~A., {et~al.} 2014, in
  Society of Photo-Optical Instrumentation Engineers (SPIE) Conference Series,
  Vol. 9147, Ground-based and Airborne Instrumentation for Astronomy V, ed.
  S.~K. {Ramsay}, I.~S. {McLean}, \& H.~{Takami}, 91471F

\bibitem[{{Quirrenbach} {et~al.}(2020){Quirrenbach}, {CARMENES Consortium},
  {Amado}, {Ribas}, {Reiners}, {Caballero}, {Aceituno}, {Alacid},
  {Alonso-Floriano}, {Anglada-Escud{\'e}}, {Azzaro}, {Baroch}, {Bauer},
  {Becerril}, {B{\'e}jar}, {Bluhm}, {Calvo Ortega}, {Cardona Guill{\'e}n},
  {Casasayas-Barris}, {Chaturvedi}, {Cifuentes}, {Colom{\'e}}, {Conte},
  {Cort{\'e}s-Contreras}, {Czesla}, {D{\'\i}ez-Alonso}, {Dom{\'\i}nguez
  Fern{\'a}ndez}, {Dreizler}, {Duque-Arribas}, {Espinoza}, {Fuhrmeister},
  {Galad{\'\i}-Enr{\'\i}quez}, {Gara Quintana}, {Gonz{\'a}lez-Alvare},
  {Gonz{\'a}lez Cuesta}, {Gonz{\'a}lez Hern{\'a}ndez}, {Guenther}, {de
  Guindos}, {Hatzes}, {Henning}, {Herbort}, {Herrero}, {Hintz},
  {Iglesias-P{\'a}ra}, {Jeffers}, {Johnson}, {de Juan}, {Kaminski}, {Kemmer},
  {Khaimova}, {Khalafinejad}, {Klahr}, {Kossakowski}, {Kreidberg},
  {K{\"u}rster}, {Labarga}, {Lafarga}, {Lamp{\'o}n}, {Lara}, {Lillo-Box},
  {Lodieu}, {L{\'o}pez Gallifa}, {L{\'o}pez Gonz{\'a}lez}, {L{\'o}pez-Puertas},
  {Luque}, {Marfil}, {Mart{\'\i}n-Ruiz}, {Matth{\'e}}, {Molaverdikhani},
  {Montes}, {Morales}, {Morales-Calder{\'o}on}, {Nagel}, {Nortmann}, {Nowak},
  {Ofir}, {Oshaghi}, {Pall{\'e}}, {Passegger}, {Pavlov}, {Pedraz},
  {Perdelwitz}, {Perger}, {Reffert}, {Revilla}, {Rodr{\'\i}guez},
  {Rodr{\'\i}guez L{\'o}pez}, {Sabotta}, {Sadegi}, {Sairam}, {Salz},
  {S{\'a}nchez-L{\'o}pez}, {Sanz-Forcada}, {Sarkis}, {Sch{\"a}fer}, {Schiller},
  {Schlecker}, {Schmitt}, {Sch{\"o}fer}, {Schweitzer}, {Seiferta}, {Shan},
  {Shulyak}, {Skrzypinski}, {Solano}, {Soto}, {Stahl}, {Stangret}, {Stock},
  {Strachan}, {Stuber}, {St{\"u}rmer}, {Tabernero}, {Tal-Or}, {Tala-Pinto},
  {Trifonov}, {Vanaverbeke}, {Yan}, {Zapatero Osorio}, \&
  {Zechmeister}}]{Quirrenbach_2020}
{Quirrenbach}, A., {CARMENES Consortium}, {Amado}, P.~J., {et~al.} 2020, in
  Society of Photo-Optical Instrumentation Engineers (SPIE) Conference Series,
  Vol. 11447, Society of Photo-Optical Instrumentation Engineers (SPIE)
  Conference Series, 114473C

\bibitem[{{Ribas} {et~al.}(2018){Ribas}, {Tuomi}, {Reiners}, {Butler},
  {Morales}, {Perger}, {Dreizler}, {Rodr{\'\i}guez-L{\'o}pez}, {Gonz{\'a}lez
  Hern{\'a}ndez}, {Rosich}, {Feng}, {Trifonov}, {Vogt}, {Caballero}, {Hatzes},
  {Herrero}, {Jeffers}, {Lafarga}, {Murgas}, {Nelson}, {Rodr{\'\i}guez},
  {Strachan}, {Tal-Or}, {Teske}, {Toledo-Padr{\'o}n}, {Zechmeister},
  {Quirrenbach}, {Amado}, {Azzaro}, {B{\'e}jar}, {Barnes}, {Berdi{\~n}as},
  {Burt}, {Coleman}, {Cort{\'e}s-Contreras}, {Crane}, {Engle}, {Guinan},
  {Haswell}, {Henning}, {Holden}, {Jenkins}, {Jones}, {Kaminski}, {Kiraga},
  {K{\"u}rster}, {Lee}, {L{\'o}pez-Gonz{\'a}lez}, {Montes}, {Morin}, {Ofir},
  {Pall{\'e}}, {Rebolo}, {Reffert}, {Schweitzer}, {Seifert}, {Shectman},
  {Staab}, {Street}, {Su{\'a}rez Mascare{\~n}o}, {Tsapras}, {Wang}, \&
  {Anglada-Escud{\'e}}}]{Ribas_2018Natur}
{Ribas}, I., {Tuomi}, M., {Reiners}, A., {et~al.} 2018, \nat, 563, 365

\bibitem[{{Ricker} {et~al.}(2015){Ricker}, {Winn}, {Vanderspek}, {Latham},
  {Bakos}, {Bean}, {Berta-Thompson}, {Brown}, {Buchhave}, {Butler}, {Butler},
  {Chaplin}, {Charbonneau}, {Christensen-Dalsgaard}, {Clampin}, {Deming},
  {Doty}, {De Lee}, {Dressing}, {Dunham}, {Endl}, {Fressin}, {Ge}, {Henning},
  {Holman}, {Howard}, {Ida}, {Jenkins}, {Jernigan}, {Johnson}, {Kaltenegger},
  {Kawai}, {Kjeldsen}, {Laughlin}, {Levine}, {Lin}, {Lissauer}, {MacQueen},
  {Marcy}, {McCullough}, {Morton}, {Narita}, {Paegert}, {Palle}, {Pepe},
  {Pepper}, {Quirrenbach}, {Rinehart}, {Sasselov}, {Sato}, {Seager},
  {Sozzetti}, {Stassun}, {Sullivan}, {Szentgyorgyi}, {Torres}, {Udry}, \&
  {Villasenor}}]{TESS_Ricker}
{Ricker}, G.~R., {Winn}, J.~N., {Vanderspek}, R., {et~al.} 2015, Journal of
  Astronomical Telescopes, Instruments, and Systems, 1, 014003

\bibitem[{{Santerne} {et~al.}(2019){Santerne}, {Malavolta}, {Kosiarek}, {Dai},
  {Dressing}, {Dumusque}, {Hara}, {Lopez}, {Mortier}, {Vanderburg},
  {Adibekyan}, {Armstrong}, {Barrado}, {Barros}, {Bayliss}, {Berardo},
  {Boisse}, {Bonomo}, {Bouchy}, {Brown}, {Buchhave}, {Butler}, {Collier
  Cameron}, {Cosentino}, {Crane}, {Crossfield}, {Damasso}, {Deleuil}, {Delgado
  Mena}, {Demangeon}, {D{\'\i}az}, {Donati}, {Figueira}, {Fulton}, {Ghedina},
  {Harutyunyan}, {H{\'e}brard}, {Hirsch}, {Hojjatpanah}, {Howard}, {Isaacson},
  {Latham}, {Lillo-Box}, {L{\'o}pez-Morales}, {Lovis}, {Martinez Fiorenzano},
  {Molinari}, {Mousis}, {Moutou}, {Nava}, {Nielsen}, {Osborn}, {Petigura},
  {Phillips}, {Pollacco}, {Poretti}, {Rice}, {Santos}, {S{\'e}gransan},
  {Shectman}, {Sinukoff}, {Sousa}, {Sozzetti}, {Teske}, {Udry}, {Vigan},
  {Wang}, {Watson}, {Weiss}, {Wheatley}, \& {Winn}}]{Density_HIP_f}
{Santerne}, A., {Malavolta}, L., {Kosiarek}, M.~R., {et~al.} 2019, arXiv
  e-prints, arXiv:1911.07355

\bibitem[{{Smith} {et~al.}(2012){Smith}, {Stumpe}, {Van Cleve}, {Jenkins},
  {Barclay}, {Fanelli}, {Girouard}, {Kolodziejczak}, {McCauliff}, {Morris}, \&
  {Twicken}}]{PDC_1}
{Smith}, J.~C., {Stumpe}, M.~C., {Van Cleve}, J.~E., {et~al.} 2012, \pasp, 124,
  1000

\bibitem[{{Soubiran} {et~al.}(2018){Soubiran}, {Jasniewicz}, {Chemin},
  {Zurbach}, {Brouillet}, {Panuzzo}, {Sartoretti}, {Katz}, {Le Campion},
  {Marchal}, {Hestroffer}, {Th{\'e}venin}, {Crifo}, {Udry}, {Cropper},
  {Seabroke}, {Viala}, {Benson}, {Blomme}, {Jean-Antoine}, {Huckle}, {Smith},
  {Baker}, {Damerdji}, {Dolding}, {Fr{\'e}mat}, {Gosset}, {Guerrier}, {Guy},
  {Haigron}, {Jan{\ss}en}, {Plum}, {Fabre}, {Lasne}, {Pailler}, {Panem},
  {Riclet}, {Royer}, {Tauran}, {Zwitter}, {Gueguen}, \& {Turon}}]{GAIA_RV_2018}
{Soubiran}, C., {Jasniewicz}, G., {Chemin}, L., {et~al.} 2018, \aap, 616, A7

\bibitem[{{Southworth}(2011)}]{Southworth_database}
{Southworth}, J. 2011, \mnras, 417, 2166

\bibitem[{{Speagle}(2020)}]{dynesty}
{Speagle}, J.~S. 2020, \mnras, 493, 3132

\bibitem[{{Stassun} {et~al.}(2018){Stassun}, {Oelkers}, {Pepper}, {Paegert},
  {De Lee}, {Torres}, {Latham}, {Charpinet}, {Dressing}, {Huber}, {Kane},
  {L{\'e}pine}, {Mann}, {Muirhead}, {Rojas-Ayala}, {Silvotti}, {Fleming},
  {Levine}, \& {Plavchan}}]{TIC_Catalog_Stassun2018}
{Stassun}, K.~G., {Oelkers}, R.~J., {Pepper}, J., {et~al.} 2018, \aj, 156, 102

\bibitem[{{Stassun} \& {Torres}(2018)}]{offset_DR3}
{Stassun}, K.~G. \& {Torres}, G. 2018, \apj, 862, 61

\bibitem[{{Stumpe} {et~al.}(2014){Stumpe}, {Smith}, {Catanzarite}, {Van Cleve},
  {Jenkins}, {Twicken}, \& {Girouard}}]{PDC_2}
{Stumpe}, M.~C., {Smith}, J.~C., {Catanzarite}, J.~H., {et~al.} 2014, \pasp,
  126, 100

\bibitem[{{Stumpe} {et~al.}(2012){Stumpe}, {Smith}, {Van Cleve}, {Twicken},
  {Barclay}, {Fanelli}, {Girouard}, {Jenkins}, {Kolodziejczak}, {McCauliff}, \&
  {Morris}}]{stumpe_2012}
{Stumpe}, M.~C., {Smith}, J.~C., {Van Cleve}, J.~E., {et~al.} 2012, \pasp, 124,
  985

\bibitem[{{Thorngren} {et~al.}(2016){Thorngren}, {Fortney}, {Murray-Clay}, \&
  {Lopez}}]{thorngren2016}
{Thorngren}, D.~P., {Fortney}, J.~J., {Murray-Clay}, R.~A., \& {Lopez}, E.~D.
  2016, \apj, 831, 64

\bibitem[{{Trifonov}(2019)}]{2019ascl.soft06004T}
{Trifonov}, T. 2019, {The Exo-Striker: Transit and radial velocity interactive
  fitting tool for orbital analysis and N-body simulations}

\bibitem[{{Trifonov} {et~al.}(2020){Trifonov}, {Tal-Or}, {Zechmeister},
  {Kaminski}, {Zucker}, \& {Mazeh}}]{Trifonov2020_nzp}
{Trifonov}, T., {Tal-Or}, L., {Zechmeister}, M., {et~al.} 2020, \aap, 636, A74

\bibitem[{{Trotta}(2008)}]{Trotta_2008}
{Trotta}, R. 2008, Contemporary Physics, 49, 71

\bibitem[{{Tsiaras} {et~al.}(2018){Tsiaras}, {Waldmann}, {Zingales},
  {Rocchetto}, {Morello}, {Damiano}, {Karpouzas}, {Tinetti}, {McKemmish},
  {Tennyson}, \& {Yurchenko}}]{tsiaras2018}
{Tsiaras}, A., {Waldmann}, I.~P., {Zingales}, T., {et~al.} 2018, \aj, 155, 156

\bibitem[{{Tuomi} {et~al.}(2019){Tuomi}, {Jones}, {Butler}, {Arriagada},
  {Vogt}, {Burt}, {Laughlin}, {Holden}, {Shectman}, {Crane}, {Thompson},
  {Keiser}, {Jenkins}, {Berdi{\~n}as}, {Diaz}, {Kiraga}, \&
  {Barnes}}]{GJ273_de_Tuomi}
{Tuomi}, M., {Jones}, H.~R.~A., {Butler}, R.~P., {et~al.} 2019, arXiv e-prints,
  arXiv:1906.04644

\bibitem[{{Van Eylen} \& {Albrecht}(2015)}]{VanEylen_ecc}
{Van Eylen}, V. \& {Albrecht}, S. 2015, \apj, 808, 126

\bibitem[{{Van Eylen} {et~al.}(2019){Van Eylen}, {Albrecht}, {Huang},
  {MacDonald}, {Dawson}, {Cai}, {Foreman-Mackey}, {Lundkvist}, {Silva Aguirre},
  {Snellen}, \& {Winn}}]{VanEylen_ecc_2019}
{Van Eylen}, V., {Albrecht}, S., {Huang}, X., {et~al.} 2019, \aj, 157, 61

\bibitem[{{van Leeuwen}(2007)}]{HIP_Catalog}
{van Leeuwen}, F. 2007, \aap, 474, 653

\bibitem[{{Vanderburg} {et~al.}(2016){Vanderburg}, {Becker}, {Kristiansen},
  {Bieryla}, {Duev}, {Jensen-Clem}, {Morton}, {Latham}, {Adams}, {Baranec},
  {Berlind}, {Calkins}, {Esquerdo}, {Kulkarni}, {Law}, {Riddle}, {Salama}, \&
  {Schmitt}}]{HIP41378f}
{Vanderburg}, A., {Becker}, J.~C., {Kristiansen}, M.~H., {et~al.} 2016, \apjl,
  827, L10

\bibitem[{{Vissapragada} {et~al.}(2020){Vissapragada}, {Jontof-Hutter},
  {Shporer}, {Knutson}, {Liu}, {Thorngren}, {Lee}, {Chachan}, {Mawet},
  {Millar-Blanchaer}, {Nilsson}, {Tinyanont}, {Vasisht}, \&
  {Wright}}]{KOI1783c_Vissa2020}
{Vissapragada}, S., {Jontof-Hutter}, D., {Shporer}, A., {et~al.} 2020, \aj,
  159, 108

\bibitem[{{Vogt} {et~al.}(2014){Vogt}, {Radovan}, {Kibrick}, {Butler},
  {Alcott}, {Allen}, {Arriagada}, {Bolte}, {Burt}, {Cabak}, {Chloros},
  {Cowley}, {Deich}, {Dupraw}, {Earthman}, {Epps}, {Faber}, {Fischer}, {Gates},
  {Hilyard}, {Holden}, {Johnston}, {Keiser}, {Kanto}, {Katsuki}, {Laiterman},
  {Lanclos}, {Laughlin}, {Lewis}, {Lockwood}, {Lynam}, {Marcy}, {McLean},
  {Miller}, {Misch}, {Peck}, {Pfister}, {Phillips}, {Rivera}, {Sandford},
  {Saylor}, {Stover}, {Thompson}, {Walp}, {Ward}, {Wareham}, {Wei}, \&
  {Wright}}]{APF_Vogt2014}
{Vogt}, S.~S., {Radovan}, M., {Kibrick}, R., {et~al.} 2014, \pasp, 126, 359

\bibitem[{{Weiss} {et~al.}(2013){Weiss}, {Marcy}, {Rowe}, {Howard}, {Isaacson},
  {Fortney}, {Miller}, {Demory}, {Fischer}, {Adams}, {Dupree}, {Howell},
  {Kolbl}, {Johnson}, {Horch}, {Everett}, {Fabrycky}, \& {Seager}}]{KOI-94}
{Weiss}, L.~M., {Marcy}, G.~W., {Rowe}, J.~F., {et~al.} 2013, \apj, 768, 14

\bibitem[{{Wolfgang} {et~al.}(2016){Wolfgang}, {Rogers}, \&
  {Ford}}]{Wolfgang_2016}
{Wolfgang}, A., {Rogers}, L.~A., \& {Ford}, E.~B. 2016, \apj, 825, 19

\bibitem[{{Xie} {et~al.}(2016){Xie}, {Dong}, {Zhu}, {Huber}, {Zheng}, {De Cat},
  {Fu}, {Liu}, {Luo}, {Wu}, {Zhang}, {Zhang}, {Zhou}, {Cao}, {Hou}, {Wang}, \&
  {Zhang}}]{Xie_ecc}
{Xie}, J.-W., {Dong}, S., {Zhu}, Z., {et~al.} 2016, Proceedings of the National
  Academy of Science, 113, 11431

\bibitem[{{Yu} {et~al.}(2021){Yu}, {He}, {Zhang}, {H{\"o}rst}, {Dymont},
  {McGuiggan}, {Moses}, {Lewis}, {Fortney}, {Gao}, {Kempton}, {Moran},
  {Morley}, {Powell}, {Valenti}, \& {Vuitton}}]{yu2021}
{Yu}, X., {He}, C., {Zhang}, X., {et~al.} 2021, Nature Astronomy, 5, 822

\bibitem[{{Zechmeister} \& {K{\"u}rster}(2009)}]{GLS_paper}
{Zechmeister}, M. \& {K{\"u}rster}, M. 2009, \aap, 496, 577

\bibitem[{{Zechmeister} {et~al.}(2018){Zechmeister}, {Reiners}, {Amado},
  {Azzaro}, {Bauer}, {B{\'e}jar}, {Caballero}, {Guenther}, {Hagen}, {Jeffers},
  {Kaminski}, {K{\"u}rster}, {Launhardt}, {Montes}, {Morales}, {Quirrenbach},
  {Reffert}, {Ribas}, {Seifert}, {Tal-Or}, \& {Wolthoff}}]{SERVAL}
{Zechmeister}, M., {Reiners}, A., {Amado}, P.~J., {et~al.} 2018, \aap, 609, A12

\bibitem[{{Zeng} {et~al.}(2019){Zeng}, {Jacobsen}, {Sasselov}, {Petaev},
  {Vanderburg}, {Lopez-Morales}, {Perez-Mercader}, {Mattsson}, {Li}, {Heising},
  {Bonomo}, {Damasso}, {Berger}, {Cao}, {Levi}, \& {Wordsworth}}]{Zeng_models}
{Zeng}, L., {Jacobsen}, S.~B., {Sasselov}, D.~D., {et~al.} 2019, Proceedings of
  the National Academy of Science, 116, 9723

\end{thebibliography}


\begin{appendix}
\label{Sec:Appendix}

\section{Additional figures and tables}

\begin{table}
\caption[width=\textwidth]{
\label{table - TESS GP}
Prior and posterior distributions for PDC-SAP detrending fit. Prior labels $\mathcal{F}$, $\mathcal{N}$, and $\mathcal{J}$ represent fixed, normal, and Jeffrey's distributions, respectively.
}
\centering

\begin{tabular}{lcc}

\hline \hline 
\noalign{\smallskip} 

Parameter & Prior & Posterior \vspace{0.05cm}\\
\hline
\noalign{\smallskip}

$D_{\textit{TESS}}$ & $\mathcal{F}(1)$ & -- \vspace{0.05cm} \\ 
$\mu_{\textit{TESS}}$ (ppm) & $\mathcal{N}(0,0.1)$ & 20$^{+110}_{-100}$ \vspace{0.05cm} \\ 
$\sigma_{\textit{TESS}}$ (ppm) & $\mathcal{J}(10^{-6},10^{6})$ & 135$^{+1}_{-2}$ \vspace{0.05cm} \\ 
$\sigma_{\textit{GP}}$ (ppm) & $\mathcal{J}(10^{-6},10^{6})$ & 970$^{+55}_{-35}$ \vspace{0.05cm} \\ 
$\rho_{\textit{GP}}$ [d] & $\mathcal{J}(10^{-3},10^{3})$ & 10.2$\pm$0.5 \vspace{0.05cm} \\ 

\noalign{\smallskip}
\hline
\end{tabular}

\end{table}

\begin{figure*}
    \centering
    \includegraphics[width=0.49\hsize]{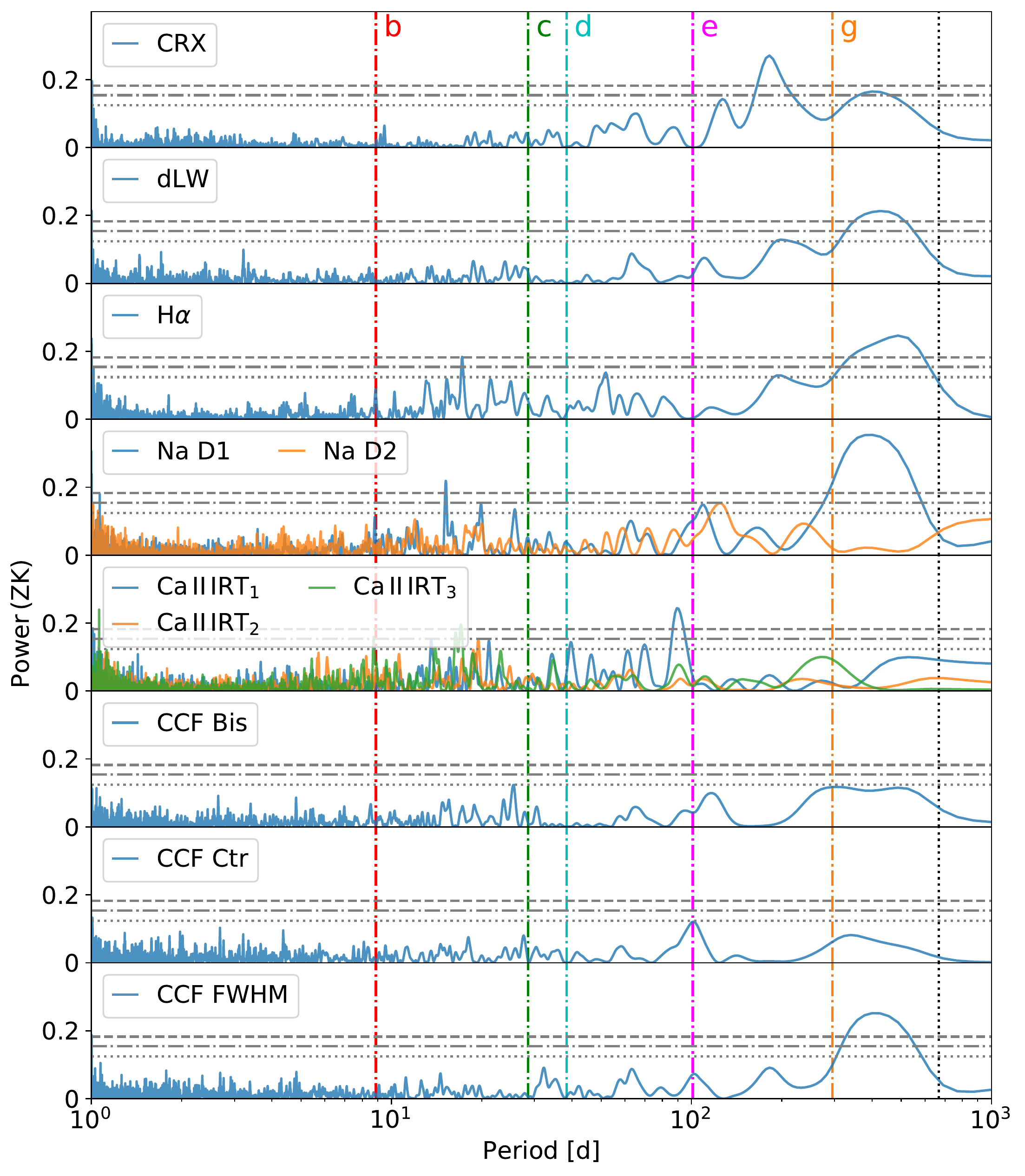}
    \includegraphics[width=0.49\hsize]{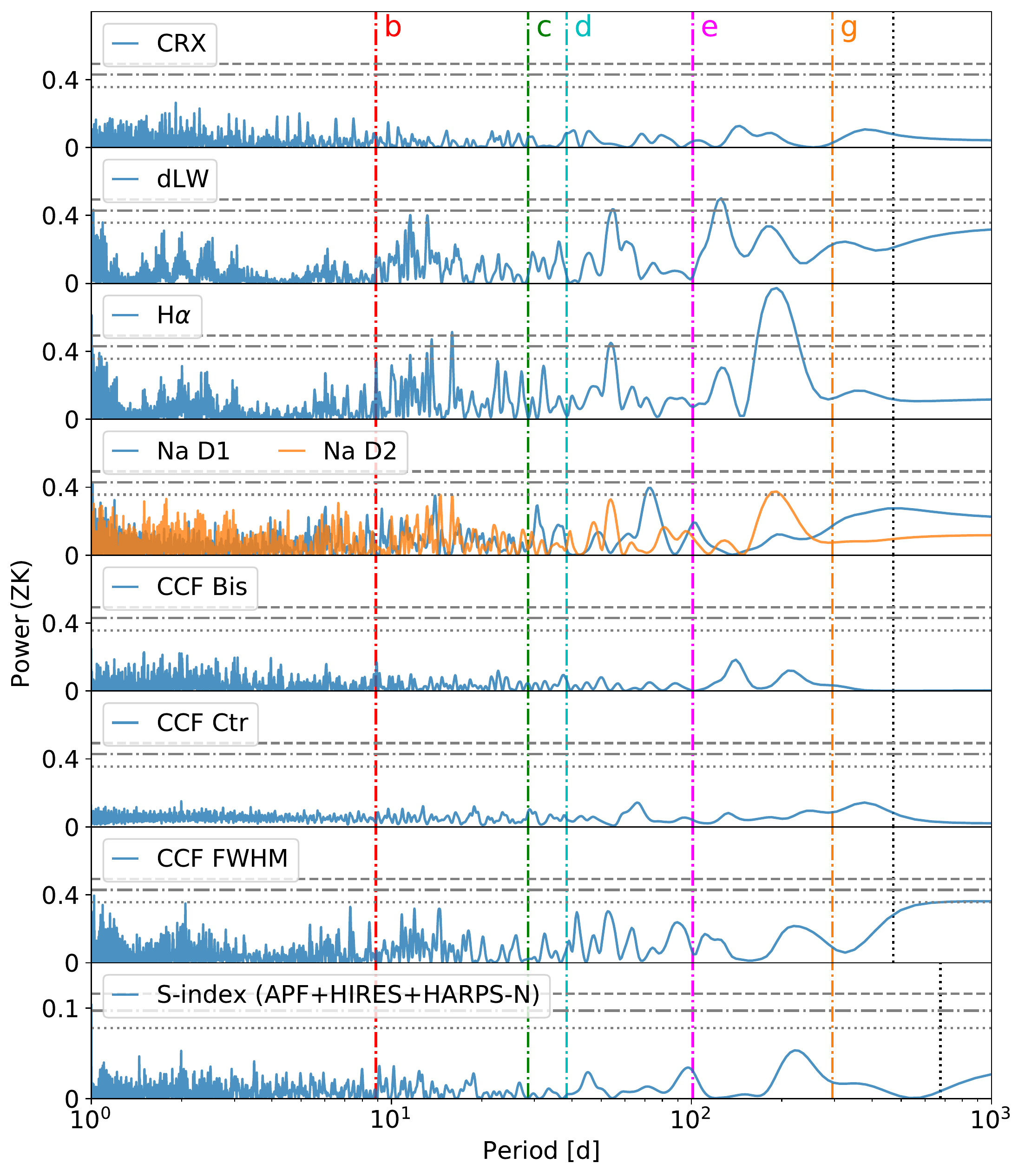}
    \caption{\label{fig: GLS Activity}
    Generalised Lomb-Scargle periodograms of the activity indices from CARMENES (\textit{left}) and HARPS-N (\textit{right}), and S-index (\textit{bottom right}) from APF, HIRES, and HARPS-N. In all panels, the broken vertical lines indicate the planetary signals at 8.9 (red), 28.6 (green), 38.4 (cyan), 101 (magenta), and 280 (orange) days. In all panels, the 10\%, 1\%, and 0.1\% FAP levels are indicated by dotted, dash-dotted, and dashed  grey lines, respectively. The vertical black dotted line marks the baseline for each dataset. We highlight the different scale in the y axis in each panel.
    }
\end{figure*}

\begin{figure*}
    \centering
    \includegraphics[width=0.49\hsize]{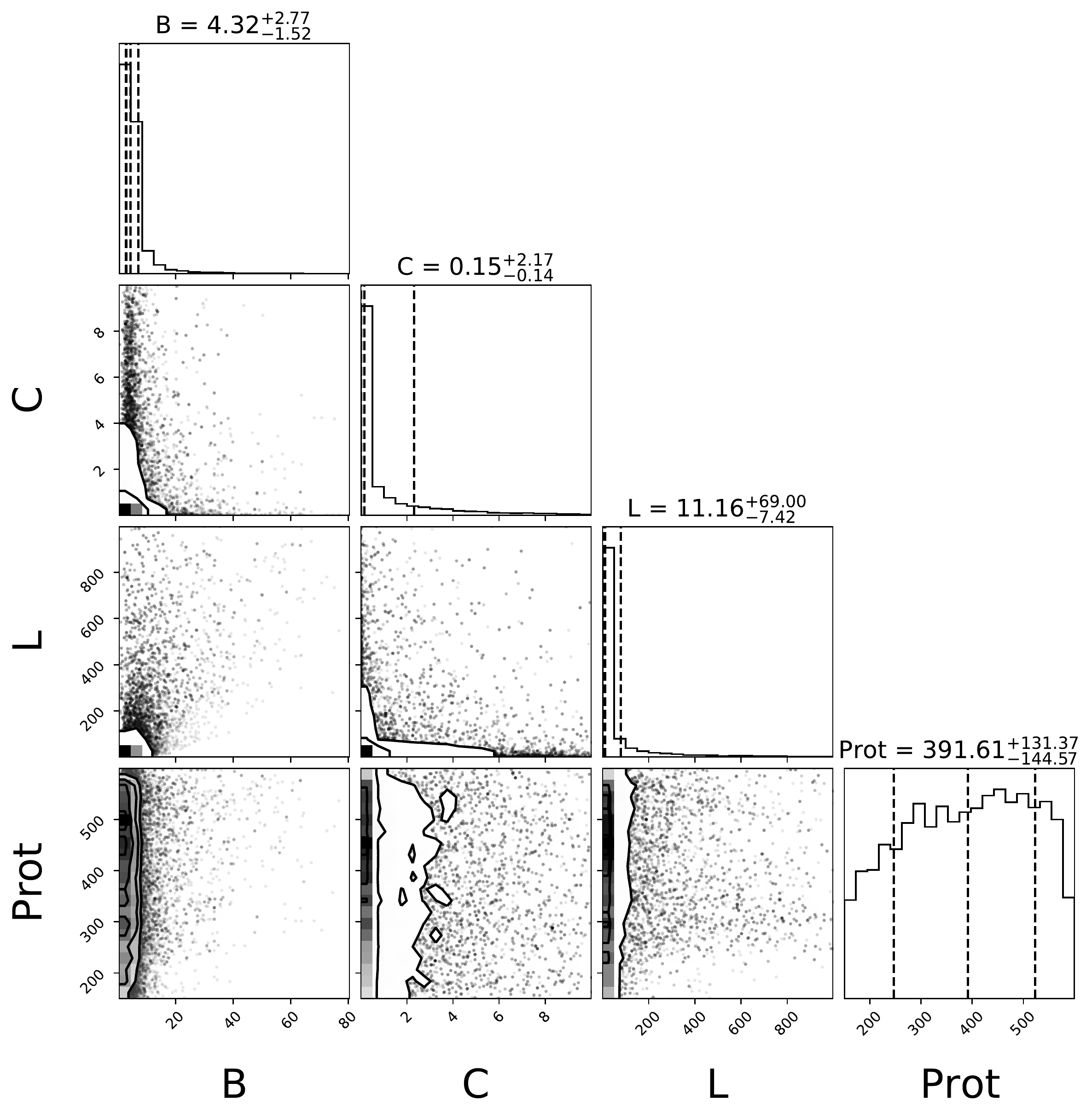}
    \includegraphics[width=0.49\hsize]{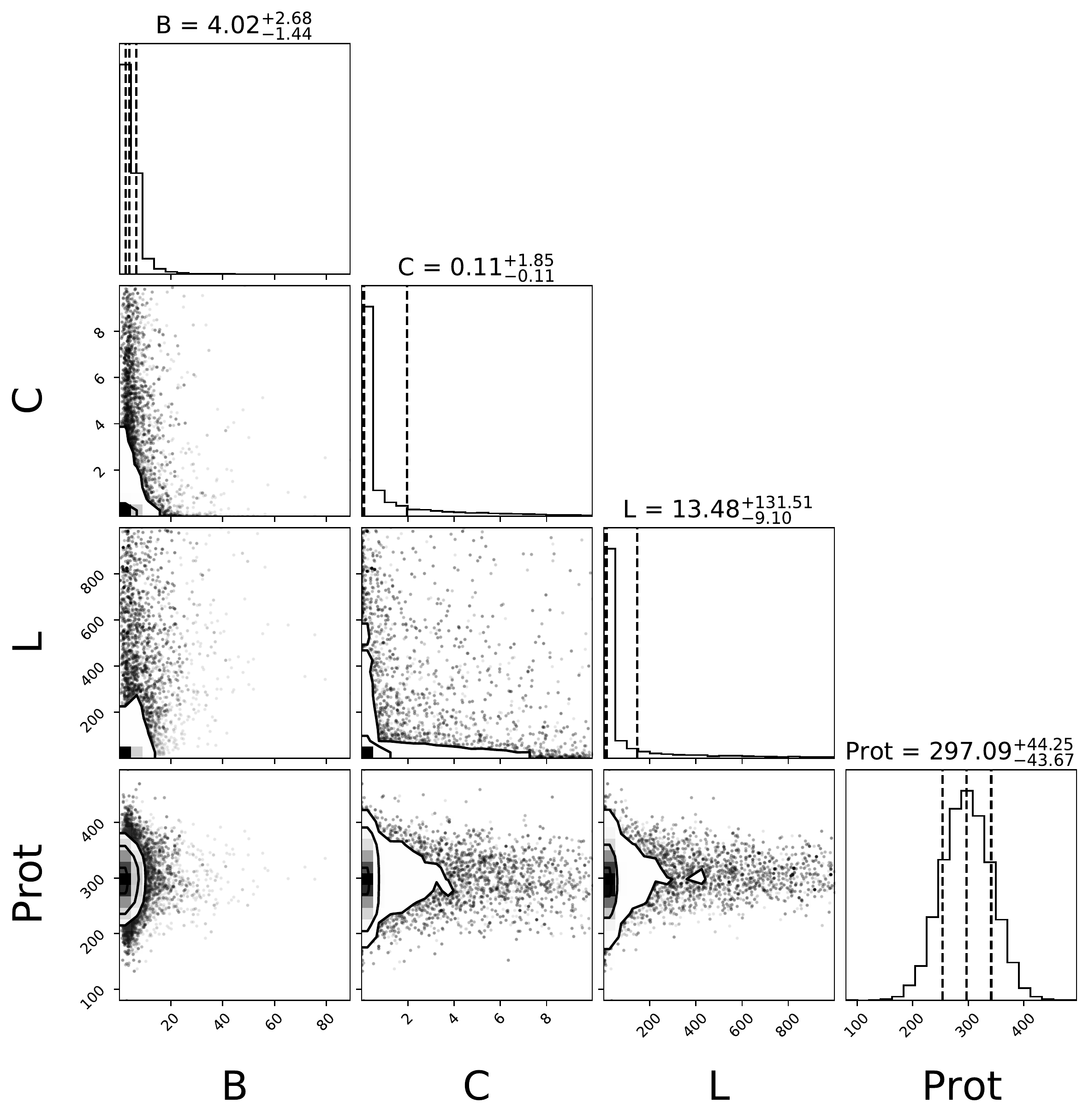}
    \caption{\label{fig: GP CORNER PLOTS}
    Posterior distribution of hyperparameters from the uninformative (left) and constrained (right) GP$_{\mathrm{qp}}$ models used in Sect.\,\ref{sect: RV ANALISIS}. Prior distributions are in Table\,\ref{table - GP hyperparameters}.
    }
\end{figure*}

\begin{figure}
    \centering
    \includegraphics[width=\hsize]{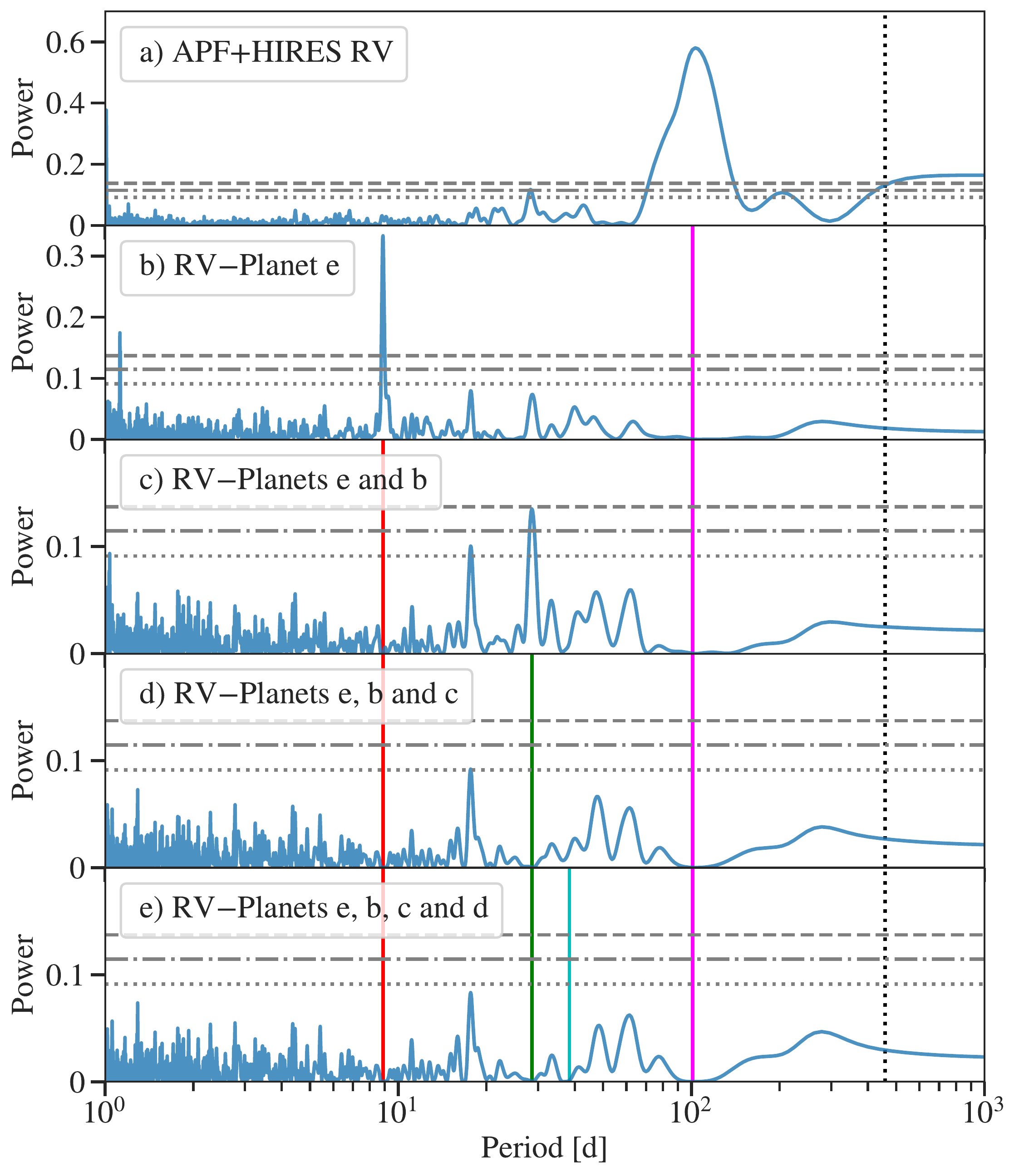}
    \caption{\label{fig: GLS LUBIN} Reproduction of the RV analyses presented in Fig.\,1 from \citetalias{Lubin_HD191939}. All the models include quadratic and linear terms to account for the long-term trend.
    $_{}(a)$ GLS periodogram of APF and HIRES datasets.
    $(b)$ GLS periodogram of the RV residuals after fitting the 101\,d signal (vertical magenta line).
    $(c)$ GLS periodogram of the RV residuals after simultaneously fitting the 8.8\,d (vertical red line) and 101\,d signals.
    $(d)$ GLS periodogram of the RV residuals after simultaneously fitting the 8.8\,d, 28.6\,d (vertical green line), and 101\,d signals.
    $(e)$ GLS periodogram of the RV residuals after simultaneously fitting the 8.8\,d, 28.6\,d, 38\,d (vertical cyan line), and 101\,d signals.
    In all panels, the 10\%, 1\%, and 0.1\% FAP levels are indicated by dotted, dash-dotted, and dashed 
grey  horizontal lines, respectively. The vertical black dotted line indicates the dataset baseline. We highlight the different scale in the y axis in each panel.
    }
\end{figure}

\begin{table}
\caption[width=\textwidth]{
\label{table - GP hyperparameters}
Prior and posterior distributions for the RV GP models explored in Sect.\,\ref{sect: RV ANALISIS}. Prior labels $\mathcal{U}$, $\mathcal{N}$, and $\mathcal{J}$ represent uniform, normal, and Jeffrey's distributions, respectively. Corner plots for the quasi-periodic model hyperparameters are shown in Fig.\,\ref{fig: GP CORNER PLOTS}.
}
\centering

\begin{tabular}{lcc}

\hline \hline 
\noalign{\smallskip} 

Hyperparameter & Prior & Posterior \vspace{0.05cm}\\
\hline
\noalign{\smallskip}

\multicolumn{3}{c}{Exponential model}\\
\noalign{\smallskip} 
\noalign{\smallskip} 

$\sigma$ [$\mathrm{m\,s^{-1}}$] & $\mathcal{U}(0,100)$ &  3.3$^{+ 0.8 }_{- 0.7 }$  \vspace{0.05cm}\\
$\tau$ [d] & $\mathcal{J}(0,10^{3})$ &  1.3$^{+ 0.8 }_{- 0.3 }$   \vspace{0.05cm}\\

\noalign{\smallskip} 
\multicolumn{3}{c}{Matern model}\\
\noalign{\smallskip} 
\noalign{\smallskip} 

$\sigma$ [$\mathrm{m\,s^{-1}}$] & $\mathcal{U}(0,100)$ &  1.7$^{+ 0.5 }_{- 0.4 }$  \vspace{0.05cm}\\
$\rho$ [d] & $\mathcal{J}(0,10^{3})$ &  15$^{+ 27 }_{- 11 }$  \vspace{0.05cm}\\

\noalign{\smallskip} 
\multicolumn{3}{c}{Quasi-periodic models}\\
\noalign{\smallskip} 
\noalign{\smallskip} 

B [$\mathrm{m\,s^{-1}}$] & $\mathcal{U}(0,100)$ &  4.02$^{+ 2.7 }_{- 1.4 }$  \vspace{0.05cm}\\
C [$\mathrm{m\,s^{-1}}$] & $\mathcal{J}(10^{-3},10)$ & 0.11 $^{+1.85 }_{-0.11}$   \vspace{0.05cm}\\
L [d]& $\mathcal{J}(0.1,10^{3})$ & 13 $^{+131 }_{- 10 }$   \vspace{0.05cm}\\
P$_{\mathrm{rot}}$ [d] & $\mathcal{N}(300,50)$ &  297$\pm$44  \vspace{0.05cm}\\

\noalign{\smallskip} 
\noalign{\smallskip} 

B [$\mathrm{m\,s^{-1}}$] & $\mathcal{U}(0,100)$ &  4.2$^{+ 2.8 }_{- 1.5 }$  \vspace{0.05cm}\\
C [$\mathrm{m\,s^{-1}}$] & $\mathcal{J}(10^{-3},10)$ & 0.15 $^{+2.17 }_{-0.14}$   \vspace{0.05cm}\\
L [d]& $\mathcal{J}(0.1,10^{3})$ & 11 $^{+70 }_{- 7 }$   \vspace{0.05cm}\\
P$_{\mathrm{rot}}$ [d] & $\mathcal{U}(150,600)$ &  390$^{+130}_{-145}$    \vspace{0.05cm}\\

\noalign{\smallskip}
\noalign{\smallskip}
\hline
\end{tabular}

\end{table}

\begin{table*}
\caption[width=\textwidth]{
\label{table - Joint Fit PRIORS}
Priors for each parameter used in the \texttt{juliet} joint fit model for HD 191939 planetary system. Prior labels $\mathcal{F}$, $\mathcal{U}$, $\mathcal{N}$, $\mathcal{B}$, and $\mathcal{J}$ represent fixed, uniform, normal, beta and Jeffrey's distributions, respectively.
The parametrization for $(p,b)$ using $(r_1,r_2)$ \citep{Espinoza2018} and $(q_1,q_2)$ quadratic limb darkening \citep{Kipping2013} are both explained in Section\,\ref{sect: JOINT FIT ANALISIS}.
}
\centering

\begin{tabular}{lccccc}

\hline \hline 
\noalign{\smallskip}
Parameter & \textbf{Planet b}  & \textbf{Planet c}  & \textbf{Planet d}  & \textbf{Planet e}  & \textbf{Planet g} \\ 
\noalign{\smallskip} 
\hline 
\noalign{\smallskip} 
\noalign{\smallskip} 
$P$ [d]  & $\mathcal{N}(8.8803,0.0005)$  & $\mathcal{N}(28.5795,0.0005)$ & $\mathcal{N}(38.3531,0.0005)$  & $\mathcal{N}(101.0,2.0)$ & $\mathcal{N}(300.0,20.0)$ \vspace{0.05cm} \\ 
$t_{0}$\,$^{(a)}$ & $\mathcal{N}(2443.5414,0.005)$  & $\mathcal{N}(2440.5455,0.005)$ & $\mathcal{N}(2433.9098,0.005)$  & $\mathcal{N}(2347.5,2.0)$ & $\mathcal{N}(2400.0,50.0)$ \vspace{0.05cm} \\ 
$K$ [$\mathrm{m\,s^{-1}}$]  & $\mathcal{U}(0,25)$ & $\mathcal{U}(0,25)$ & $\mathcal{U}(0,25)$ & $\mathcal{U}(0,25)$ & $\mathcal{U}(0,25)$ \vspace{0.05cm} \\  
\textit{ecc} & $\mathcal{B}(1.52,29)$ & $\mathcal{B}(1.52,29)$ & $\mathcal{B}(1.52,29)$ & $\mathcal{B}(1.52,29)$ & $\mathcal{B}(1.52,29)$ \vspace{0.05cm} \\ 
$\omega$ (deg) & $\mathcal{U}(-180,180)$ & $\mathcal{U}(-180,180)$ & $\mathcal{U}(-180,180)$ & $\mathcal{U}(-180,180)$ & $\mathcal{U}(-180,180)$ \vspace{0.05cm} \\ 
$r_{1}$  & $\mathcal{U}(0,1)$  & $\mathcal{U}(0,1)$ & $\mathcal{U}(0,1)$  & -- & -- \vspace{0.05cm} \\  
$r_{2}$  & $\mathcal{U}(0,1)$  & $\mathcal{U}(0,1)$ & $\mathcal{U}(0,1)$  & -- & -- \vspace{0.05cm} \\ 

\noalign{\smallskip} 
\hline \hline 
\noalign{\smallskip} 
\multicolumn{6}{c}{\textbf{ Model Parameters }} \\ 
\noalign{\smallskip} 
\multicolumn{6}{c}{\textit{ Stellar density }} \\ 
$\rho_{\star}$ [kg\,m$^{-3}$] & \multicolumn{4}{c}{ $\mathcal{N}(1370,150)$ } \vspace{0.20cm} \\ 
 
\noalign{\smallskip} 
\multicolumn{6}{c}{\textit{ Photometry parameters }} \\ \noalign{\smallskip}
$D_{\textit{TESS}}$ & \multicolumn{4}{c}{ $\mathcal{F}(1)$ } \vspace{0.05cm} \\ 
$\mu_{\textit{TESS}}$ (ppm) & \multicolumn{4}{c}{ $\mathcal{N}(0,0.1)$ } \vspace{0.05cm} \\ 
$\sigma_{\textit{TESS}}$ (ppm) & \multicolumn{4}{c}{ $\mathcal{J}(0.1,1000)$ } \vspace{0.05cm} \\ 
$q_{1,\textit{TESS}}$ & \multicolumn{4}{c}{ $\mathcal{U}(0,1)$ } \vspace{0.05cm} \\ 
$q_{2,\textit{TESS}}$ & \multicolumn{4}{c}{ $\mathcal{U}(0,1)$ } \vspace{0.20cm} \\ 
 
\noalign{\smallskip} 
\multicolumn{6}{c}{\textit{ RV parameters }} \\ \noalign{\smallskip} 
Intercept {$\gamma$} [$\mathrm{m\,s^{-1}}$] & \multicolumn{4}{c}{ $\mathcal{F}(-300)$ } \vspace{0.05cm} \\
Slope {$\dot \gamma$} [$\mathrm{m\,s^{-1}\,d^{-1}}$] & \multicolumn{4}{c}{ $\mathcal{U}(-1,1)$ } \vspace{0.05cm} \\ 
Curve {$\ddot \gamma$} [$\mathrm{m\,s^{-1}\,d^{-2}}$] & \multicolumn{4}{c}{ $\mathcal{U}(-0.1,0.1)$ } \vspace{0.05cm} \\ 
$\gamma_{\mathrm{APF}}$ [$\mathrm{m\,s^{-1}}$] & \multicolumn{4}{c}{ $\mathcal{U}(-10,10)$ } \vspace{0.05cm} \\ 
$\sigma_{\mathrm{APF}}$ [$\mathrm{m\,s^{-1}}$] & \multicolumn{4}{c}{ $\mathcal{J}(0.1,10)$ } \vspace{0.05cm} \\ 
$\gamma_{\mathrm{HIRES}}$ [$\mathrm{m\,s^{-1}}$] & \multicolumn{4}{c}{ $\mathcal{U}(-10,10)$ } \vspace{0.05cm} \\ 
$\sigma_{\mathrm{HIRES}}$ [$\mathrm{m\,s^{-1}}$] & \multicolumn{4}{c}{ $\mathcal{J}(0.1,10)$ } \vspace{0.05cm} \\ 
$\gamma_{\mathrm{CARMENES}}$ [$\mathrm{m\,s^{-1}}$] & \multicolumn{4}{c}{ $\mathcal{U}(-10,10)$ } \vspace{0.05cm} \\ 
$\sigma_{\mathrm{CARMENES}}$ [$\mathrm{m\,s^{-1}}$] & \multicolumn{4}{c}{ $\mathcal{J}(0.1,10)$ } \vspace{0.05cm} \\ 
$\gamma_{\mathrm{HARPS-N}}$ [$\mathrm{m\,s^{-1}}$] & \multicolumn{4}{c}{ $\mathcal{U}(-10,10)$ } \vspace{0.05cm} \\ 
$\sigma_{\mathrm{HARPS-N}}$ [$\mathrm{m\,s^{-1}}$] & \multicolumn{4}{c}{ $\mathcal{J}(0.1,10)$ } \vspace{0.05cm} \\ 
 
\noalign{\smallskip} 
\hline \hline

\end{tabular}
\tablefoot{
$^{(a)}$ Central time of transit ($t_0$) units are BJD\,$-$\,2\,457\,000.
}
\end{table*}

\begin{figure*}
    \centering
    \includegraphics[width=\textwidth]{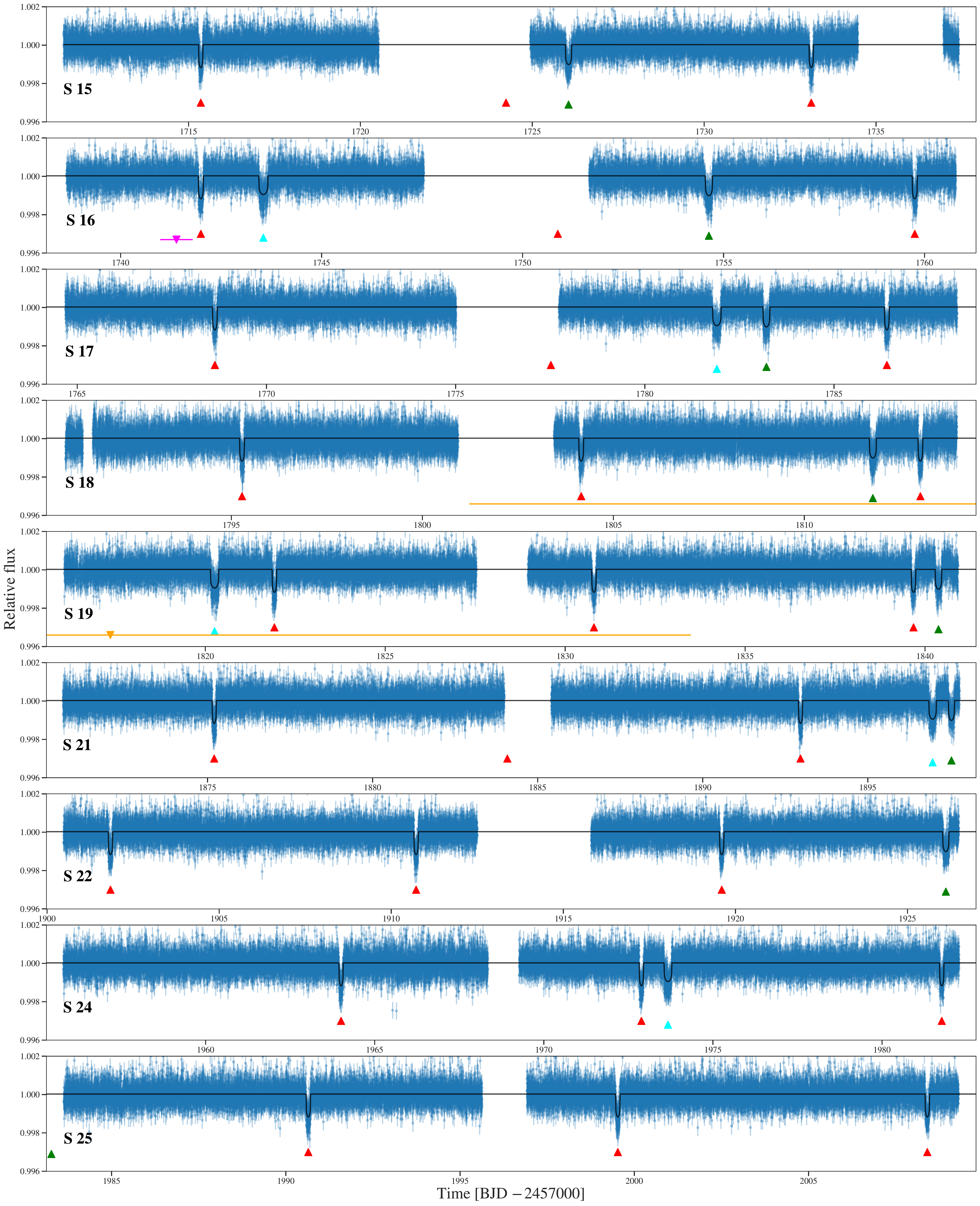}
    \caption{\label{fig: TESS SECTORS} \textit{TESS} photometry from Sectors 15--19, 21, 22, 24, and 25 along with the best-fit model (see Fig.\,\ref{fig: TESS SECTORS 41 48} for Sectors 41 and 48). Upward-pointing triangles mark the transits for HD 191939\,b (red), c (cyan), and d (green). Downward-pointing triangles with error bars mark the expected $t_0$ and $\pm \, 1\sigma$ uncertainty for the non-transiting planets HD 191939\,e (magenta) and g (orange).}
\end{figure*}

\begin{figure*}
    \centering
    \includegraphics[width=\textwidth]{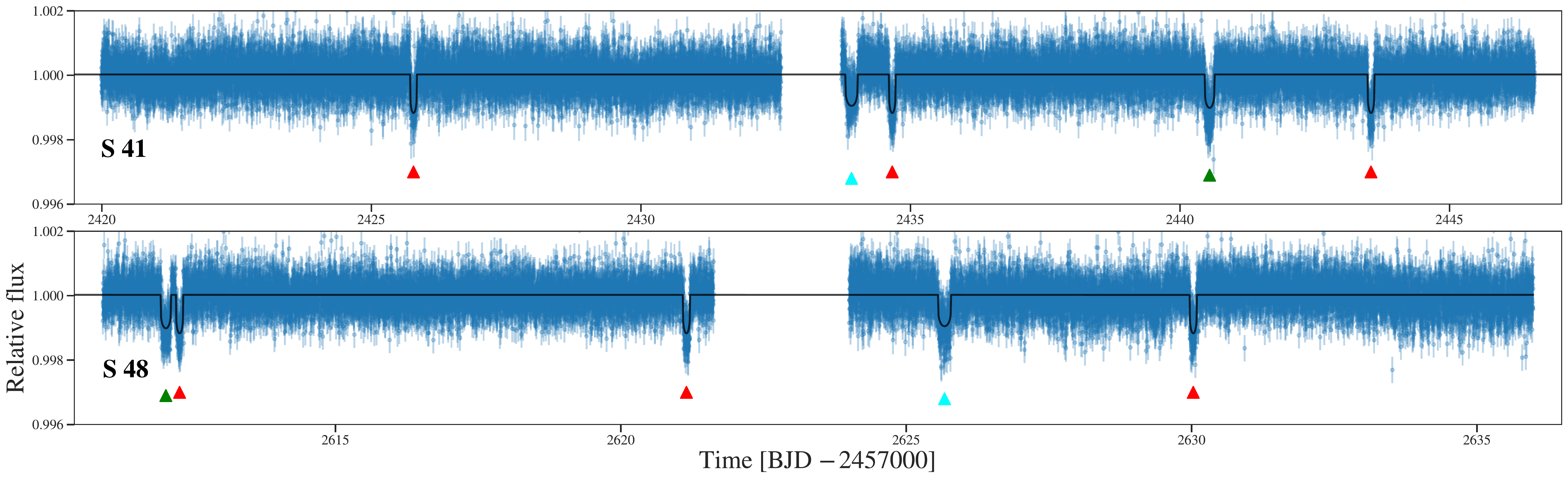}
    \caption{\label{fig: TESS SECTORS 41 48}
    As in Fig.\,\ref{fig: TESS SECTORS} but for Sectors 41 and 48.}
\end{figure*}

\begin{figure*}
\includegraphics[width=1\linewidth]{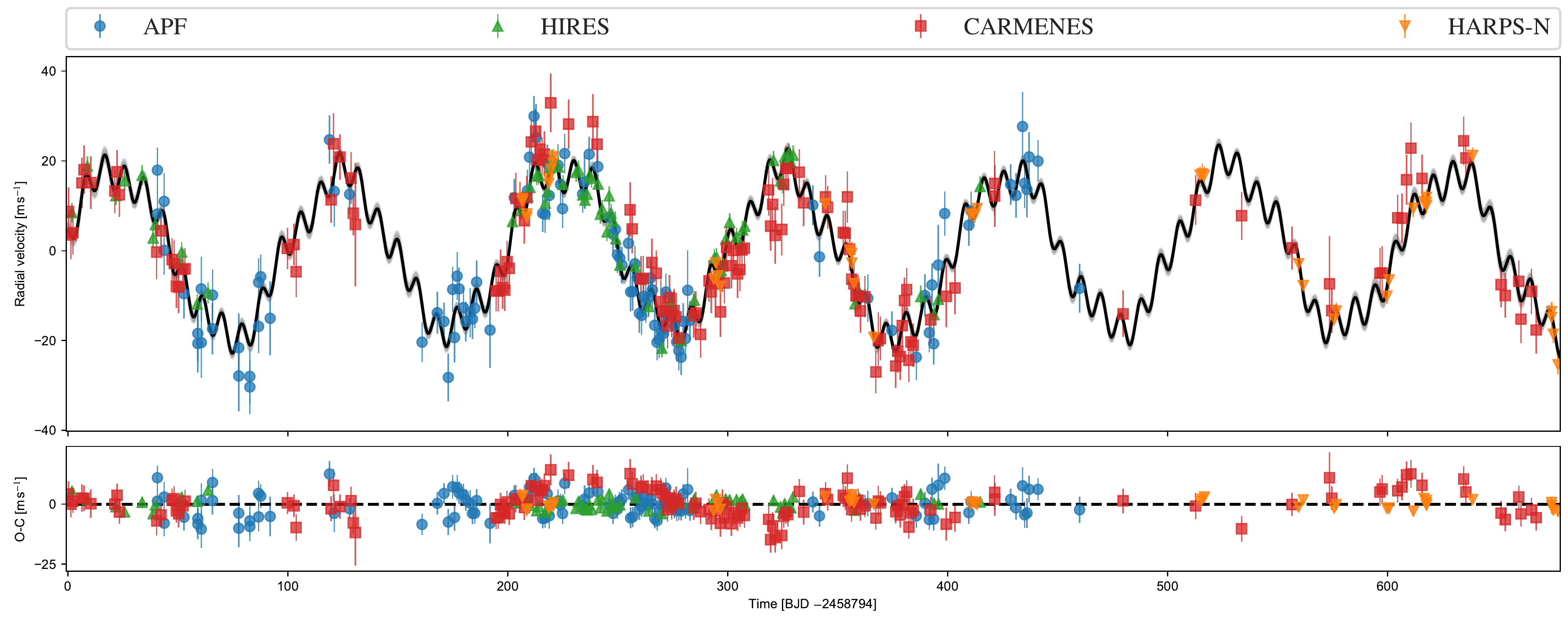}
\caption{
\label{fig: RV MODEL JF} RV results from the joint fit model. Detrended RV time series (\textit{top panel}) of APF (blue circles), HIRES (green up triangles), CARMENES (red squares), and HARPS-N (orange down triangles) along with the best-fit Keplerian model (black line) and the $3\sigma$ confidence interval (shaded grey area). The error bars include the instrumental jitter term added in quadrature.
}
\end{figure*}

\begin{figure*}
\includegraphics[width=1\linewidth]{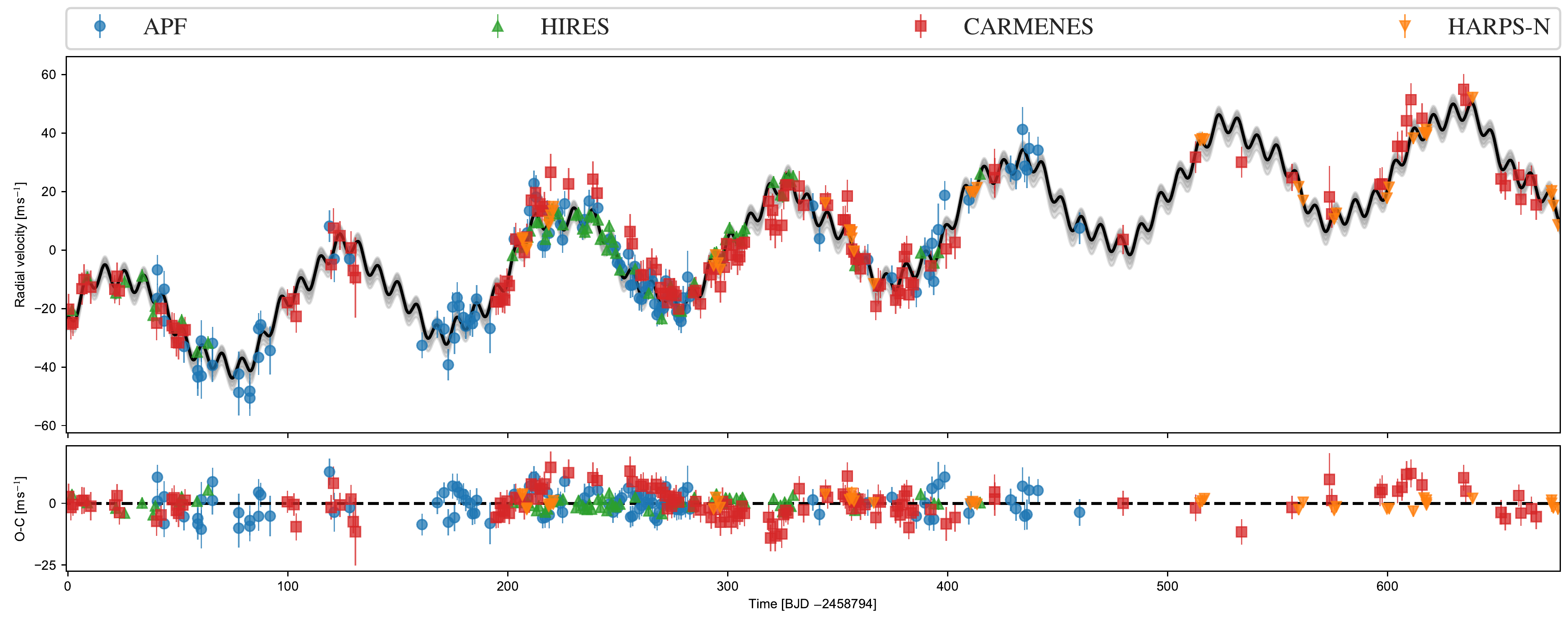}
\caption{
\label{fig: PLANET F fit model}
As in Fig.\,\ref{fig: RV MODEL JF} but considering a Keplerian signal for planet f (see Sect.\,\ref{sect: Planet f}).
}
\end{figure*}

\begin{figure}
    \centering
    \includegraphics[width=\hsize]{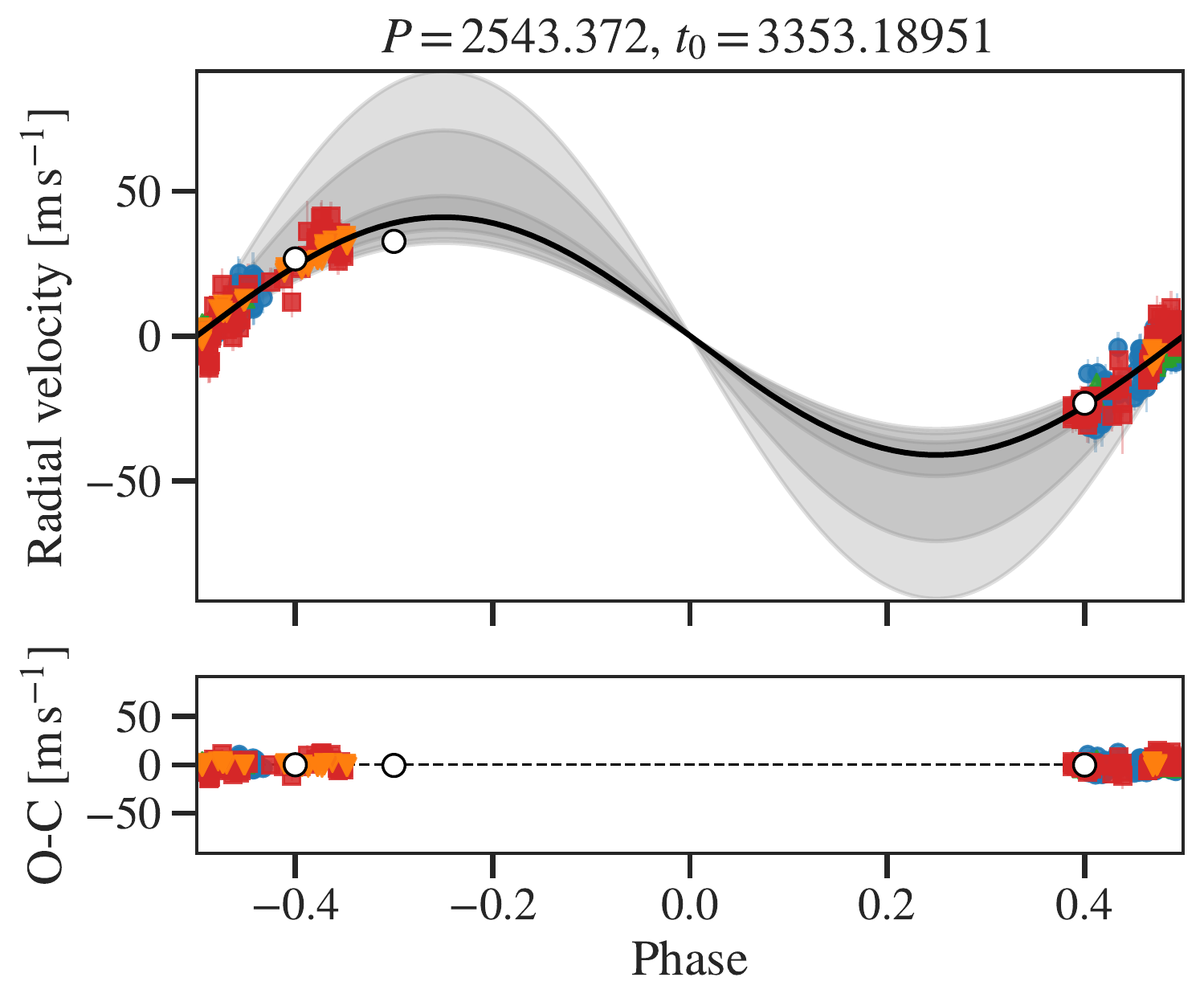}
    \caption{\label{fig: PLANET F phase-folded}
    RVs phase-folded to the $P$ and $t_0$ (shown above the panel, $P$ units are days and $t_0$ units are BJD$-2\,457\,000$) for planet f along with the best-fit model (black line) and the $1\sigma$, $2\sigma$, and $3\sigma$ confidence intervals (shaded grey areas).
    }
\end{figure}

\begin{figure}
    \centering
    \includegraphics[width=\hsize]{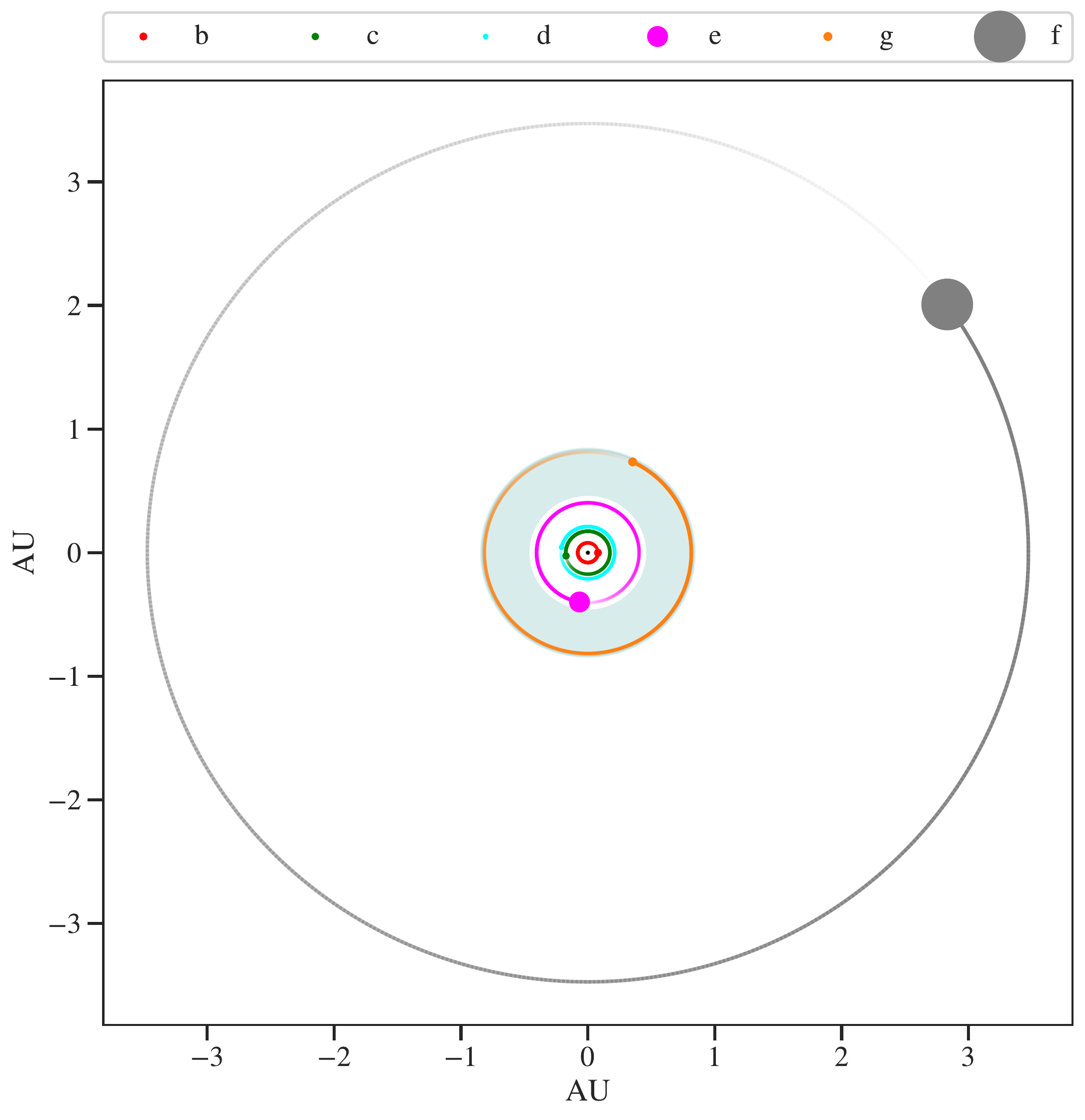}
    \caption{\label{fig: HZ planet F}
    As in Fig.\,\ref{fig: HZ diagram} but including planet f. Planet f parameters are from Table\,\ref{table - Joint Fit RESULTS}.
    }
\end{figure}

\end{appendix}

\end{document}